\documentclass[12pt]{article}

\usepackage[a4paper, margin=1in]{geometry}
\usepackage{enumitem}
\usepackage{titlesec}
\usepackage{kotex} 
\usepackage{datetime} 
\usepackage{amsmath, amssymb, amsthm, bm}
\usepackage{bbm}
\usepackage{mathtools}
\usepackage{graphicx}
\usepackage[colorlinks=true,linkcolor=blue!50!black,citecolor=blue!50!black,urlcolor=blue!50!black,linktoc=all]{hyperref}
\usepackage{setspace}
\usepackage[authoryear]{natbib} 
\usepackage{algorithm}
\usepackage{algorithmic}
\usepackage{xcolor}
\usepackage{float}
\usepackage{subcaption}
\usepackage{caption}
\usepackage{makecell}
\usepackage{multirow}
\usepackage{booktabs}
\usepackage{threeparttable}
\usepackage{diagbox}
\usepackage{tikz}
\usetikzlibrary{arrows.meta}
\usepackage{indentfirst} 
\usepackage{mathrsfs} 
\usepackage{ulem, cancel} 
\usepackage{mdframed}
\makeatletter
\newcommand{\manualitemlabel}[2]{%
  \def\@currentlabel{#1}%
  \label{#2}%
}

\makeatother

\theoremstyle{plain}
\newtheorem{theorem}{Theorem}

\newtheorem{corollary}{Corollary}

\definecolor{TrailCenter}{HTML}{284B63}
\definecolor{TrailRetained}{HTML}{2D746A}
\definecolor{TrailActive}{HTML}{E2EEEB}
\definecolor{TrailInk}{HTML}{283F4D}
\definecolor{TrailLine}{HTML}{B2C8C8}

\tikzset{
  trail/center/.style={
    draw=TrailCenter,
    fill=TrailCenter,
    rounded corners=1.5pt,
    line width=0.4pt
  },
  trail/retained/.style={
    draw=TrailRetained,
    fill=TrailRetained,
    rounded corners=1.5pt,
    line width=0.4pt
  },
  trail/active/.style={
    draw=TrailLine,
    fill=TrailActive,
    rounded corners=1.5pt,
    line width=0.4pt
  },
  trail/stage/.style={
    anchor=east,
    font=\scriptsize\bfseries,
    text=TrailInk
  },
  trail/white/.style={
    font=\scriptsize\bfseries,
    text=white
  },
  trail/dark/.style={
    font=\scriptsize,
    text=TrailInk
  }
}

\title{\textbf{Empirical Null Estimation via Adaptive Shifting and Tail Reconstruction for Large-Scale Multiple Testing}}

\author{
Kyuram Kim$^{1}$, 
Junsik Kim$^{2}$\thanks{Corresponding author: Department of Statistics, Duksung Women's University, 33, Samyang-ro 144-gil, Dobong-gu, Seoul, Republic of Korea. E-mail: \texttt{junsik@duksung.ac.kr}}, 
Junyong Park$^{1}$\\[0.5em]
$^{1}$Seoul National University, 
$^{2}$Duksung Women's University
}

\date{}

\begin{document}

\maketitle

\begin{abstract}
Dependence among test statistics can produce complex empirical null shapes that are poorly captured by conventional parametric models. We propose \texttt{TRAIL} (Tail Reconstruction via Adaptive Iterative Learning), a method that estimates the pooled conditional null under a Gaussian factor model using a Gaussian--Hermite expansion without requiring the null correlation matrix as input. The procedure progressively reconstructs the potentially contaminated tail by generating pseudo-observations and retaining them in successive blocks from the initial cutoff outward. Its implementation, \texttt{TRAIL\_BIC}, jointly selects the Gaussian reference location and Hermite truncation order using a BIC-type criterion. We characterize its limiting behavior and establish pooled-null consistency, together with asymptotic false discovery rate control and recovery of pooled oracle power under suitable conditions. Simulations across diverse dependence structures and null shapes show generally more stable false discovery behavior than competing empirical null methods with competitive power. An application to 48 coronary artery disease studies illustrates the method's flexibility in capturing complicated empirical shapes and its ability to identify more discoveries on average than the competing empirical null procedures.
\end{abstract}


\section{Introduction}

Large-scale multiple testing has become a fundamental component of modern scientific analysis, with prominent applications in genomics, neuroimaging, and other high-dimensional settings where a large number of hypotheses are examined simultaneously. The goal is to identify nonnull effects while controlling the false discovery rate (FDR), the expected proportion of false rejections among all rejections. A standard approach is to regard these statistics as arising from a mixture of null and non-null populations and to construct $p$-values using the null distribution. In practice, however, the theoretical null distribution may not adequately describe the observed null behavior. This motivates estimating the null distribution directly from the data and using the estimated distribution for subsequent multiple testing.

The need for a flexible empirical null becomes particularly important when the test statistics are dependent. Even when each null statistic marginally follows standard normal distribution, dependence can cause the empirical distribution within a single realization to differ substantially from the standard normal distribution \citep{Efron2007,Schwartzman2010,AzrielSchwartzman2015}. Such discrepancies may involve not only changes in location and scale but also more complicated features including asymmetry and multimodality, which are difficult to capture using conventional Gaussian empirical-null models \citep{Efron2004} or generalized-normal distributions \citep{LiangLiuWang2007}.

To describe such behavior under dependence, we adopt a Gaussian factor model with heterogeneous loadings, following the conditional distribution perspective of \citet{AzrielSchwartzman2015}. Our estimation target is the pooled conditional null, defined as the average of the individual null densities conditional on the common factor realized in the experiment. The resulting pool is a Gaussian mixture whose shape depends on the realized factor and the loading configuration. We represent this pooled density through a Gaussian--Hermite expansion, connecting the factor-based target to the Gaussian-derivative representation of \citet{Schwartzman2010}, also employed by \citet{SunStephens2018} and \citet{tailbic}. Estimating the Hermite coefficients directly provides a flexible model for the realized pooled null without requiring the null correlation matrix, individual factor loadings, or realized factor scores as inputs.

A remaining challenge is to estimate the null-tail probabilities while limiting the influence of nonnull observations. Conditional likelihood fitting uses a predominantly null central region as in the Gaussian empirical-null approach of \citet{Efron2007SizePower}. Alternatively, \citet{tailbic} repeatedly replace the designated tail observations with pseudo-observations generated from the current null estimate. However, under a flexible Gaussian--Hermite representation, a good conditional fit to the center does not guarantee stable tail extrapolation. Repeated replacement of the entire pseudo-tail may also propagate finite-sample fitting errors, potentially distorting the estimated right-tail shape and the resulting null-tail probabilities. These difficulties are particularly relevant to multiple testing, where estimated null-tail probabilities determine individual $p$-values and affect FDR control.

We propose \texttt{TRAIL} (Tail Reconstruction via Adaptive Iterative Learning), a progressive reconstruction method for flexible null density estimation. This method retains successive blocks of pseudo-observations from the initial cutoff outward and regenerates only the remaining active tail. This partial-replacement strategy is designed to stabilize the contribution of generated observations to subsequent fits. Furthermore, in order to enrich the finite Gaussian--Hermite approximation, we also allow the location of the Gaussian reference to be shifted. The implementation, \texttt{TRAIL\_BIC}, jointly selects the reference location and model complexity order using a BIC--type criterion evaluated on the reconstructed data, with the aim of improving the accuracy of the resulting null density estimator in finite samples.

Our theoretical analysis establishes reconstruction tracking and pooled-null consistency for a regularized version of \texttt{TRAIL}. Under suitable conditions, the resulting Benjamini--Hochberg procedure \citep{BenjaminiHochberg1995} achieves asymptotic FDR control and matches the asymptotic power of an oracle procedure that knows the true pooled conditional null, while using the same observations for estimation and testing. We also provide sufficient conditions for power to converge to one.

This paper is organized as follows. Section~\ref{sec:model} introduces the Gaussian factor model, the pooled conditional null, and its Gaussian--Hermite representation. Section~\ref{sec:method} presents progressive tail reconstruction, joint shift--order selection, and the associated multiple-testing procedure. Section~\ref{sec:theoretical-rationale} establishes reconstruction tracking, pooled-null consistency, and pooled-oracle recovery. Section~\ref{sec:numerical} evaluates the methods through numerical experiments under diverse dependence structures and null shapes. Section~\ref{sec:real-data} applies the methods to coronary artery disease association statistics, and Section~\ref{sec:discussion} concludes with a discussion. Detailed conditions, technical lemmas, and proofs are provided in the supplementary note.

\section{Model Setup}
\label{sec:model}

Suppose that we observe test statistics $Z_1,\cdots,Z_n$ arising from a large-scale multiple testing problem, where each test statistic corresponds to a hypothesis test on a population mean. We throughout focus on right-sided tests, for which larger values of $Z_i$ provide stronger evidence against the corresponding null hypothesis.
\[
H_{0i} : \mu_i = 0 \quad \text{vs} \quad H_{1i}:\mu_i>0, \qquad i = 1, \cdots , n
\]

A common working model for the distribution of a randomly selected test statistic $Z_i$ is the two-group mixture
\begin{equation}
 Z_i \sim \pi_0 f_0 + (1-\pi_0) f_1,
 \label{eq:two_group}
\end{equation}
where $f_0$ is the null density, $f_1$ is the non-null density, and $\pi_0$ is the proportion of true nulls. In our setting, each null statistic has marginal distribution $N(0,1)$. Nevertheless, the empirical null shape observed across dependent statistics within a single experiment may differ from this marginal specification. To describe this distinction, we introduce a common Gaussian factor and consider the pooled null distribution conditional on its realization.

\subsection{Pooled conditional null distribution}

Let $\mathcal H_{0}$ and $\mathcal H_{1}$ denote the fixed sets of null and non-null indices, respectively. Set $n_0=|\mathcal H_{0}|>0$. Following the conditional Gaussian factor representation of \citet{AzrielSchwartzman2015}, we assume that, for a fixed factor dimension $d$, the null statistics admit the representation
\[
  Z_{i}
  =\ell_{i}^{\mathsf T}U+\varepsilon_{i},
  \quad i\in\mathcal H_{0},
\]
where $U\sim N_d(0,I_d)$, $\boldsymbol{\varepsilon}_{0} = (\varepsilon_i)_{i \in \mathcal{H}_0}\sim N_{n_0}(0,B_{0})$, $U$ and $\varepsilon_{0}$ are independent, and $\ell_{i}^{\mathsf T}$ is the corresponding row of the loading matrix $L \in \mathbb{R}^{n_0 \times d}$. The unknown null correlation matrix $R_{0}$ satisfies
\begin{equation}
 R_{0}=LL^{\mathsf T}+B_{0},
 \qquad
 \sigma_{i}^2:=(B_{0})_{ii}
 =1-\|\ell_{i}\|^2\ge \sigma_{0,*}^2>0.
 \label{eq:model-residual-variance}
\end{equation}
Hence, each null statistic marginally follows $N(0,1)$, and the model allows heterogeneous loadings and residual correlations. The common factor $U$ represents random influences shared across tests, while the loadings describe how strongly each test responds to them. The marginal specification averages over possible factor realizations, whereas the conditional description holds the factor realized in the current experiment fixed.

Now let $\phi$ and $\Phi$ be the standard normal density and cumulative distribution function, respectively. Conditional on $U$, the $i$th null statistic has distribution $N(\ell_{i}^{\mathsf T}U,\sigma_{i}^2)$. Motivated by the factor-dependent Gaussian mixture characterization of the empirical distribution in \citet{AzrielSchwartzman2015}, we define the pooled conditional null density and CDF by averaging the corresponding conditional distributions as
\begin{equation}
  \overline f_{0,U}(z)
  =\frac1{n_0}\sum_{i\in\mathcal H_{0}}
    \frac1{\sigma_i}
    \phi\!\left(
      \frac{z-\ell_{i}^{\mathsf T}U}{\sigma_i}
    \right), \qquad
  \overline F_{0,U}(z)
  =\frac1{n_0}\sum_{i\in\mathcal H_{0}}
    \Phi\!\left(
      \frac{z-\ell_{i}^{\mathsf T}U}{\sigma_i}
    \right).
 \label{eq:model-finite-pool}
\end{equation}
This is the conditional distribution of a statistic whose index is selected uniformly from $\mathcal H_{0}$, independently of the observations and $U$. It is also the conditional expectation of the null empirical CDF, as $\overline F_{0,U}(z)= \mathbb E[\mathbb F_{0}(z)\mid U]$ for $\mathbb F_{0}(z)=n_0^{-1}\sum_{i\in\mathcal H_{0}} \mathbf 1\{Z_{i}\le z\}$. This pooled distribution describes the average null behavior conditional on the realized factor.

The pooled interpretation is consistent with \citet{Efron2007}, who notes that false discovery calculations do not require identical null densities, but only a suitable representation of their average. More recently, \citet{BarberSamworth2026} connected the averaging of null tail probabilities to compound $p$-values, which require superuniformity only in aggregate across the true nulls rather than for each null individually. 

\subsection{Gaussian--Hermite representation}
Now we represent the pooled null density as a Gaussian reference multiplied by an expansion of Hermite polynomials. The factor model provides a justification for this representation. For every fixed factor realization $u$, the residual-variance bound in \eqref{eq:model-residual-variance} implies
\[
 0<\frac{\overline f_{0,u}(z)}{\phi(z)}
 \le \sigma_{0,*}^{-1}\exp\!\left(\frac{\|u\|^2}{2}\right),
 \qquad z\in\mathbb R.
\]
Hence the density ratio belongs to $L^2(\phi) = \{h : \int h(z)^2\phi(z)\,dz<\infty \}$. Now let
\[
 \psi_j(z):=\frac{\operatorname{He}_j(z)}{\sqrt{j!}},
 \qquad j\ge0,
\]
where $\operatorname{He}_j$ denotes the probabilists' Hermite polynomial and $\psi_0=1$. These functions form an orthonormal basis of $L^2(\phi)$. Consequently, the pooled null density in \eqref{eq:model-finite-pool} can be represented as
\begin{equation}
 \overline f_{0,U}(z)
 =\phi(z)\left\{
    1+\sum_{j=1}^{\infty}w_{j}^{\circ}(U)\psi_j(z)
   \right\},
 \label{eq:model-hermite}
\end{equation}
\[
 w_{j}^{\circ}(U)
 :=\int_{\mathbb R}\psi_j(z)\overline f_{0,U}(z)\,dz
 =\frac1{n_0}\sum_{i\in\mathcal H_{0}}
   \mathbb E[\psi_j(Z_{i})\mid U],
 \qquad j\ge1,
\]
with convergence of the ratio series in $L^2(\phi)$. Since $\phi^{(j)}(z)=(-1)^j\operatorname{He}_j(z)\phi(z)$, the density can equivalently be expressed as
\[
 \overline f_{0,U}(z)
 =\phi(z)+\sum_{j=1}^{\infty}
   \frac{w_{j}^{\circ}(U)}{\sqrt{j!}}
   (-1)^j\phi^{(j)}(z).
\]
This has the same form as the Gram--Charlier representation considered by \citet{Schwartzman2010}, with $W_j=w_{j}^{\circ}(U)/\sqrt{j!}$. In the present model, these coefficients are determined by the realized factor and the null loading configuration.

Furthermore, square integrability  justifies finite truncation. By orthonormality, for each fixed $u$,
\[
 \left\|
  \frac{\overline f_{0,u}}{\phi}
  -
  \left(
    1+\sum_{j=1}^{k}w_{j}^{\circ}(u)\psi_j
   \right)
 \right\|_{L^2(\phi)}^2
 =
 \sum_{j>k}\{w_{j}^{\circ}(u)\}^2
 \longrightarrow 0
 \quad \text{as }k \to \infty.
\]
Thus the cumulative contribution of the omitted terms vanishes in this weighted $L^2$ sense. Therefore, we use the finite-dimensional truncation of \eqref{eq:model-hermite} for estimation with even $k \ge 2$ as
\begin{equation}
 R_{\boldsymbol{w}_k}(z)=1+\sum_{j=1}^{k}w_j\psi_j(z),
 \qquad
 f(z;\boldsymbol{w}_k)=\phi(z)R_{\boldsymbol{w}_k}(z),
 \qquad
 \boldsymbol{w}_k=(w_1,\ldots,w_k)^{\mathsf T}.
 \label{truncated_density}
\end{equation}
Orthogonality ensures that this family integrates to one. Because finite truncation does not preserve nonnegativity, we restrict the coefficients so that $R_{\boldsymbol{w}_k}(z)\ge 0$ for all $z\in\mathbb R$, implying $f(z;\boldsymbol{w}_k)$ becomes a valid probability density.

Directly evaluating the factor-mixture density in \eqref{eq:model-finite-pool} requires the loadings, residual variances, and realized factor scores. \citet{AzrielSchwartzman2015} describe a covariance-based reconstruction that derives the factor loadings and residual variances from a known or estimated correlation matrix and estimates the realized factor scores from the observed statistics.  However, the Hermite representation provides coordinates for estimating the realized pooled density directly. Accordingly, our estimation procedure fits the coefficients $\boldsymbol{w}_k$ without requiring the null correlation matrix, the loading matrix, or the realized factor scores as inputs. The factor model specifies the dependence mechanism and the target density, while the fitted Hermite coefficients describe the realized pooled conditional null distribution in the observed experiment.

Under the model in \eqref{truncated_density}, the key inferential problem is to estimate $\boldsymbol{w}_k$, and then to construct the $p$-values. Once the $p$-values are obtained, they can be combined with a multiple testing procedure such as the Benjamini--Hochberg procedure \citep{BenjaminiHochberg1995}. The main difficulty is that direct fitting of (\ref{truncated_density}) using full observed data may be unstable since the observed sample may be contaminated by non-null effects. This is the point at which our proposed estimation method becomes useful.

\section{Proposed Algorithm}
\label{sec:method}

\subsection{Motivation}
\label{subsec:motivation}

Let $\mathcal X=\{Z_1,\ldots,Z_n\}$ denote the observed test statistics, with order statistics $Z_{(1)}\le\cdots\le Z_{(n)}$. For right-sided testing, the upper part of the sample tends to be more contaminated by nonnull effects. We therefore choose a fraction $q\in(0,1)$, set $N_n=\lfloor nq\rfloor$ and $\tau_n=n-N_n$, with $1\le N_n<n$, and partition the observations as
\begin{equation}
 \mathcal X
 =\mathcal X_0\sqcup\mathcal X_+,
 \qquad
 \mathcal X_0
 =\{Z_{(j)}:1\le j\le N_n\},
 \qquad
 \mathcal X_+
 =\{Z_{(j)}:N_n<j\le n\}.
\label{eq:initial-split}
\end{equation}
Here $\sqcup$ denotes concatenation of observation collections, preserving multiplicities. We refer to $\mathcal X_0$ as the initial center and write $b_n=Z_{(N_n)}$ for its upper cutoff. The original center $\mathcal{X}_0$ is treated as predominantly null and provides the basis for estimating the null density.

A natural approach to estimating the null is to use the conditional likelihood of $\mathcal X_0$. For a candidate density $f$ with CDF $F_f$, the corresponding working log-likelihood is
\begin{equation}
 \ell_{\mathrm C}(f)
 =
 \sum_{z\in\mathcal X_0}
 \log\left\{\frac{f(z)}{F_f(b_n)}\right\}
 =
 \sum_{z\in\mathcal X_0}\log f(z)
 -
 N_n\log F_f(b_n).
 \label{eq:conditional-likelihood}
\end{equation}
The normalization by $F_f(b_n)$ accounts for the restriction of the fitted observations to $(-\infty,b_n]$. Thus, the method estimates the null from the central region without requiring a model for the potentially contaminated values in $\mathcal X_+$.

This reasoning is closely related to the empirical-null approach of \citet{Efron2004}. In particular, \citet{Efron2007SizePower} uses a truncated-normal likelihood component to fit a Gaussian null within a central interval, together with a binomial component for the number of observations in that interval. Under a Gaussian null $N(\mu,\sigma^2)$, the retained observations inform the same two parameters that determine the entire density. Once $\mu$ and $\sigma$ have been estimated, the right-tail probability at any $x$ is specified by $1-\Phi\{(x-\mu)/\sigma\}$. The Gaussian assumption therefore supplies a strong structural connection between the center and the tail.

The Gaussian--Hermite representation in \eqref{truncated_density}, motivated by \citet{Schwartzman2010}, provides the flexibility needed to accommodate asymmetry and multiple modes in the pooled conditional null. For a fixed truncation order, the log-likelihood is concave in these coefficients when the observation weights are held fixed. The conditional log-likelihood in \eqref{eq:conditional-likelihood} also admits an equivalent concave formulation after the coefficients are normalized by the initial central probability, with a corresponding convex admissible set. These properties facilitate coefficient estimation, but the accuracy of the extrapolated tail remains a separate statistical question.

The conditional criterion evaluates the candidate density after normalization over the initial center. For example, consider that the two candidate densities $f$ and $\tilde{f}$ are nearly proportional throughout the center as $f(z) \approx a \tilde{f}(z), \; z \le b_n$. This implies $F_{f}(b_n) \approx aF_{\tilde{f}}(b_n)$, and
\[
\ell_{\mathrm C}(f)
 =
 \sum_{z\in\mathcal X_0}
 \log\left\{\frac{f(z)}{F_f(b_n)}\right\}
 \approx
 \sum_{z\in\mathcal X_0}
 \log\left\{\frac{\tilde{f}(z)}{F_{\tilde{f}}(b_n)}\right\}
 =
 \ell_{\mathrm C}(\tilde{f})
\]
Hence, the corresponding change in their center probabilities largely cancels the change in their fitted heights. Similar conditional fits can therefore correspond to substantially different probabilities assigned to the center. Since each candidate integrates to one, this difference changes the total probability assigned to the excluded tail. The density family links these quantities to the conditional center shape, but the retained observations $\mathcal{X}_0$ may weakly determine that connection.

Furthermore, within a finite Hermite expansion, different combinations of coefficients may produce nearly identical normalized shapes over the initial center while yielding substantially different densities beyond the cutoff. Contributions from different orders may offset one another within the center without doing so in the tail, where the fitted densities can differ in curvature or exhibit additional modes. For a fixed order $k$, exact knowledge of the conditional density on an interval identifies the full density within this family. With finitely many observations, however, similar conditional fits need not imply similar center probabilities or tail shapes. Accurate maximization of the conditional log-likelihood therefore need not ensure accurate null-tail probabilities.

An alternative fitting strategy represents the excluded region through the current null model. In the procedure of \citet{tailbic}, pseudo-observations generated from the current estimated tail replace the original upper observations and contribute to subsequent coefficient updates together with the initial center. To describe this construction, define the conditional right-tail density
\[
 (\mathcal T_a h)(y)
 =
 \frac{h(y)\mathbf 1\{y>a\}}{1-F_h(a)}
\]
for a working density $h$ and cutoff $a$. Given the current null estimate $h$, generate
\[
 Y_1,\cdots,Y_{\tau_n}
 \overset{\mathrm{iid}}{\sim}
 \mathcal T_{b_n}h,
\]
and form $\widetilde{\mathcal X} = \mathcal X_0\sqcup\{Y_1,\cdots,Y_{\tau_n}\}$. The original upper observations $\mathcal{X}_+$ are thereby replaced by $\{Y_1, \cdots , Y_{\tau_n}\}$, so that the null density is refitted using this reconstructed sample $\widetilde{\mathcal{X}}$, and the updated estimate is used as the sampling distribution for the next iteration. At the next iteration, a new pseudo-sample is generated from the updated estimate conditional on exceeding $b_n$, replacing all pseudo-observations from the preceding iteration. The density is then refitted using $\mathcal X_0$ together with this new pseudo-sample that contributes to subsequent coefficient updates. This cycle of regenerating the entire pseudo-tail and refitting the density is repeated over successive iterations.

However, a difficulty remains since all pseudo-samples are discarded and regenerated at every iteration. In many right-tail settings, the null density decreases as one moves away from the initial cutoff $b_n$. Consequently, sampling from the entire tail region naturally produces substantially more pseudo-observations near the cutoff $b_n$ than in the more distant tail. Since each updated estimate serves as the sampling distribution for the next iteration, finite-sample estimation errors near the cutoff may be carried into subsequent updates. Such feedback may distort the shape of the estimated density not only near $b_n$, but also farther into the right tail, potentially affecting the accuracy of estimated null-tail probabilities and the resulting right-sided $p$-values. This issue is particularly important in our multiple testing settings. While \citet{tailbic} primarily considered global testing, large-scale multiple testing relies on accurate null-tail probabilities to calibrate individual tests and support reliable FDR control.

To address this issue, we propose reconstructing the tail progressively rather than repeatedly replacing it in its entirety. Specifically, pseudo-observations are retained in successive portions of the tail, starting near the initial cutoff $b_n$, while fresh sampling is directed toward the remaining outer region. We note that retention fixes the pseudo-observations values, not the fitted density on the corresponding region. In this way, previously reconstructed portions continue to contribute to subsequent fits, allowing the reconstruction to proceed outward without repeatedly discarding the entire pseudo-tail. Combining progressive reconstruction with joint shift--order selection yields \texttt{TRAIL\_BIC}, as described in Section~\ref{subsec:selection}.

\begin{figure}[t]
    \centering
    \includegraphics[width=0.95\linewidth]{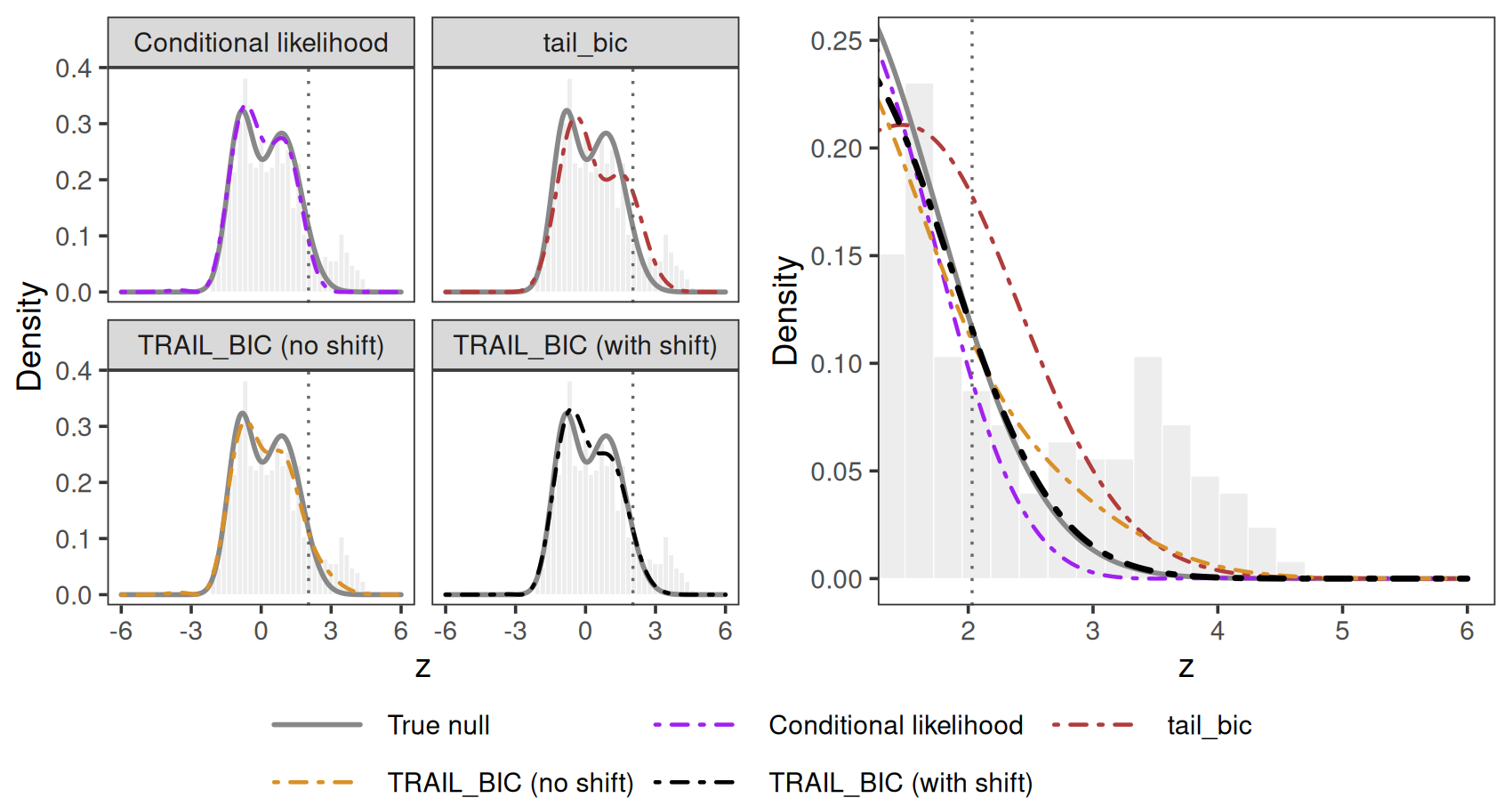}
    \caption{Null-tail reconstruction under alternative fitting methods (Example 1)}
    \label{fig:cl1}
\end{figure}
\begin{figure}[t]
    \centering
    \includegraphics[width=0.95\linewidth]{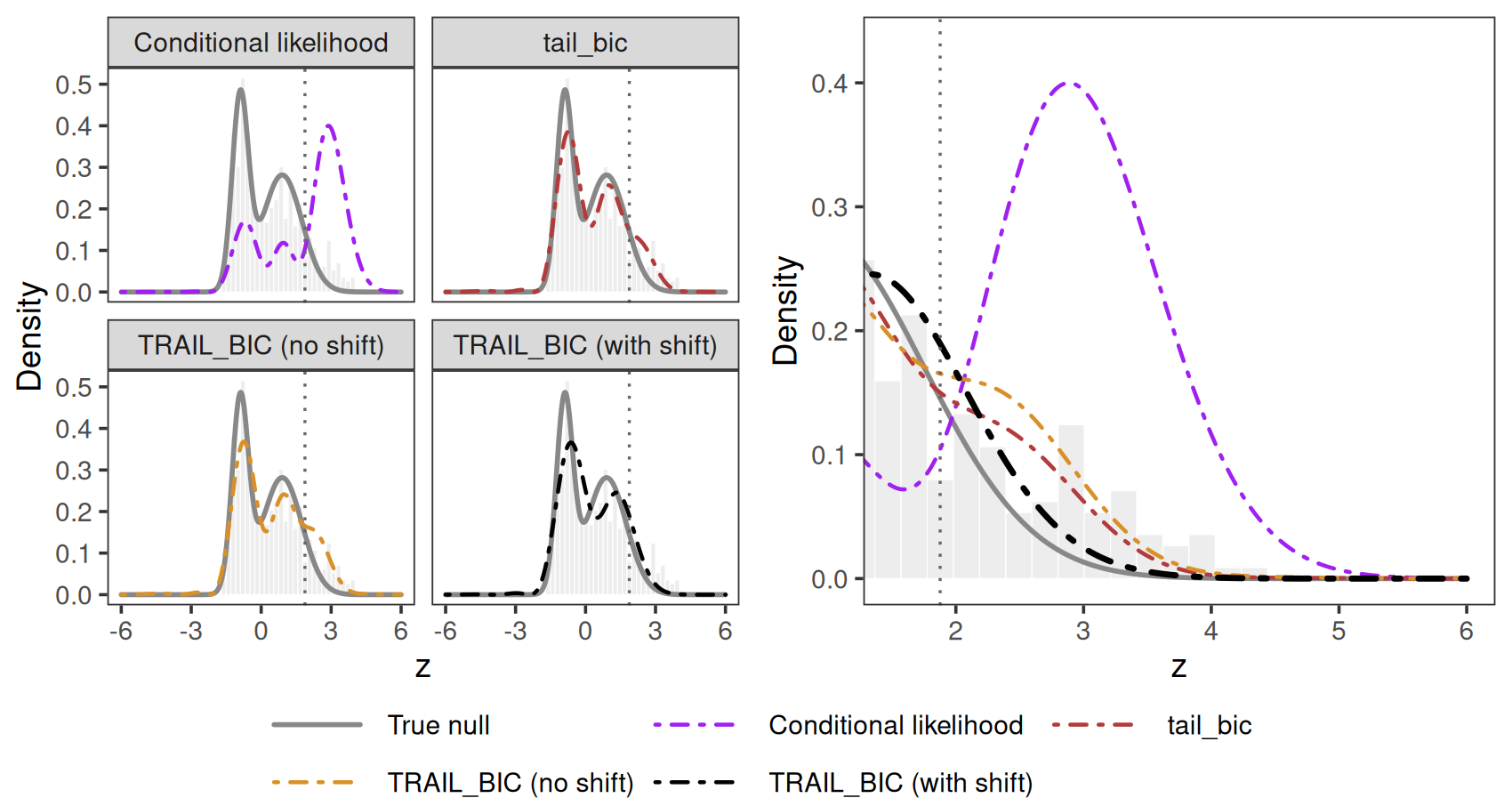}
    \caption{Null-tail reconstruction under alternative fitting methods (Example 2)}
    \label{fig:cl2}
\end{figure}

Figures \ref{fig:cl1} and \ref{fig:cl2} illustrate these issues through two examples. In each example, the true null density is an unequally weighted mixture of two Gaussian distributions, while nonnull signals form an additional peak to its right, with the signal-generation mechanism described in Section \ref{sec:numerical}. We consider four fitting configurations within the shared Gaussian--Hermite representation: direct maximization of the conditional likelihood in \eqref{eq:conditional-likelihood}; the full-tail replacement benchmark based on \citet{tailbic}, labeled \texttt{tail\_bic}; progressive reconstruction without shifting; and progressive reconstruction with joint shift--order selection. The latter two are labeled \texttt{TRAIL\_BIC} without and with shifting, respectively. For this comparison, a common BIC-based rule on initial center observations is used to select the truncation order $k$. The gray solid curve represents the true null density, and the vertical dotted line marks the initial cutoff $b_n$. The right panel provides an enlarged view of the right-tail region.

The flexibility of this common density model allows all four methods to capture the central bimodal structure in both examples, although the estimated peak heights differ. Their right-tail estimates, however, exhibit substantial differences. Conditional likelihood produces a substantially thinner right tail than the true null in Figure \ref{fig:cl1}, whereas in Figure~\ref{fig:cl2} it develops a pronounced spurious right-tail mode that is higher than either fitted central mode. These contrasting outcomes illustrate the finite-sample extrapolation sensitivity discussed above, that recovering the central bimodal pattern does not by itself ensure an accurate null-tail shape.

For \texttt{tail\_bic}, Figure \ref{fig:cl1} shows substantial overestimation of the null density near and beyond $b_n$. In Figure \ref{fig:cl2}, the fitted density decreases more gradually beyond the cutoff, forming a shoulder and remaining above the true null over much of the displayed tail. \texttt{TRAIL\_BIC} without shifting reduces the discrepancy near the cutoff in Figure~\ref{fig:cl1}, but still yields an overly elevated tail. A similar shoulder remains in Figure~\ref{fig:cl2}, indicating that progressive reconstruction alone does not fully eliminate the observed distortions.

Among the displayed fits, \texttt{TRAIL\_BIC} with shifting provides the closest overall visual agreement with the true right-tail shape in both examples. These comparisons illustrate the complementary benefits of progressive reconstruction and shifting for recovering the null-tail shape relevant to the calculation of right-sided $p$-values.

\subsection{Progressive tail reconstruction}
\label{subsec:reconstruction}

\begin{figure}[t]
\centering
\resizebox{0.96\linewidth}{!}{%
\begin{tikzpicture}[x=0.76cm,y=0.76cm]


\draw[
  -{Latex[length=2.5mm]},
  line width=0.9pt,
  TrailRetained
]
(5.85,4.02) -- (12.90,4.02);

\node[
  font=\scriptsize\bfseries,
  text=TrailRetained,
  anchor=south
] at (9.375,4.13)
{Sampling boundary $a_{n,t}$ moves right};


\node[trail/stage] at (-0.35,3.30) {$t=0$};
\node[trail/stage] at (-0.35,2.30) {$t=1$};
\node[trail/stage] at (-0.35,1.30) {$t=2$};
\node[trail/stage] at (-0.35,0.30) {$t=T$};


\draw[trail/center]
  (0,2.96) rectangle (5.70,3.64);

\draw[trail/active]
  (5.85,2.96) rectangle (12.90,3.64);

\node[trail/white] at (2.85,3.30)
  {$\mathcal X_0$ \ (initial center)};

\node[trail/dark] at (9.375,3.30)
  {$\mathcal X_+$ \ (initial tail)};


\draw[trail/center]
  (0,1.96) rectangle (5.70,2.64);

\draw[trail/retained]
  (5.85,1.96) rectangle (7.40,2.64);

\draw[trail/active]
  (7.50,1.96) rectangle (12.90,2.64);

\node[trail/white] at (2.85,2.30)
  {$\mathcal X_0$};

\node[trail/white] at (6.625,2.30)
  {$\mathcal X_0^{(1)}$};

\node[trail/dark] at (10.20,2.30)
  {$\mathcal X_+^{(1)}$};


\draw[trail/center]
  (0,0.96) rectangle (5.70,1.64);

\draw[trail/retained]
  (5.85,0.96) rectangle (7.40,1.64);

\draw[trail/retained]
  (7.50,0.96) rectangle (9.05,1.64);

\draw[trail/active]
  (9.15,0.96) rectangle (12.90,1.64);

\node[trail/white] at (2.85,1.30)
  {$\mathcal X_0$};

\node[trail/white] at (6.625,1.30)
  {$\mathcal X_0^{(1)}$};

\node[trail/white] at (8.275,1.30)
  {$\mathcal X_0^{(2)}$};

\node[trail/dark] at (11.025,1.30)
  {$\mathcal X_+^{(2)}$};


\draw[trail/center]
  (0,-0.04) rectangle (5.70,0.64);

\draw[trail/retained]
  (5.85,-0.04) rectangle (7.40,0.64);

\draw[trail/retained]
  (7.50,-0.04) rectangle (9.05,0.64);

\draw[trail/retained]
  (9.15,-0.04) rectangle (11.30,0.64);

\draw[trail/retained]
  (11.40,-0.04) rectangle (12.95,0.64);

\node[trail/white] at (2.85,0.30)
  {$\mathcal X_0$};

\node[trail/white] at (6.625,0.30)
  {$\mathcal X_0^{(1)}$};

\node[trail/white] at (8.275,0.30)
  {$\mathcal X_0^{(2)}$};

\node[trail/white] at (10.225,0.30)
  {$\cdots$};

\node[trail/white] at (12.175,0.30)
  {$\mathcal X_0^{(T)}$};


\draw[trail/center]
  (0.25,-0.79) rectangle (0.65,-0.49);

\node[
  anchor=west,
  font=\scriptsize,
  text=TrailInk
] at (0.80,-0.64)
{Original center};

\draw[trail/retained]
  (4.00,-0.79) rectangle (4.40,-0.49);

\node[
  anchor=west,
  font=\scriptsize,
  text=TrailInk
] at (4.55,-0.64)
{Retained pseudo-observations};

\draw[trail/active]
  (10.30,-0.79) rectangle (10.70,-0.49);

\node[
  anchor=west,
  font=\scriptsize,
  text=TrailInk
] at (10.85,-0.64)
{Active tail};

\end{tikzpicture}%
}

\caption{
Schematic illustration of progressive tail reconstruction
in \texttt{TRAIL}.
}
\label{fig:progressive-tail-reconstruction}

\end{figure}
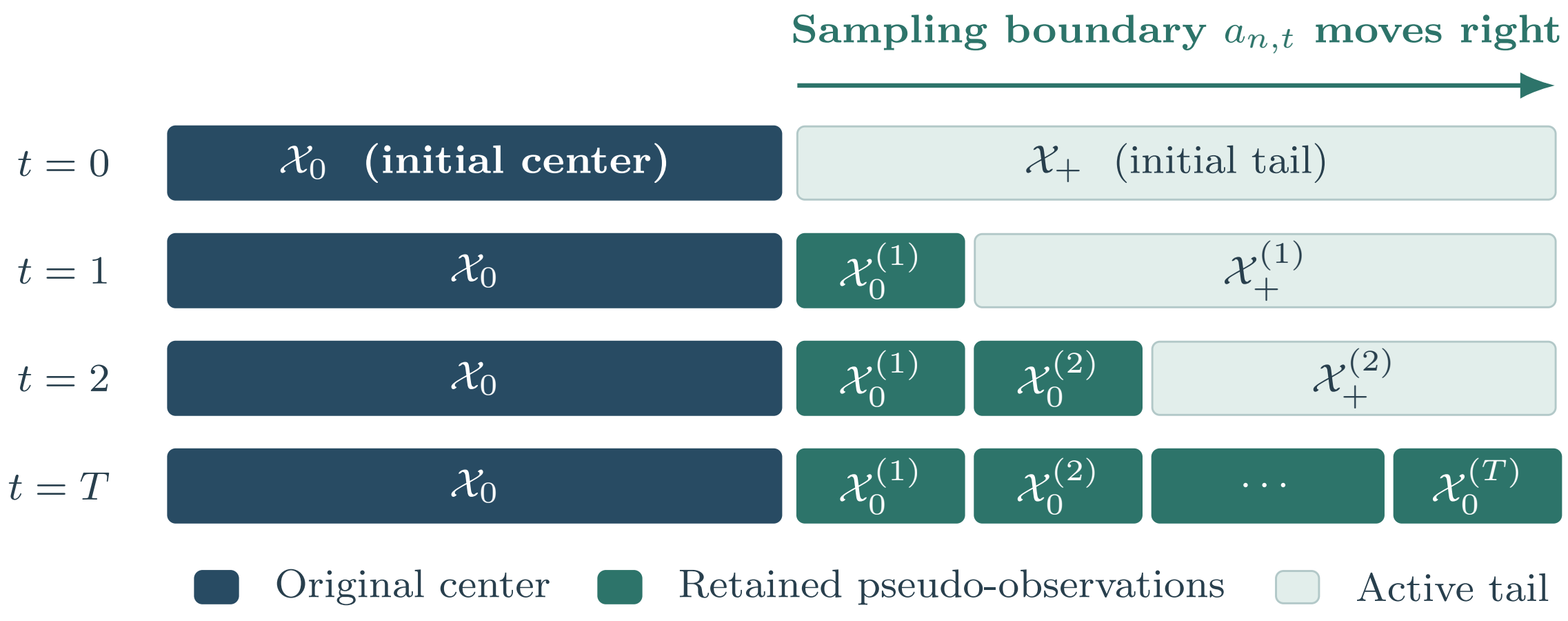

We now describe \texttt{TRAIL} (Tail Reconstruction via Adaptive Iterative Learning) at a prescribed even Hermite order $k$. Starting with the partition in \eqref{eq:initial-split}, the $N_n=\lfloor nq\rfloor$ smallest observations form $\mathcal X_0$, and the remaining $\tau_n=n-N_n$ observations form $\mathcal X_+$. Let $b_n=Z_{(N_n)}$ and choose an admissible initial density $\hat f^{(0)}$ in the family \eqref{truncated_density}. The initial center $\mathcal X_0$ remains unchanged, while the initial tail $\mathcal X_+$ is replaced by pseudo-observations for fitting. Initially, no pseudo-observations have been retained, and the lower sampling boundary is $a_{n,0}=b_n$.

Choose $T\ge2$ reconstruction stages and positive
retention counts
$M_{n,1},\ldots,M_{n,T}$
satisfying
\begin{equation}
 \sum_{t=1}^{T}M_{n,t}=\tau_n,
 \qquad
 m_{n,t}
 =\tau_n-\sum_{s=1}^{t-1}M_{n,s}.
 \label{eq:reconstruction-schedule}
\end{equation}
Here $m_{n,t}$ is the size of the fresh batch at stage $t$, and $M_{n,t}$ is the number of its values to retain for subsequent stages. We first describe the procedure using one generated batch per stage, before introducing replicated updates. Figure \ref{fig:progressive-tail-reconstruction} provides a schematic overview of this construction.

At the first stage, generate $m_{n,1}=\tau_n$ pseudo-observations from the initial estimate above $b_n$ as
\[
 Y_{1,1},\ldots,Y_{1,m_{n,1}}
 \overset{\mathrm{iid}}{\sim}
 \mathcal T_{b_n}\hat f^{(0)},
\]
where $\mathcal T_a f$ denotes the density $f$ conditional on a value exceeding $a$. Sort the generated values as $Y_{1,(1)}\le\cdots\le Y_{1,(\tau_n)}$. The smallest $M_{n,1}$ values form the first retained block, while the others form the active block,
\[
 \mathcal X_0^{(1)}
 =\{Y_{1,(j)}:1\le j\le M_{n,1}\},
 \qquad
 \mathcal X_+^{(1)}
 =\{Y_{1,(j)}:M_{n,1}<j\le\tau_n\}.
\]
Refitting uses $\mathcal X_0\sqcup\mathcal X_0^{(1)} \sqcup\mathcal X_+^{(1)}$ to obtain $\hat f^{(1)}$. Both generated blocks contribute to this fit, and the distinction is that $\mathcal X_0^{(1)}$ will be preserved, whereas $\mathcal X_+^{(1)}$ will be replaced. The fitting criterion is specified below.

At the second stage, keep $\mathcal X_0$ and $\mathcal X_0^{(1)}$ and generate only the $m_{n,2}=\tau_n-M_{n,1}$ observations in the active block. Sampling no longer starts at the original cutoff. Instead, move the boundary to the smallest value in $\mathcal X_+^{(1)}$ and draw from the updated density as
\[
 a_{n,1}
 =\min\mathcal X_+^{(1)}
   =Y_{1,(M_{n,1}+1)},
   \qquad
 Y_{2,1},\ldots,Y_{2,m_{n,2}}
 \overset{\mathrm{iid}}{\sim}
   \mathcal T_{a_{n,1}}\hat f^{(1)}.
\]
The smallest $M_{n,2}$ values of this fresh batch form $\mathcal X_0^{(2)}$, and the remaining values form $\mathcal X_+^{(2)}$. The next fit therefore uses $\mathcal X_0\sqcup\mathcal X_0^{(1)} \sqcup\mathcal X_0^{(2)}\sqcup\mathcal X_+^{(2)}$.

Continue in this manner at each stage $t$. Generate $m_{n,t}$ fresh values above $a_{n,t-1}$ from $\hat f^{(t-1)}$, retain their lowest $M_{n,t}$ ranks as $\mathcal X_0^{(t)}$, and designate the remaining values as $\mathcal X_+^{(t)}$. Thus, $\mathcal X_0^{(t)}$ denotes the block retained at stage $t$. The fitting dataset at iteration $t$ is
\[
 \mathcal X^{(t)}
 =\mathcal X_0
  \sqcup\left(\bigsqcup_{s=1}^{t}\mathcal X_0^{(s)}\right)
  \sqcup\mathcal X_+^{(t)},
\]
with block sizes
\[
 |\mathcal X_0^{(t)}|=M_{n,t},
 \qquad
 |\mathcal X_+^{(t)}|
 =m_{n,t}-M_{n,t}.
\]
For $t<T$, the active-block size is $m_{n,t+1}$, and its smallest value becomes the next sampling boundary, $a_{n,t}=\min\mathcal X_+^{(t)}$. Because each fresh batch lies above the preceding boundary, the boundary advances toward the outer tail. Meanwhile, all previously retained blocks remain in the fitting dataset, which always contains $N_n$ original observations and $\tau_n$ pseudo-observations.

At the final stage, the schedule gives $M_{n,T}=m_{n,T}$, so the entire last batch is retained and $\mathcal X_+^{(T)}=\varnothing$. The final fit uses
\[
 \mathcal X^{(T)}
 =\mathcal X_0
  \sqcup\mathcal X_0^{(1)}
  \sqcup\cdots\sqcup\mathcal X_0^{(T)}.
\]
The resulting estimate is $\hat f^{(T)}$.

To reduce sensitivity to a single generated completion, we perform $M$ independent replications within each stage. All replications share the incoming density, sampling boundary, and previously retained blocks, but use conditionally independent fresh batches as
\begin{equation}
 Y_{t,1}^{(m)},\ldots,Y_{t,m_{n,t}}^{(m)}
 \overset{\mathrm{iid}}{\sim}
 \mathcal T_{a_{n,t-1}}\hat f^{(t-1)},
 \qquad m=1,\ldots,M.
 \label{eq:progressive-sampling}
\end{equation}
Within replication $m$, sort the fresh values and define
\begin{equation}
 \mathcal X_0^{(t,m)}
 =\{Y_{t,(j)}^{(m)}:
       1\le j\le M_{n,t}\},
       \qquad
 \mathcal X_+^{(t,m)}
 =\{Y_{t,(j)}^{(m)}:
       M_{n,t}<j\le m_{n,t}\}.
\label{eq:replicate-blocks}
\end{equation}
Both blocks enter the replication-specific fitting dataset
\begin{equation}
 \mathcal X^{(t,m)}
 =\mathcal X_0
  \sqcup\left(\bigsqcup_{s=1}^{t-1}\mathcal X_0^{(s)}\right)
  \sqcup\mathcal X_0^{(t,m)}
  \sqcup\mathcal X_+^{(t,m)},
 \label{eq:reconstructed-data}
\end{equation}
where the collection of earlier retained blocks is empty when $t=1$. Denote the fitted coefficient vector by $\widehat{\boldsymbol w}_k^{(t,m)}$. Average these vectors to define the next density estimate by
\begin{equation}
 \overline{\widehat{\boldsymbol w}}_k^{(t)}
 =\frac1M\sum_{m=1}^{M}\widehat{\boldsymbol w}_k^{(t,m)},
 \qquad
 \hat f^{(t)}(z)
 =f\!\left(z;\overline{\widehat{\boldsymbol w}}_k^{(t)}\right).
 \label{eq:averaged-fit}
\end{equation}
Since the density is affine in the Hermite coefficients, this also averages the fitted densities and preserves nonnegativity and normalization.

We likewise average corresponding ranks of the same fresh batches as
\begin{equation}
 \overline Y_{t,(j)}
 =\frac1M\sum_{m=1}^{M}Y_{t,(j)}^{(m)},
 \qquad j=1,\ldots,m_{n,t}.
 \label{eq:averaged-order-statistics}
\end{equation}
Then, the shared retained and active blocks are
\begin{equation}
 \mathcal X_0^{(t)}
 =\{\overline Y_{t,(j)}:
       1\le j\le M_{n,t}\},
       \qquad
 \mathcal X_+^{(t)}
 =\{\overline Y_{t,(j)}:
       M_{n,t}<j\le m_{n,t}\}.
\label{eq:freezing-update}
\end{equation}
The averaged retained block $\mathcal X_0^{(t)}$ is carried into all subsequent fits. For $t<T$, the averaged active block determines the next sampling boundary,
\begin{equation}
 a_{n,t}
 =\min\mathcal X_+^{(t)}
 =\overline Y_{t,(M_{n,t}+1)}.
 \label{eq:active-boundary}
\end{equation}
Each current fit uses its own fresh batch before rank averaging, and the averaged retained values enter fitting only at later stages.

Finally at $t=T$, every replication has an empty active block, and the final fitting datasets are
\begin{equation}
 \mathcal X^{(T,m)}
 =\mathcal X_0
  \sqcup\left(\bigsqcup_{s=1}^{T-1}\mathcal X_0^{(s)}\right)
  \sqcup\mathcal X_0^{(T,m)}.
 \label{eq:terminal-fitting-data}
\end{equation}
The terminal estimate $\hat f^{(T)}$ averages these replicated fits, and no additional fit is performed after forming $\mathcal X_0^{(T)}$ by rank averaging. For $M=1$, replication-specific and averaged blocks coincide, recovering the construction described above.

It remains to specify how the original and generated observations contribute to each refit. Each refit assigns separate weights to the original center and the reconstructed pseudo-observations. Using the incoming density estimate, compute
\[
 c_{n,t}=F_{\hat f^{(t-1)}}(b_n),
\]
which represents the center probability under the current null estimate. Each observation in $\mathcal X_0$ receives weight $c_{n,t}/N_n$, whereas every pseudo-observation, whether retained or active, receives weight $(1-c_{n,t})/\tau_n$. The corresponding total weights are therefore $c_{n,t}$ and $1-c_{n,t}$. Retaining a pseudo-observation fixes its value, but its weight continues to change across stages.

Accordingly, at stage $t$ and replication $m$, the Hermite coefficients are estimated by
\begin{equation}
\begin{aligned}
 \widehat{\boldsymbol w}^{(t,m)}_k
 \in\operatorname*{argmax}_{\boldsymbol w_k}\Bigg[
 &\frac{c_{n,t}}{N_n}
   \sum_{z\in\mathcal X_0}\log f(z;\boldsymbol w_k)\\
 &+\frac{1-c_{n,t}}{\tau_n}\Bigg\{
   \sum_{s=1}^{t-1}\sum_{z\in\mathcal X_0^{(s)}}
        \log f(z;\boldsymbol w_k) +
   \sum_{z\in\mathcal X_0^{(t,m)} \sqcup \mathcal{X}_{+}^{(t,m)}}\log f(z;\boldsymbol w_k)
   \Bigg\}\Bigg].
\end{aligned}
\label{eq:replicate-fit}
\end{equation}
Optimization is over the admissible coefficient vectors in \eqref{truncated_density}, ensuring $R_{\boldsymbol w_k}(z)\ge0$ for all $z\in\mathbb R$. We use a fixed measurable maximizer selection when needed. The weight $c_{n,t}$ is computed once from the incoming density and held fixed across replications and during coefficient optimization.

This progressive reconstruction is primarily motivated by the need to maintain adequate coverage across the tail, but it is also related in spirit to several established computational ideas. Incremental and blockwise EM methods update only part of the uncertain information at a time while retaining previously incorporated contributions \citep{NealHinton1998,ThiessonMeekHeckerman2001}. Stratified resampling reduces resampling variability by allocating uniform draws across strata of the cumulative-probability interval \citep{DoucCappeMoulines2005}, while multilevel rare-event methods replace a single difficult sampling problem with a sequence of more manageable intermediate problems \citep{AuBeck2001}. These methods provide useful precedents for progressively decomposing an unstable global update into more localized steps. Algorithm~\ref{alg:trail} summarizes the full procedure.

\begin{algorithm}[!htbp]
\footnotesize
\caption{\texttt{TRAIL} reconstruction at a prescribed
Hermite order $k$}
\label{alg:trail}
\begin{algorithmic}[1]
\STATE Form $(\mathcal X_0,\mathcal X_+)$ and $b_n$
       using \eqref{eq:initial-split}.
\STATE Specify $T$, $M$, and
       $\{M_{n,t}\}_{t=1}^{T}$;
       compute $\{m_{n,t}\}_{t=1}^{T}$ using
       \eqref{eq:reconstruction-schedule}.
\STATE Initialize an admissible $\hat f^{(0)}$ and set $a_{n,0}=b_n$.
\FOR{$t=1,\ldots,T$}
 \STATE Compute $c_{n,t}=F_{\hat f^{(t-1)}}(b_n)$.
 \FOR{$m=1,\ldots,M$}
  \STATE Generate $m_{n,t}$ fresh values from
         $\mathcal T_{a_{n,t-1}}\hat f^{(t-1)}$
         using \eqref{eq:progressive-sampling}.
  \STATE Sort the fresh values and form
         $\mathcal X_0^{(t,m)}$ and
         $\mathcal X_+^{(t,m)}$ using
         \eqref{eq:replicate-blocks}.
  \STATE Combine both blocks with $\mathcal X_0$ and
         all earlier retained blocks to form
         $\mathcal X^{(t,m)}$ using
         \eqref{eq:reconstructed-data}.
  \STATE Obtain $\widehat{\boldsymbol w}^{(t,m)}_k$
         in \eqref{eq:replicate-fit}.
 \ENDFOR
 \STATE Average the fitted coefficients and obtain
        $\hat f^{(t)}$ using \eqref{eq:averaged-fit}.
 \STATE Average matching order statistics of the same
        fresh batches using
        \eqref{eq:averaged-order-statistics}.
 \STATE Form $\mathcal X_0^{(t)}$ and
        $\mathcal X_+^{(t)}$ using
        \eqref{eq:freezing-update}; preserve
        $\mathcal X_0^{(t)}$ together with all
        earlier retained blocks.
 \IF{$t<T$}
  \STATE Set $a_{n,t}=\min\mathcal X_+^{(t)}$
         using \eqref{eq:active-boundary}; replace
         the active block at the next stage.
 \ENDIF
\ENDFOR
\STATE \textbf{Output:} $\hat f^{(T)}$ and $\{\mathcal X^{(T,m)}\}_{m=1}^{M}$.
\end{algorithmic}
\end{algorithm}

\subsection{Shift--order selection and multiple testing}
\label{subsec:selection}

Progressive tail reconstruction specifies how the tail region is reconstructed, but the number of Hermite terms remains to be determined. A larger truncation order $k$ allows the fitted density to represent more complicated distributions. However, it also introduces more coefficients to estimate and increases the computational cost of the constrained fitting problem. An appropriate order should therefore balance the flexibility of the approximation against its complexity.

This balance is difficult to determine in advance because the order needed to describe the pooled null depends on its unknown shape and the information available in the retained  observations. Following the BIC-based order selection of \citet{tailbic}, we compare fits over a prespecified set of candidate orders. We further extend this approach by allowing the location of the Gaussian reference to vary. At a fixed order, changing the reference location offers an additional way to enrich the available density shapes without introducing more Hermite terms.

\begin{figure}[t]
    \centering
    \includegraphics[width=0.95\linewidth]{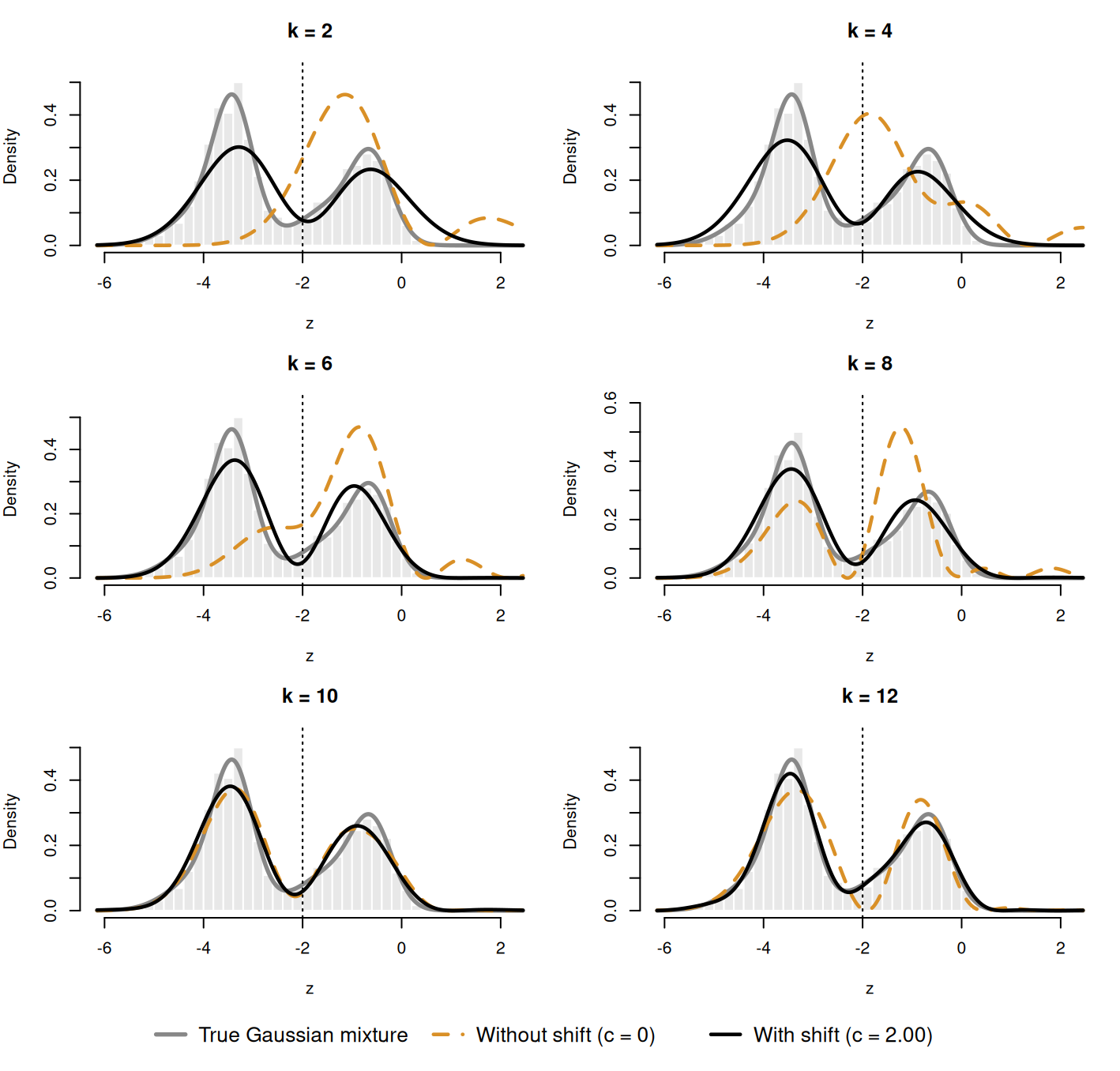}
    \caption{Effect of the Gaussian reference location on finite order Gaussian-Hermite fitting}
    \label{fig:shift-effect}
\end{figure}

When the Gaussian reference is poorly aligned with the target density, the Hermite expansion must represent both their location difference and the remaining departures from Gaussian shape. A location mismatch can affect coefficients across multiple orders, increasing the number of terms needed to reproduce important distributional features. Choosing a more suitable reference location can reduce this mismatch and allow a finite expansion to describe the dominant shape more effectively.

This idea is consistent with approaches developed for other finite approximations. In the Gram--Charlier setting, \citet{PelliccioniAngeloni2007} consider adjusting the location and scale of the Gaussian reference to improve finite--order approximations. Related ideas appear in adaptive Hermite spectral methods. \citet{ChouShaoXia2023} study the adjustment of basis displacement, scale, and expansion order, while \citet{PagliantiniDelzannoMarkidis2023} adapt Hermite representations through scaling and shifting. At a broader level, saddlepoint approximations use exponential tilting to  evaluate an expansion around the mean of a tilted distribution before recovering the density on the original probability scale \citep{Daniels1954, Reid1988}. These approaches illustrate how the choice of reference distribution or expansion coordinates can improve the approximation quality. Here, we keep the Gaussian scale fixed and select its location together with the Hermite order. 

Specifically, we consider nonnegative data shifts $c\ge0$, corresponding to leftward displacements of the Gaussian reference on the original scale. This directional restriction is motivated by the anticipated concentration of nonnull effects in the upper tail. For each candidate $c$, we apply Algorithm~\ref{alg:trail} to $Z_1+c,\cdots,Z_n+c$, with initial cutoff $b_n+c$. The sampling, retention, averaging, and weighting steps are performed on the shifted scale with $c$ held fixed throughout the reconstruction. The fitted density and reconstructed datasets are then translated back to the original scale. On this scale, the corresponding density family is
\begin{equation}
 f_{c,k}(x;\boldsymbol w_k)
 =
 f(x+c;\boldsymbol w_k)
 =
 \phi(x+c)
 \left\{
  1+\sum_{j=1}^{k}w_j\psi_j(x+c)
 \right\}.
 \label{eq:shifted-hermite-family}
\end{equation}
The Gaussian reference is now centered at $-c$ on the original scale, and $c=0$ recovers the unshifted family.

Figure~\ref{fig:shift-effect} illustrates the role of the Gaussian reference location using a mixture of five Gaussian components. For each fixed pair $(c,k)$, the Hermite coefficients are estimated directly from the same full sample, without progressive reconstruction or pseudo-sample generation. We compare $c=0$ and $c=2$, corresponding to Gaussian reference locations $0$ and $-2$ on the original scale.

At $k=2$ and $k=4$, the shifted fits already represent the two principal modal regions, although their peak heights remain underestimated. The unshifted fits miss the prominent left mode, and substantial distortions in the relative peak heights remain at $k=6$ and $k=8$. The two fits nearly coincide at $k=10$, whereas the unshifted fit at $k=12$ exhibits an overly deep valley between the modes and an elevated right peak. These comparisons illustrate that a suitable reference location can reduce the order needed to represent the dominant shape. By reducing the location mismatch that the polynomial correction must accommodate, the shifted reference allows a limited number of Hermite terms to describe the modal structure more effectively. This interaction between reference location and approximation order motivates their joint selection.

Let $\mathscr K$ be a finite set of positive even orders and $\mathscr C$ a finite set of nonnegative data shifts containing zero. For each candidate pair $(c,k)\in\mathscr C\times\mathscr K$, denote the terminal averaged density on the original scale by $\hat f_{c,k}^{(T)}$ and its final replicated fitting datasets by $\mathcal X_{c,k}^{(T,m)}$, $m=1,\ldots,M$. These are the datasets in \eqref{eq:terminal-fitting-data}, before rank averaging of the terminal pseudo-observations. We compare candidate pairs using a BIC-type criterion defined as
\begin{equation}
 \overline{\mathrm{BIC}}_{c,k}
 =
 \frac{1}{M}
 \sum_{m=1}^{M}
 \sum_{z\in\mathcal X_{c,k}^{(T,m)}}
     \log\hat f_{c,k}^{(T)}(z)
 -
 \frac{k}{2}\log n.
 \label{bick}
\end{equation}
Larger values are preferred. The first term measures fit to the reconstructed datasets, while the second penalizes the number of Hermite coefficients. Unlike the weighted objective used for coefficient estimation, this score uses the full, unweighted log-density, including the Gaussian contribution.

Finally, select the data shift and truncation order jointly as
\begin{equation}
 (\hat c,\hat k)
 \in
 \operatorname*{argmax}_{(c,k)\in\mathscr C\times\mathscr K}
     \overline{\mathrm{BIC}}_{c,k},
 \qquad
 \hat f=\hat f_{\hat c,\hat k}^{(T)},
 \label{eq:selected-shift-order}
\end{equation}
with a fixed tie-breaking rule. We call the procedure combining progressive reconstruction with this joint selection \texttt{TRAIL\_BIC}. Writing $\hat F$ for the CDF of the selected density, compute the right-sided values
\[
 \hat p_i=1-\hat F(Z_i),
 \qquad i=1,\ldots,n,
\]
and apply the Benjamini--Hochberg procedure \citep{BenjaminiHochberg1995} at level $\alpha$. Note that all hypotheses are tested using the original observations, whereas the pseudo-observations are used only for fitting and selection. Algorithm~\ref{alg:selection} summarizes the resulting procedure.

\begin{algorithm}[!htbp]
\footnotesize
\caption{\texttt{TRAIL\_BIC}: joint shift--order selection
and multiple testing}
\label{alg:selection}
\begin{algorithmic}[1]
\STATE Specify the candidate sets $\mathscr C$ and $\mathscr K$, the reconstruction settings, and the significance level $\alpha$.
\FOR{$(c,k)\in\mathscr C\times\mathscr K$}
 \STATE Run Algorithm~\ref{alg:trail} at order $k$ on $Z_1 + c, \cdots, Z_n + c$, using an admissible initializer on the shifted scale.
 \STATE Translate the terminal density and final fitting
        datasets back to the original scale, obtaining
        $\hat f_{c,k}^{(T)}$ and
        $\{\mathcal X_{c,k}^{(T,m)}\}_{m=1}^{M}$.
 \STATE Compute $\overline{\mathrm{BIC}}_{c,k}$ using \eqref{bick}.
\ENDFOR
\STATE Select $(\hat c,\hat k)$ using
       \eqref{eq:selected-shift-order} and set
       $\hat f=\hat f_{\hat c,\hat k}^{(T)}$.
\STATE Compute $\hat p_i=1-\hat F(Z_i)$ for every original
       observation.
\STATE Apply the Benjamini--Hochberg procedure at level $\alpha$.
\end{algorithmic}
\end{algorithm}

\section{Theoretical Rationale}
\label{sec:theoretical-rationale}

In this section, we study the theoretical properties of \texttt{TRAIL}, including reconstruction tracking, consistency for the pooled conditional null, and the resulting multiple-testing properties. The analysis considers prescribed Hermite orders and reconstruction iterations, with BIC-type selection outside the scope of the results below. To describe the asymptotic behavior as the number of tests increases, we make the dependence on $n$, the number of test statistics, explicit in the notation. The conditions and proofs are given in the supplementary note.

For $n$ tests, write $Z_{n,i}$ for the $i$th statistic and $\overline f_{0,n,U}$ and $\overline F_{0,n,U}$ for the pooled conditional null density and CDF. Let $\mathcal H_{0,n}$ and $\mathcal H_{1,n}$ be the null and nonnull index sets, with $n_j=|\mathcal H_{j,n}|$ and $\pi_{0,n}=n_0/n\to\pi_0\in(0,1)$. All conditional conclusions hold for almost every $U$.

Across sample sizes, write $L_n$ for the deterministic null loading matrix, with rows $\ell_{n,i}^{\mathsf T}$ for $i\in\mathcal H_{0,n}$. The factor dimension $d$ remains fixed, and a common factor $U$ is used throughout the sequence. Define the empirical distribution of the null loading vectors by
\[
 H_{0,n}
 =\frac{1}{n_0}\sum_{i\in\mathcal H_{0,n}}\delta_{\ell_{n,i}},
\]
where $\delta_{\ell}$ denotes the unit point mass at $\ell$. This is a deterministic probability measure describing the loading configuration across the true nulls.

We consider reconstruction at $c=0$ and $M=1$, with prescribed order $k$ and $T$ reconstruction iterations. The refits use the following density family
\[
 \mathcal F_k=\left\{f(z;\boldsymbol w_k):
 R_{\boldsymbol w_k}(z)\ge e^{-k^2}\ (z\in\mathbb R),\;\;
 \int_{\mathbb R}z^2f(z;\boldsymbol w_k)\,dz\le1+\log k\right\},
 \quad k\ge2\ \text{even},
\]
with positivity and second moment constraints.

Let $\widehat f_{n;k}^{(T)}$ denote the terminal density obtained from $Z_{n,1},\ldots,Z_{n,n}$ after $T$ reconstruction stages at Hermite order $k$, under the settings above and with refits over $\mathcal F_k$. Let $\widehat F_{n;k}^{(T)}$ denote its CDF. When the order and the number of stages depend on $n$, we write $\widehat f_{n;k_n}^{(T_n)}$ and $\widehat F_{n;k_n}^{(T_n)}$. Conditions S1--S4 in the supplementary note specify nondegenerate Gaussian residuals, polynomial decay of residual dependence, vanishing weighted nonnull contamination in the initial center, and exact reconstruction operations. Throughout, the center fraction satisfies $0<q<\pi_0$.

\subsection{Behavior of progressive reconstruction}
\label{subsec:theory-tracking}

We first investigate how the terminal estimate behaves as the number of reconstruction stages increases. The resulting limiting behavior can be expressed through the conditional likelihood of the retained center. For a probability measure $\mathsf P$ supported on $(-\infty,b]$, define the conditional log-likelihood as
\begin{eqnarray*}
 L_{\mathsf P,b}(f)&=&\int\log(f(z)/\phi(z))\,\mathsf P(dz)-\log F_f(b) \nonumber \\
 &=& \int\log\left(\frac{f(z)\mathbf{1}(z\leq b)}{F_f(b)}\right)\,\mathsf P(dz) + \rm{constant} 
\end{eqnarray*}
and let
\begin{eqnarray*}
p_{k,\mathsf P,b}\in\operatorname*{argmax}_{f\in\mathcal F_k}L_{\mathsf P,b}(f).
\end{eqnarray*} 

For the initial center, let $N_n=\lfloor qn\rfloor$, $b_n=Z_{n,(N_n)}$, $\mathsf P_n=N_n^{-1}\sum_{z\in\mathcal X_0}\delta_z$, and $p_{n,k}=p_{k,\mathsf P_n,b_n}$, with multiplicities preserved. We note that $p_{n,k}$ is defined by exact maximization of the empirical conditional likelihood over $\mathcal{F}_k$. Sampling variation may separate this fit from its population counterpart which, at a fixed order, is a conditional Kullback-Leibler projection within the chosen family and need not equal the true pooled null. The following theorem shows that the conditional fit $p_{k,\mathsf P,b}$ arises as the limit of population reconstruction and bounds the discrepancy between the empirical terminal estimate and $p_{n,k}$ under the stated schedule for $T_n$. Recovery of the true pooled conditional null is addressed in the next subsection.

\begin{theorem}[Reconstruction tracking]
\label{thm:theory-tracking}
Fix an even $k\ge2$ and an initializer $g\in\mathcal F_k$. Use the retaining and refitting operations in Condition S4 in the supplementary note. If $\int\log(2+|z|)\,\mathsf P(dz)<\infty$ and $\mathsf P$ has at least $k+1$ distinct support points in $(-\infty,b]$, then $p_{k,\mathsf P,b}$ is unique. For the population reconstruction $f^{(T)}_{k,\mathrm{pop}}$, obtained by replacing fresh samples and retained rank blocks with their exact probability measures,
\begin{eqnarray*}
 \|f^{(T)}_{k,\mathrm{pop}}-p_{k,\mathsf P,b}\|_1=O(T^{-1}).
\end{eqnarray*}
Under Conditions S1--S4 in the supplementary note, $p_{n,k}$ is unique with conditional probability tending to one. For $T_n=\lfloor n^\upsilon\rfloor$, $0<\upsilon<1/2$,
\begin{eqnarray}
 \|\widehat f_{n;k}^{(T_n)}-p_{n,k}\|_1
 =O_{\mathbb P(\cdot \mid U)}\!\left(T_n^{-1}+T_n n^{-1/2}\right).
 \label{eq:conv_rate}
\end{eqnarray}
The population bound may depend on $\mathsf P,b,k$ and the fixed reconstruction settings. Constants in the empirical bound may depend on $U,k$ and the fixed reconstruction settings.
\end{theorem}
\begin{proof}
    See supplementary note Section S3.
\end{proof}

The empirical bound separates finite-iteration reconstruction error from accumulated Monte Carlo variation. Under the stated schedule for $T_n$, both contributions vanish, so the terminal estimate approaches the conditional likelihood target $p_{n,k}$. This conclusion concerns tracking of a target estimated from the initial center, and the accuracy relative to the true pooled null additionally depends on the statistical error of that target. Consequently, more accurate tracking of $p_{n,k}$ need not guarantee more accurate null-tail estimation at a finite sample size. Theorem \ref{thm:theory-tracking} therefore does not imply that increasing the number of reconstruction stages improves finite-sample accuracy.

\subsection{Pooled-null consistency}
\label{subsec:theory-consistency}

The tracking result provides the algorithmic component of consistency. We next combine this result with vanishing nonnull contamination in the initial center and an increasing Hermite order to establish recovery of the pooled conditional null.

\begin{theorem}[Moving-target consistency]
\label{thm:theory-consistency}
Under Conditions S1--S4 in the supplementary note, choose even orders $k_n\to\infty$ and iteration counts satisfying
\[
 k_n^3=o(\log\log n),\qquad
 T_n=\lfloor n^\upsilon\rfloor,\qquad 0<\upsilon<1/2.
\]
Then
\[
 \|\widehat f_{n;k_n}^{(T_n)}-\overline f_{0,n,U}\|_1
 +\|\widehat F_{n;k_n}^{(T_n)}-\overline F_{0,n,U}\|_\infty
 \xrightarrow{\mathbb P(\cdot \mid U)}0.
\]
In addition, if $H_{0,n}$ converges weakly to a probability measure $H_0$ on $\mathbb{R}^d$, the same conclusion holds with the limiting density
\[
 f_{0,U}(z)=\int
 \frac{1}{\sqrt{1-\|\ell\|^2}}
 \phi\!\left(\frac{z-\ell^{\mathsf T}U}{\sqrt{1-\|\ell\|^2}}\right)
 H_0(d\ell)
\]
and its CDF $F_{0,U}$ in place of $\overline f_{0,n,U}$ and $\overline F_{0,n,U}$. Both conclusions also hold in unconditional probability. 
\end{theorem}
\begin{proof}
    See supplementary note in Section S4.
\end{proof}

The estimator tracks the pooled null for the current collection of tests as its shape varies with $n$.  Stabilization of the distribution of  loadings  
(i.e., $H_{0,n}$ converges weakly to $H_0$)
additionally gives a fixed conditional target. The consistency analysis controls residual dependence through a polynomial decay rate for the normalized Frobenius norm of the residual covariance. The following corollary gives a sufficient condition for this rate directly in terms of the null correlation matrix. It constructs a factor representation from the leading eigen-components and uses the remaining spectrum to bound the residual covariance norm and the average absolute residual correlation.

\begin{corollary}[Covariance-level sufficient condition]
\label{cor:theory-covariance}
For a jointly Gaussian null vector with correlation matrix $R_{0,n}$, let
$\lambda_{1,n}\ge\cdots\ge\lambda_{n_0,n}\ge0$ be its ordered eigenvalues and $e_{r,n}$ the corresponding deterministic orthonormal eigenvectors.  For fixed $d\ge0$ and $n_0>d$, set
\[
 B_{0,n}=R_{0,n}-\sum_{r=1}^d\lambda_{r,n}e_{r,n}e_{r,n}^{\mathsf T},
 \qquad
 \vartheta_n=\frac{\|B_{0,n}\|_F}{n_0}.
\]
Write $\sigma_{n,i}^2=(B_{0,n})_{ii}$ for the null residual variances. If $\inf_{n,i}\sigma_{n,i}^2\ge \sigma_{0,*}^2>0$ and $\vartheta_n\to0$, this decomposition gives the assumed Gaussian factor model, with loading columns $\sqrt{\lambda_{r,n}}e_{r,n}$, and
\begin{equation}
 D_n:=\frac1{n_0^2}\sum_{i,j\in\mathcal H_{0,n}}
       |\operatorname{Corr}(\varepsilon_{n,i},\varepsilon_{n,j})|
 \le \frac{\vartheta_n}{\sigma_{0,*}^2}
 \le \frac1{\sigma_{0,*}^2}\sqrt{\frac{\lambda_{d+1,n}}{n_0}}.
 \label{eq:theory-residual-correlation}
\end{equation}
For $0<\rho_{\mathrm{dep}}\le1/2$, the condition $\vartheta_n=O(n^{-\rho_{\mathrm{dep}}})$ provides the normalized residual-covariance rate used in the consistency analysis and implies $D_n=O(n^{-\rho_{\mathrm{dep}}})$. The stronger spectral condition $\lambda_{d+1,n}/n_0=O(n^{-2\rho_{\mathrm{dep}}})$ is sufficient for this rate.
\end{corollary}
\begin{proof}
    See supplementary note Section S6.
\end{proof}

The leading factors can therefore retain substantial dependence, while the remaining spectrum controls the average residual correlation. Polynomial decay of the normalized covariance remainder supplies the rate used in Theorem~\ref{thm:theory-consistency}.

\subsection{Pooled-oracle multiple testing}
\label{subsec:theory-testing}

We investigate the multiple-testing properties of \texttt{TRAIL} by comparing the resulting BH procedure with an oracle procedure based on the true pooled conditional null defined in  \eqref{eq:model-finite-pool}. Let $\widehat F_n=\widehat F_{n;k_n}^{(T_n)}$. 
We define the pooled oracle as the BH procedure with the knowledge of true pooled conditional null CDF $\overline F_{0,n,U}$ 
for the realized experiment and using it to calibrate every test statistic. 
The estimated and oracle right-sided $p$-values are
\begin{equation}
 \widehat p_{n,i}=1-\widehat F_n(Z_{n,i}),\qquad
 p^o_{n,i}=1-\overline F_{0,n,U}(Z_{n,i}).
 \label{eq:theory-pvalues}
\end{equation}
The calibration is exact after averaging over the null indices as
\[
\frac{1}{n_0}\sum_{i \in \mathcal{H}_{0,n}} \mathbb{P}(p^o_{n,i} \le t \mid U) = t.
\]
Thus its signal CDF and all-test mixture are
\[
 \overline G_{1,n,U}(t)
 =\frac1{n_1}\sum_{i\in\mathcal H_{1,n}}\mathbb P(p^o_{n,i}\le t \mid U),
 \qquad
 \mathsf H_{n,U}(t)
 =\pi_{0,n}t+(1-\pi_{0,n})\overline G_{1,n,U}(t).
\]
At BH level $\alpha\in(0,1)$, let
\[
 t^\star_{n,U}=\sup\{t\in(0,\alpha]:\mathsf H_{n,U}(t)\ge t/\alpha\},
\]
with empty supremum zero. Condition S5 in the supplementary note specifies continuous signal margins with bounded loadings and scales and a weakly dependent Gaussian copula. Condition S6 in the supplementary note requires the crossing to stay positive, with a positive margin in each left neighborhood and a uniformly negative margin beyond each right neighborhood.

Apply BH \citep{BenjaminiHochberg1995} to both vectors in \eqref{eq:theory-pvalues}. Denote the rejection sets by $\widehat{\mathcal R}_n$ and $\mathcal R_n^o$, and the thresholds by $\widehat t_n=\alpha|\widehat{\mathcal R}_n|/n$ and $t_n^o=\alpha|\mathcal R_n^o|/n$.  For any rejection set $\mathcal R$, we may define the false discovery proportion (FDP) and the true discovery proportion (TDP) as follows: 
\[
 \operatorname{FDP}(\mathcal R)
 =\frac{|\mathcal R\cap\mathcal H_{0,n}|}{|\mathcal R|\vee1},
 \qquad
 \operatorname{TDP}(\mathcal R)
 =\frac{|\mathcal R\cap\mathcal H_{1,n}|}{n_1}.
\]
Their expectations are FDR and TPR, respectively. The next result combines uniform null-CDF consistency with the signal and crossing conditions in Conditions S5 and S6 
to transfer estimation accuracy to BH thresholds, false discovery rates, and power.

\begin{theorem}[Recovery of the pooled oracle]
\label{thm:theory-testing}
Suppose the regime of Theorem~\ref{thm:theory-consistency} and Conditions S5--S6 in the supplementary note hold.  For either pair
$(\mathcal R_n,t_n)=(\widehat{\mathcal R}_n,\widehat t_n)$ or
$(\mathcal R_n^o,t_n^o)$,
\[
 t_n-t^\star_{n,U} \xrightarrow{\mathbb{P}(\cdot \mid U)}0,
 \qquad
 \operatorname{FDP}(\mathcal R_n) \xrightarrow{\mathbb{P}(\cdot \mid U)}\alpha\pi_0,
 \qquad
 \operatorname{TDP}(\mathcal R_n)
 -\overline G_{1,n,U}(t^\star_{n,U}) \xrightarrow{\mathbb{P}(\cdot \mid U)}0.
\]
Moreover,
\[
 \frac{|\widehat{\mathcal R}_n\triangle\mathcal R_n^o|}{n}
 \xrightarrow{\mathbb{P}(\cdot \mid U)}0.
\]
These probability limits also hold unconditionally. For either procedure,
\[
 \operatorname{FDR}(\mathcal R_n) \longrightarrow\alpha\pi_0\le\alpha,
 \qquad
 \operatorname{TPR}(\mathcal R_n)
 -\mathbb E\{\overline G_{1,n,U}(t^\star_{n,U})\}
  \longrightarrow0.
\]
The corresponding conditional expectation differences also vanish.
\end{theorem}
\begin{proof}
    See supplementary note Section S5.
\end{proof}

The fitted procedure recovers 
the pooled oracle's asymptotic FDR and power, and the fraction of hypotheses on which their rejection decisions differ tends to zero. These conclusions accommodate the use of the same observations for null-density estimation and testing.

For the sufficient condition below, the signal margins in Condition S5 in the supplementary note are
\[
 \mathbb P(Z_{n,i}\le z \mid U)
 =K\!\left(\frac{z-\mu_{n,i}-\ell_{n,i}^{\mathsf T}U}{\sigma_{n,i}}\right),
 \qquad i\in\mathcal H_{1,n},
\]
where $K$ is a continuous proper CDF, $\mu_{n,i}\ge0$ are signal locations, the loadings $\ell_{n,i}$ are uniformly bounded, and the scales $\sigma_{n,i}$ are uniformly bounded above and away from zero. Under these signal margins, the following corollary combines full left support with initial-center purity in Condition S3 to derive Condition S6 in the supplementary note. It also identifies an explicit limiting BH threshold and establishes power converging to one for both the fitted procedure and the oracle procedure. 

\begin{corollary}[Power under full left-support signal margins]
\label{cor:theory-full-left}
Suppose the regime of Theorem~\ref{thm:theory-consistency} and Condition S5 in the supplementary note hold. If
\[
 K(x)>0\qquad\text{for every }x\in\mathbb R,
\]
then Condition S6 in the supplementary note follows. With
\[
 t_\star=\frac{\alpha(1-\pi_0)}{1-\alpha\pi_0}\in(0,\alpha),
\]
we have $t^\star_{n,U}\to t_\star$ for almost every $U$. For both the fitted and pooled-oracle procedures,
\[
 t_n\xrightarrow{\mathbb{P}(\cdot \mid U)}t_\star,\qquad
 \operatorname{TDP}(\mathcal R_n)\xrightarrow{\mathbb{P}(\cdot \mid U)}1.
\]
The unconditional probability limits are the same, and both conditional and unconditional power converge to one.
\end{corollary}
\begin{proof}
    See supplementary note Section S7.
\end{proof}

The conclusion relies on combining initial center purity with the signal margin assumptions. Since $K$ assigns positive probability below the cutoff, signals with bounded locations have a positive chance of entering the initial center. With bounded loadings and scales, this chance is uniformly positive for signals below any fixed location bound. Therefore, a nonvanishing fraction of such signals would contradict vanishing center contamination. Consequently,
\[
\frac{1}{n_1} \# \{i \in \mathcal{H}_{1,n} : \mu_{n,i} \le M\} \longrightarrow 0 \qquad \text{for every finite } M.
\]
Thus, all but a vanishing proportion of signal locations move to the right, and their oracle $p$-values concentrate near zero. Together with null CDF consistency and the positive limiting BH threshold, this implies power tending to one for both fitted and pooled-oracle procedures. For Gaussian signal margins with $K = \Phi$, the center purity assumption therefore requires that the fraction of signals below any fixed location bound tends to zero.

\section{Numerical Experiments}
\label{sec:numerical}

In this section, we evaluate the finite sample performance of \texttt{TRAIL\_BIC} by comparing the false discovery rate (FDR) and power with those of competing procedures under a range of null dependence structures and empirical null shapes. Each simulated dataset contains $n_0$ true-null statistics and $n_1$ nonnull statistics. We consider the one-sided multiple testing problem
\[
H_{0i}:\mu_i = 0 \quad \text{vs} \quad H_{1i}: \mu_i > 0, \quad i = 1, \cdots , n_0 + n_1,
\]
and evaluate the performance of the proposed method under various dependence structures of the pooled conditional null distributions. We vary the dependence structure and the realized empirical null shape while preserving the specified standard normal marginal distributions.

We consider six null configurations, beginning with Gaussian dependence structures and a spiky null with a symmetric base, followed by three settings that introduce asymmetric empirical null shapes. Throughout, individual null statistics retain marginal standard normal distributions, and the differences concern the dependence structure and the empirical distribution observed within a single simulated dataset.

\begin{itemize}
\item \textbf{Exchangeable Gaussian null.}
The null data $\boldsymbol{Z}^{\mathrm{null}}=(Z_i)_{i \in \mathcal{H}_0}$ has multivariate normal distribution,
\[  
\boldsymbol{Z}^{\mathrm{null}} \sim N_{n_0}(\boldsymbol{0}, \Sigma^{\mathrm{null}}),
\quad
\Sigma^{\mathrm{null}}=(\Sigma^{\mathrm{null}}_{ij})_{1\leq i,j\leq n_0},
\]
with 
\[
\Sigma_{ij}^{\mathrm{null}} =
\begin{cases}
    \rho, &i \neq j, \\
    1, &i = j,
\end{cases}
\]
for nonnegative correlation coefficients $\rho \ge 0$. 
Varying $\rho$ allows us to examine the effect of homogeneous positive dependence on the empirical null and the resulting testing performance.

\item \textbf{AR(1) Gaussian null.}
We next consider a multivariate Gaussian null with correlation matrix entries $\Sigma_{ij}^{\mathrm{null}}=\rho^{|i-j|}$, where positive values of $\rho$ are considered. This setting examines dependence that varies with the distance between coordinates.

\item \textbf{Leptokurtic null with a symmetric base.} 
Based on the two group mixture model \eqref{eq:two_group}, we model the pooled null density as
\[
\overline{f}_{0,U}(z) =\pi_{01} \frac{1}{\sigma_{\mathrm{base}}}\phi \left(\frac{z-\mu_{01}}{\sigma_{\mathrm{base}}}\right) + (1-\pi_{01}) \frac{1}{\sigma_{\mathrm{spike}}} \phi\left(\frac{z-\mu_{02}}{\sigma_{\mathrm{spike}}}\right),
\]
which is a mixture of two normal distributions, combining a narrow Gaussian spike with a broader Gaussian base with a shared latent-factor construction. The two components have different conditional variances, and their variance contrast $\kappa=\sigma^2_{\mathrm{base}}/\sigma^2_{\mathrm{spike}}\ge1$ controls the prominence of the spike, where $\sigma^2_{\mathrm{base}}$ and $\sigma^2_{\mathrm{spike}}$ denote the conditional variances of the base and spike, respectively, given the common factor. This setting examines the performance when a sharply concentrated peak coexists with more dispersed null observations.

\item \textbf{Skewed null from a survival Clayton copula.}
We generate dependent uniform variables from a survival Clayton copula \citep{Clayton1978} and transform each coordinate through the standard normal quantile function; {however the joint distribution is no longer multivariate normal.} The construction induces upper-tail dependence and can produce a skewed empirical null while preserving standard normal margins. Skewness strength is controlled by the parameter motivated by Kendall's tau, denoted as $\tau_{\mathrm K}$ with Clayton parameter $\theta=2\tau_{\mathrm K}/(1-\tau_{\mathrm K})$.

\item \textbf{Asymmetric bimodal null.}
The null statistics are randomly assigned to two unequal latent
blocks, with observations within each block partially sharing a
block-specific common factor. This construction produces two
unequally weighted clusters while maintaining standard normal
margins. The parameter $\pi_{\mathrm{left}}$, defined as the
proportion assigned to the left cluster, controls the relative
weights of the two clusters and the resulting empirical asymmetry.

\item \textbf{Heteroscedastic bimodal null.}
Unlike the asymmetric bimodal setting based on unequal mixing weights, we use two equally weighted latent components with different conditional variances, allowing a narrow, sharply peaked cluster to coexist with a broader, more gently sloping cluster. The signed parameter $\kappa$ controls this variance contrast, with $|\kappa|=\sigma^2_{\mathrm{broad}}/\sigma^2_{\mathrm{narrow}}\ge1$, where $\sigma^2_{\mathrm{broad}}$ and $\sigma^2_{\mathrm{narrow}}$ denote the larger and smaller component variances conditional on the latent factors. For $|\kappa|>1$, positive $\kappa$ places the narrower, steeper peak on the left, whereas negative $\kappa$ places it on the right. This setting examines performance when the two empirical peaks differ in width and steepness, with the direction of asymmetry controlled by the sign of $\kappa$.
\end{itemize}
For the signal data, we generate $n_1$ test statistics jointly from a multivariate normal distribution with an AR(1) dependence structure with autocorrelation parameter $\rho=0.7$, 
\[
\boldsymbol{Z}^{\mathrm{signal}} = (Z_i)_{i \in \mathcal{H}_1} \sim N_{n_1}(\boldsymbol{\mu}, \Sigma^{\mathrm{signal}}),
\qquad
\Sigma^{\mathrm{signal}}_{ij} = 0.7^{|i-j|},
\quad
i,j = 1, \cdots,n_1,
\]
representing a moderate level of dependence among the signal statistics. All signal coordinates share the same positive mean, so that the signals are systematically shifted toward the right tail. The magnitude of this mean is determined by the signal-strength parameter $\tau_{\mathrm{signal}}$, the total number of hypotheses $n_0 + n_1$, the number of signals $n_1$, and the overall dependence strength of the null statistics. In particular, the null dependence is summarized by the normalized Frobenius norm of the null covariance matrix, $\|\boldsymbol{\Sigma}^\mathrm{null}\|_F/n_0$, which aggregates the squared correlations among the null statistics. The common signal mean is then set to $\mu_i = \sqrt{2(1-\|\boldsymbol{\Sigma}^\mathrm{null}\|_{F}/n_0)\tau_{\mathrm{signal}}\log(n_0 + n_1)}/n_{1}^{1/3}$. This construction allows the signal magnitude to adapt to the dependence structure of the corresponding null scenario while keeping the dependence structure among the signals fixed across simulation settings.

Throughout the simulations, we generate $n_0=500$ null statistics and $n_1=50$ nonnull statistics, giving $n=550$ tests in total, with signal strength parameter $\tau_{\mathrm{signal}}=5$. The proposed \texttt{TRAIL\_BIC} method is implemented with candidate sets $\mathscr{C} = \{0, 1, \cdots, 4\}$, $\mathscr{K} = \{2,4, \cdots, 16\}$, and parameters $q = 0.85$, $T = 3$, and $M = 5$, where $c \in \mathscr{C}$ controls the degree of data shift before estimation, $k \in \mathscr{K}$ controls the complexity of the null density, $q$ determines the initial proportion of the central subset, $T$ is the number of iterations, and $M$ is the number of replications within each reconstruction stage used to stabilize the estimator. After obtaining the $p$-values, the Benjamini--Hochberg procedure is applied at significance level $\alpha = 0.1$. Each simulation configuration is evaluated using 200 simulated datasets.

We compare the FDR, power, and false discovery proportion (FDP) exceedance probability across several methods for constructing $p$-values under different null specifications: (i) the standard normal method, (ii) the conditional likelihood method considered at Section \ref{subsec:motivation}, (iii) the \texttt{locfdr} method \citep{Efron2004}, (iv) the \texttt{Sequential Bayesian} method \citep{LiangLiuWang2007}, and (v) the proposed \texttt{TRAIL\_BIC} method. In particular, we consider $\Pr(\mathrm{FDP}>0.1)$, estimated by the proportion of Monte Carlo replications in which the realized FDP exceeds 0.1, to assess the frequency of excessive false discoveries beyond what is captured by the average FDR. For the latter four methods, the null density is estimated from the observed test statistics, and the resulting estimate is used to compute the corresponding $p$-values. For the standard normal benchmark, $p$-values are computed directly under the $N(0,1)$ null distribution. The Benjamini--Hochberg (BH) procedure is then applied to these $p$-values for multiple testing. In addition, we include a version of \texttt{TRAIL\_BIC} without the data-shift step as an auxiliary benchmark to illustrate the effect of data shifting on the operating characteristics of \texttt{TRAIL\_BIC}.

\subsection{False discovery rate}

\begin{figure}[t]
    \centering
    \includegraphics[width=0.99\linewidth]{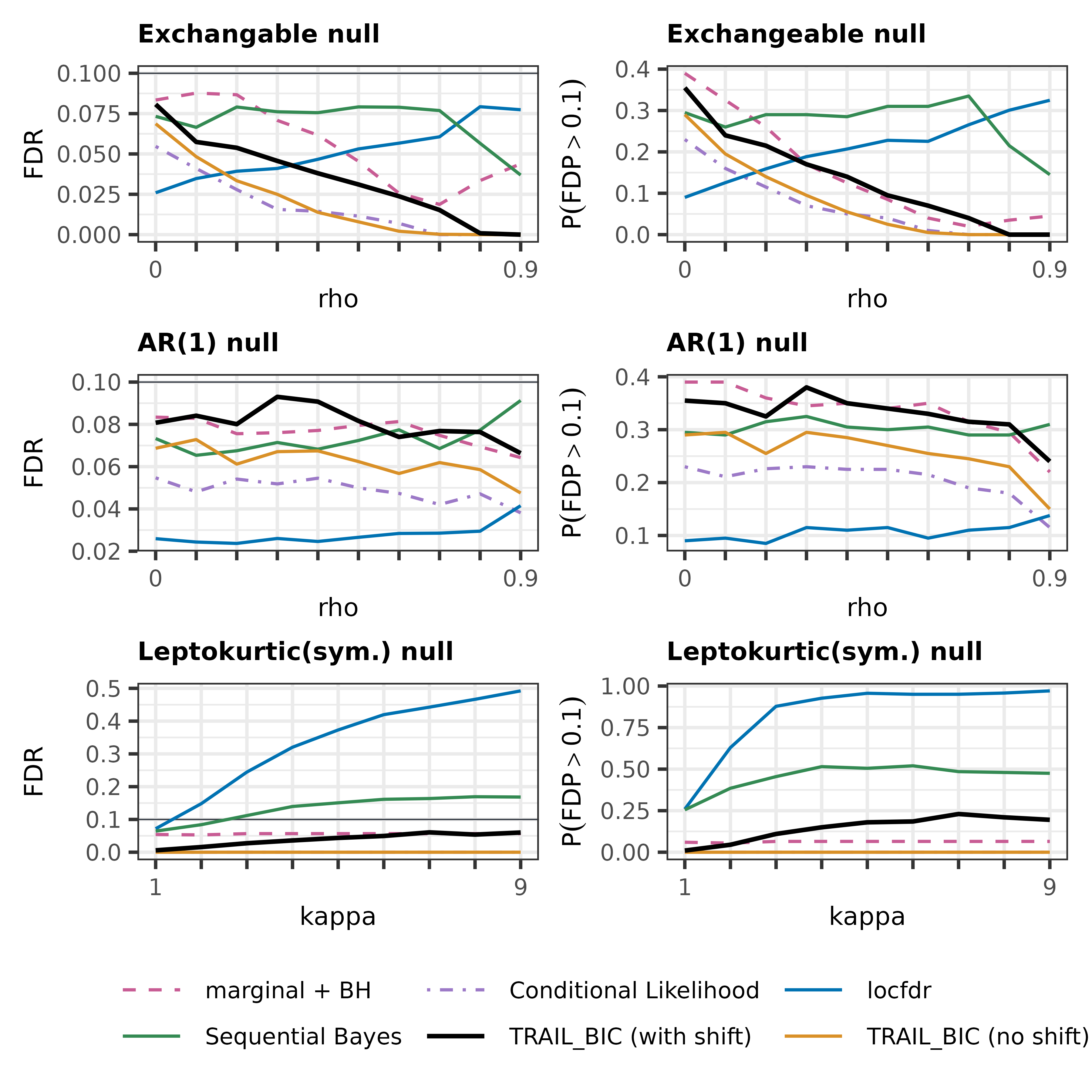}
    \caption{Estimated FDR and FDP exceedance probability
    $\Pr(\mathrm{FDP}>0.1)$ under symmetric null configurations}
    \label{fig:fdr_sym}
\end{figure}

Figure \ref{fig:fdr_sym} presents the empirical FDR and FDP exceedance probability under three symmetric null settings. All methods maintain estimated FDRs below the nominal level 0.1 under the exchangeable and AR(1) nulls. Under exchangeable dependence, the FDR of \texttt{locfdr} tends to increase with correlation, whereas that of \texttt{TRAIL\_BIC} decreases. Under AR(1) dependence, variation across correlation levels is relatively limited, with the tendency of \texttt{locfdr} being quite conservative. Under the leptokurtic null, however, \texttt{locfdr} and \texttt{Sequential Bayesian} exceed the nominal FDR level as $\kappa$ increases, while the remaining methods stay below it. The distinction becomes particularly evident in the FDP exceedance probabilities. For large $\kappa$, the FDP exceedance probability of \texttt{locfdr} approaches one, and that of \texttt{Sequential Bayesian} reaches approximately 0.5, whereas that of \texttt{TRAIL\_BIC} remains below 0.25 throughout.

\begin{figure}[t]
    \centering
    \includegraphics[width=0.99\linewidth]{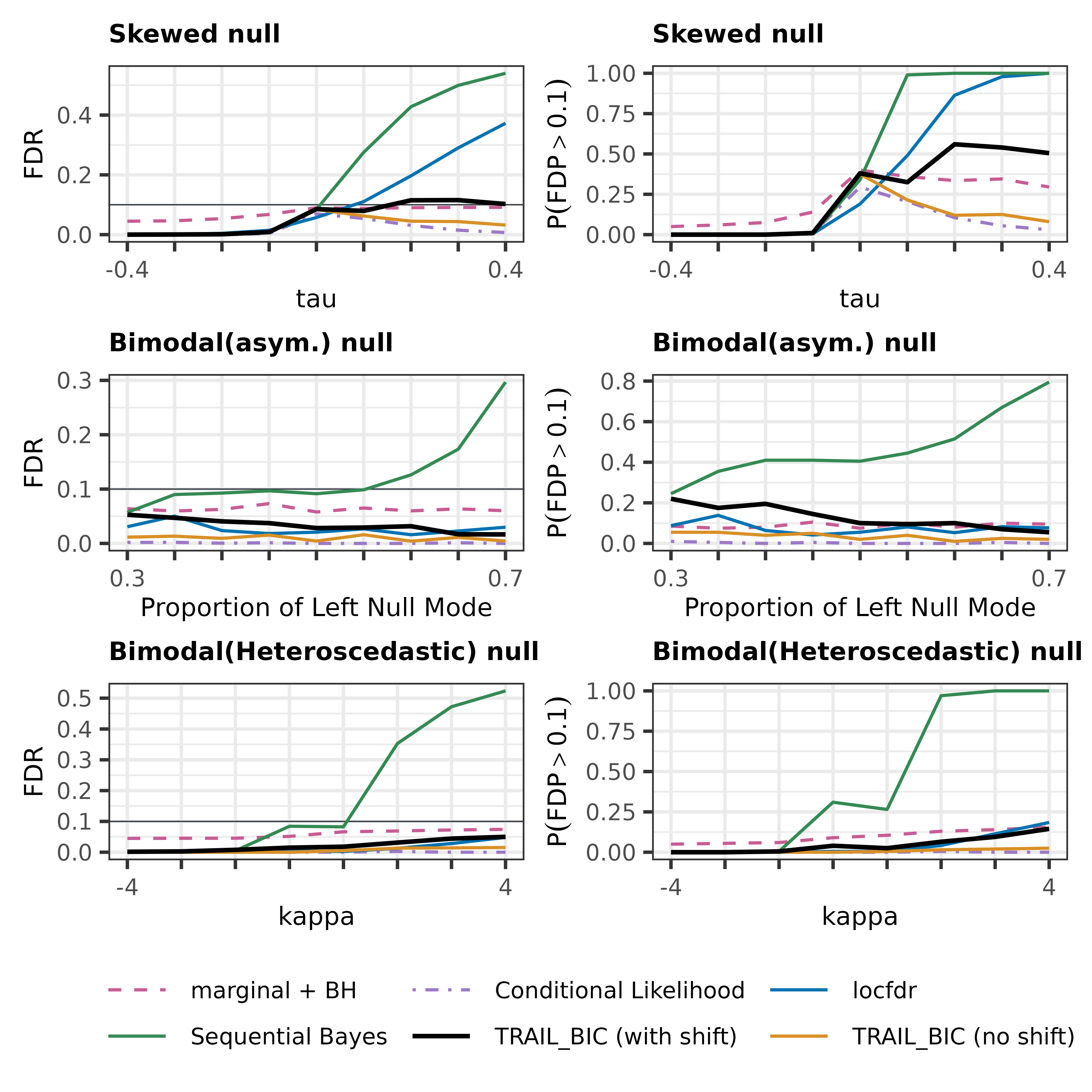}
    \caption{Estimated FDR and FDP exceedance probability
    $\Pr(\mathrm{FDP}>0.1)$ under asymmetric null configurations}
    \label{fig:fdr_asym}
\end{figure}

Figure \ref{fig:fdr_asym} reports the corresponding results for asymmetric null settings. Under the skewed null, all methods maintain their estimated FDRs below the nominal level 0.1 for $\tau_{\mathrm K}<0$. But for positive $\tau_{\mathrm K}$, several methods exhibit inflated FDR with the most pronounced for \texttt{Sequential Bayesian}, followed by \texttt{locfdr}, while \texttt{TRAIL\_BIC} shows only a slight exceedance. The corresponding FDP exceedance probability approaches one for \texttt{Sequential Bayesian} and increases sharply for \texttt{locfdr}. Although the exceedance probability of \texttt{TRAIL\_BIC} also tends to increase with $\tau_{\mathrm K}$, the increase is considerably less pronounced than those of the two competing empirical null methods.

Under the asymmetric bimodal and heteroscedastic bimodal nulls, only \texttt{Sequential Bayesian} exceeds the nominal FDR level, for $\pi_{\mathrm{left}}\geq 0.6$ and positive $\kappa$, respectively. Its FDP exceedance probabilities are also higher, approaching one when $\kappa\geq 2$ in the heteroscedastic case.

Overall, these results highlight the importance of evaluating both the average FDR and the probability of exceeding the nominal FDP threshold. While most methods maintain FDR control under conventional correlation structures, substantial differences arise when the realized null distribution exhibits more complex shapes, particularly pronounced leptokurtosis, skewness, or asymmetric bimodality. Across these settings, \texttt{TRAIL\_BIC} generally provides more stable FDR performance than the competing empirical null methods, with comparatively lower FDP exceedance probabilities.

\subsection{Power}

\begin{figure}[t]
    \centering
    \includegraphics[width=0.99\linewidth]{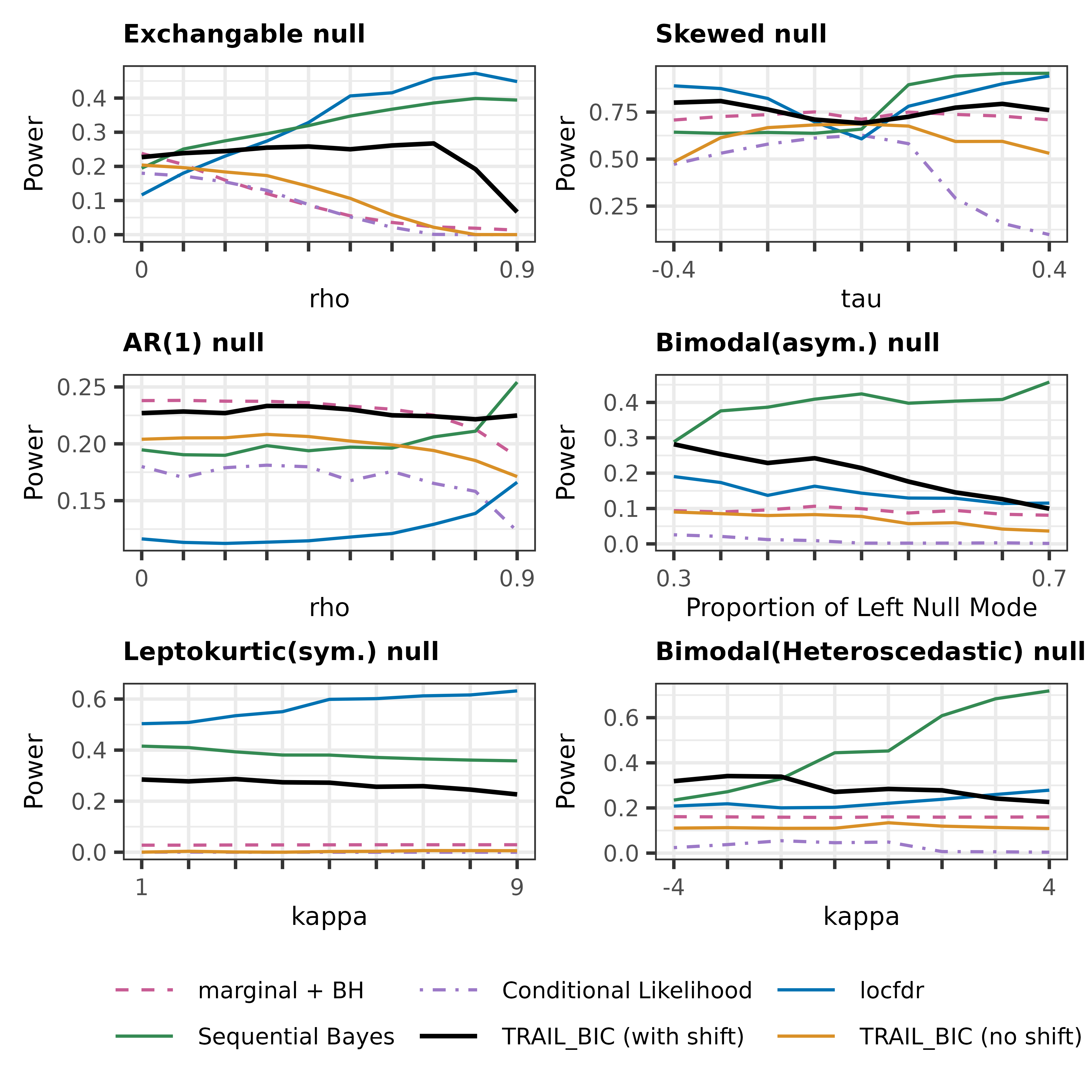}
    \caption{Power of different correlation structures (Symmetric null/Asymmetric null)}
    \label{fig:power}
\end{figure}

Figure \ref{fig:power} presents the empirical power of the competing methods under the six null settings. Under the exchangeable null, the power of \texttt{locfdr} and \texttt{Sequential Bayesian} increases with the correlation coefficient $\rho$, and achieve the highest power under moderate to strong dependence. In contrast, \texttt{TRAIL\_BIC} remains relatively stable, although its power decreases somewhat under strong dependence. Under the AR(1) null, all methods exhibit relatively stable power across different correlation levels. The standard normal method with BH correction and \texttt{TRAIL\_BIC} achieve the highest power, followed closely by \texttt{Sequential Bayesian}, while \texttt{locfdr} exhibits the lowest power among the empirical--null methods. Under the leptokurtic null, power is also relatively stable, with \texttt{locfdr} ranking highest, followed by \texttt{Sequential Bayesian} and \texttt{TRAIL\_BIC}. However, as shown in Figure \ref{fig:fdr_sym}, both \texttt{locfdr} and \texttt{Sequential Bayesian} fail to maintain FDR control as $\kappa$ increases. Consequently, their higher power should be interpreted with caution, as it is accompanied by an inflated proportion of false discoveries.

Under the skewed null, the relative performance of the methods depends substantially on the direction of skewness. For negative $\tau_{\mathrm K}$, where all methods successfully control FDR, \texttt{locfdr} achieves the highest power, followed by \texttt{TRAIL\_BIC}, and the standard normal method with BH correction. For positive $\tau_{\mathrm K}$, however, the ranking changes considerably with \texttt{Sequential Bayesian} exhibiting the highest power, followed by \texttt{locfdr} and \texttt{TRAIL\_BIC}. As observed in Figure \ref{fig:fdr_asym}, the substantial power of \texttt{Sequential Bayesian} and \texttt{locfdr} in these settings is accompanied by a failure to control FDR. Although \texttt{TRAIL\_BIC} also exhibits mild FDR inflation under sufficiently strong positive skewness, its excess is considerably smaller than those of the two competing methods.

Under the asymmetric bimodal null, \texttt{Sequential Bayesian} consistently achieves the highest power across all considered values of $\pi_{\mathrm{left}}$, followed by \texttt{TRAIL\_BIC}. Nevertheless, the FDR of \texttt{Sequential Bayesian} exceeds the nominal level when $\pi_{\mathrm{left}} \ge 0.6$, suggesting that its power advantage under pronounced modal imbalance comes at the expense of FDR control. 

For the heteroscedastic bimodal null, \texttt{TRAIL\_BIC} achieves the highest power for negative values of $\kappa$, followed by \texttt{Sequential Bayesian} and \texttt{locfdr}. As $\kappa$ increases into the positive range, the power of \texttt{Sequential Bayesian} increases, but this improvement is accompanied by FDR inflation, as shown in Figure \ref{fig:fdr_asym}.

Taken together, these results should be interpreted alongside the FDR results in Figures \ref{fig:fdr_sym} and \ref{fig:fdr_asym}. Although \texttt{locfdr} and \texttt{Sequential Bayesian} achieve higher power in several settings, their advantages are often accompanied by a failure to maintain FDR control. In contrast, \texttt{TRAIL\_BIC} generally provides a more favorable balance between power and FDR control across the considered null settings. 

Furthermore, comparing the two variants of \texttt{TRAIL\_BIC} reveals that incorporating the data-shift step generally leads to higher power than omitting it. Also, the conditional likelihood method tends to yield more conservative testing decisions than the other methods in these simulations. Together, these results highlight the importance of examining how accurately each method captures the shape of the null density in both the central and tail regions for balancing power and FDR control. The following subsection examines these aspects of density estimation in greater detail.

\subsection{Null density estimation}

\begin{figure}[t]
    \centering

    \begin{subfigure}{0.48\textwidth}
        \centering
        \includegraphics[width=\linewidth]{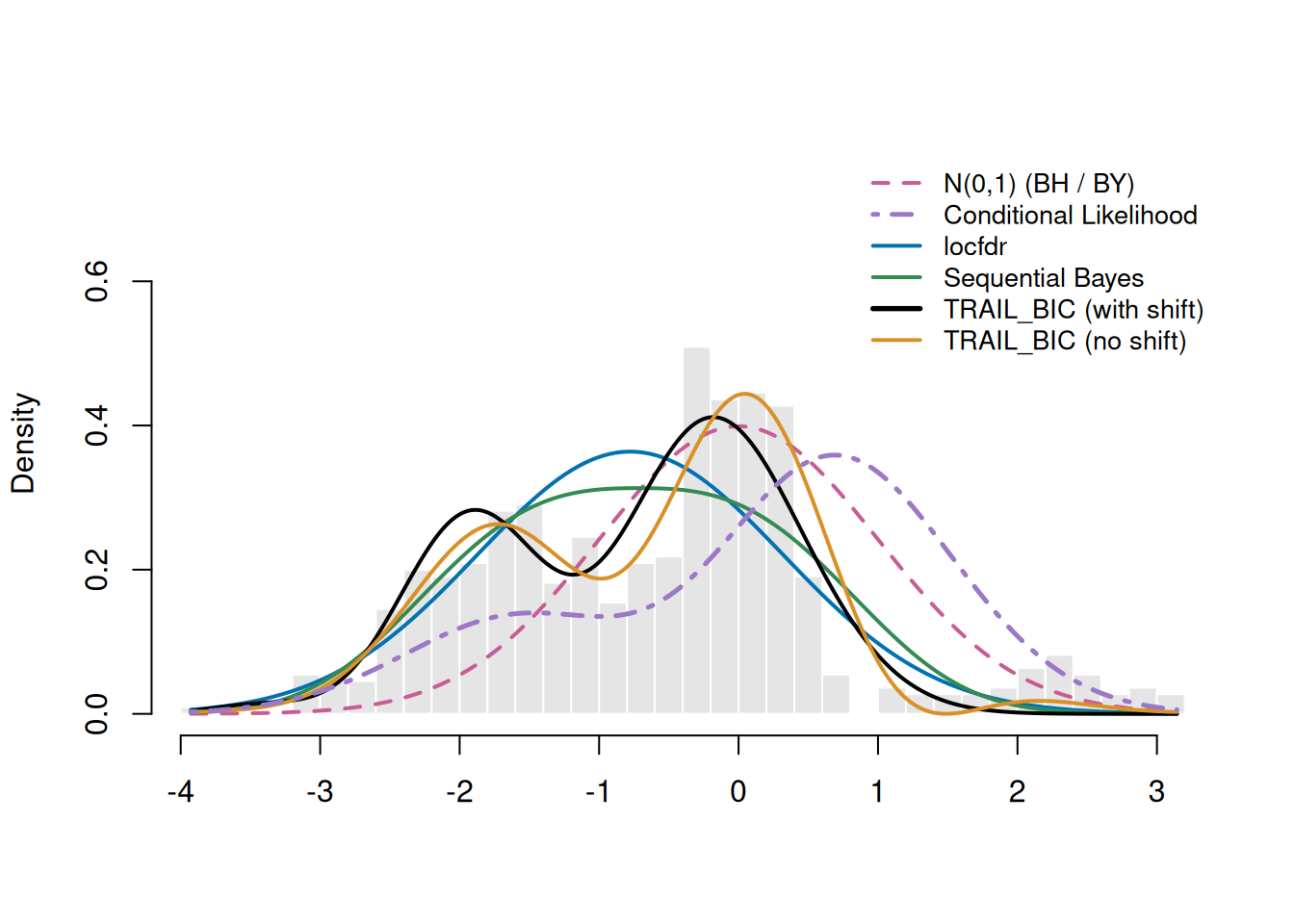}
        \caption{Example 1}
    \end{subfigure}
    \hfill
    \begin{subfigure}{0.48\textwidth}
        \centering
        \includegraphics[width=\linewidth]{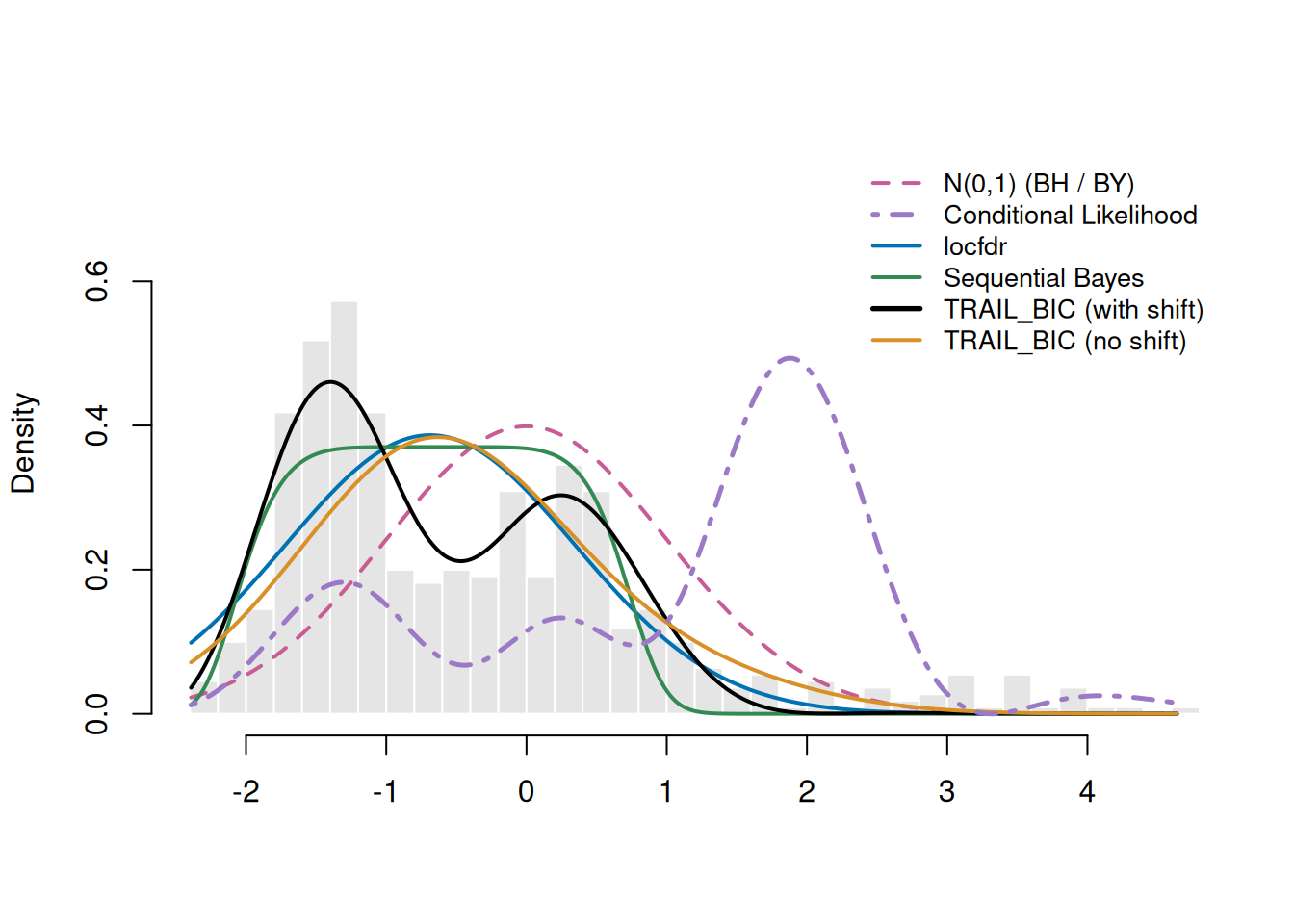}
        \caption{Example 2}
    \end{subfigure}
    \caption{Example of estimated null densities}
    \label{fig:examples}
\end{figure}

Figure \ref{fig:examples} presents two illustrative realizations from the Monte Carlo replications of our numerical experiments under the heteroscedastic bimodal null setting. The location of the realized pooled null can vary with the latent factors underlying the dependence structure, and the fixed N(0,1) reference is visibly misaligned with the null observations in these examples, compared with the empirical null based methods. Both examples exhibit clear bimodality, which is captured by \texttt{TRAIL\_BIC} but not by \texttt{Sequential Bayesian} or \texttt{locfdr}. In Example 1, the left null component is more dispersed than the right null component, whereas the opposite holds in Example 2. The density estimated by \texttt{Sequential Bayesian} method appears to reflect primarily the left flank, under its symmetric distributional assumption. Consequently, its estimated right tail, where the signals are located, is overly heavy in Example 1 and overly light in Example 2. \texttt{locfdr} exhibits a similar tendency, although less pronounced. In comparison, \texttt{TRAIL\_BIC} provides a visually more faithful representation of both the body and the right tail of the null distribution.

In both examples, the conditional likelihood estimate differs substantially from the empirical shape of the null distribution. In Example 2, it produces a pronounced spurious peak in the right tail. These discrepancies illustrate the finite-sample extrapolation problem discussed in Section \ref{subsec:motivation} that estimation based on the conditional density within the initial center can yield an unreliable extrapolation beyond that region.

Comparing the shifted and unshifted versions of \texttt{TRAIL\_BIC} further illustrates the role of the Gaussian reference location in finite-order approximation. When the fixed Gaussian reference is poorly aligned with the null distribution, the unshifted representation may require a larger truncation order $k$ to capture complex features in the central region. Because higher-order Hermite components are oscillatory and act globally, using them to improve the central fit can also introduce spurious fluctuations in the tails. Adjusting the reference location may reduce the reliance on these higher-order corrections. In Example 1, the unshifted estimate exhibits an artificial oscillation in the right tail, whereas the shifted estimate decreases smoothly. To the extent that this oscillation introduces excess probability mass beyond an observed test statistic, the unshifted estimate may produce inflated right-sided $p$-values and reduce detection power. In Example 2, the unshifted estimate fails to recover the bimodal structure, while the shifted estimate captures both modes. Together, these examples suggest that jointly selecting the reference location and truncation order can improve the representation of both the central shape and tail behavior of the null density, potentially yielding more accurate right-sided $p$-values and improved signal detection.

\section{Real Data Analysis}
\label{sec:real-data}

We considered study-level summary association statistics from the CARDIoGRAMplusC4D 1000 Genomes-based meta-analysis of Coronary Artery Disease (CAD) \citep{Magosi2017}. The dataset contains association estimates for 214 CAD-related variants across 48 contributing studies, with the number of available variants varying slightly across studies. The analyzed variants were selected for evidence of CAD association in the original meta-analysis, so the study-level collections are enriched for association signals. For each study, we constructed signed test statistics from the harmonized log-odds-ratio estimates and their genomic-control-corrected standard errors and treated the resulting collection of statistics as a separate multiple-testing problem. Each available variant corresponds to one hypothesis test, and the studies are analyzed separately. We used the same tuning parameters and implementation settings for \texttt{TRAIL\_BIC} as in Section~\ref{sec:numerical}.

Because the true null and non-null status of individual variants is not available in these observational data, quantities such as the empirical FDR and power cannot be directly evaluated. We therefore compare the methods primarily through the number of rejected hypotheses within each study and summarize how frequently each procedure yields the largest number of discoveries across the 48 studies. This analysis is intended to illustrate how the competing null-calibration strategies translate into different discovery patterns across heterogeneous real GWAS datasets, rather than to regard a larger number of rejections by itself as evidence of superior error control.

\begin{figure}[t]
    \centering    
    \includegraphics[width=0.9\linewidth]{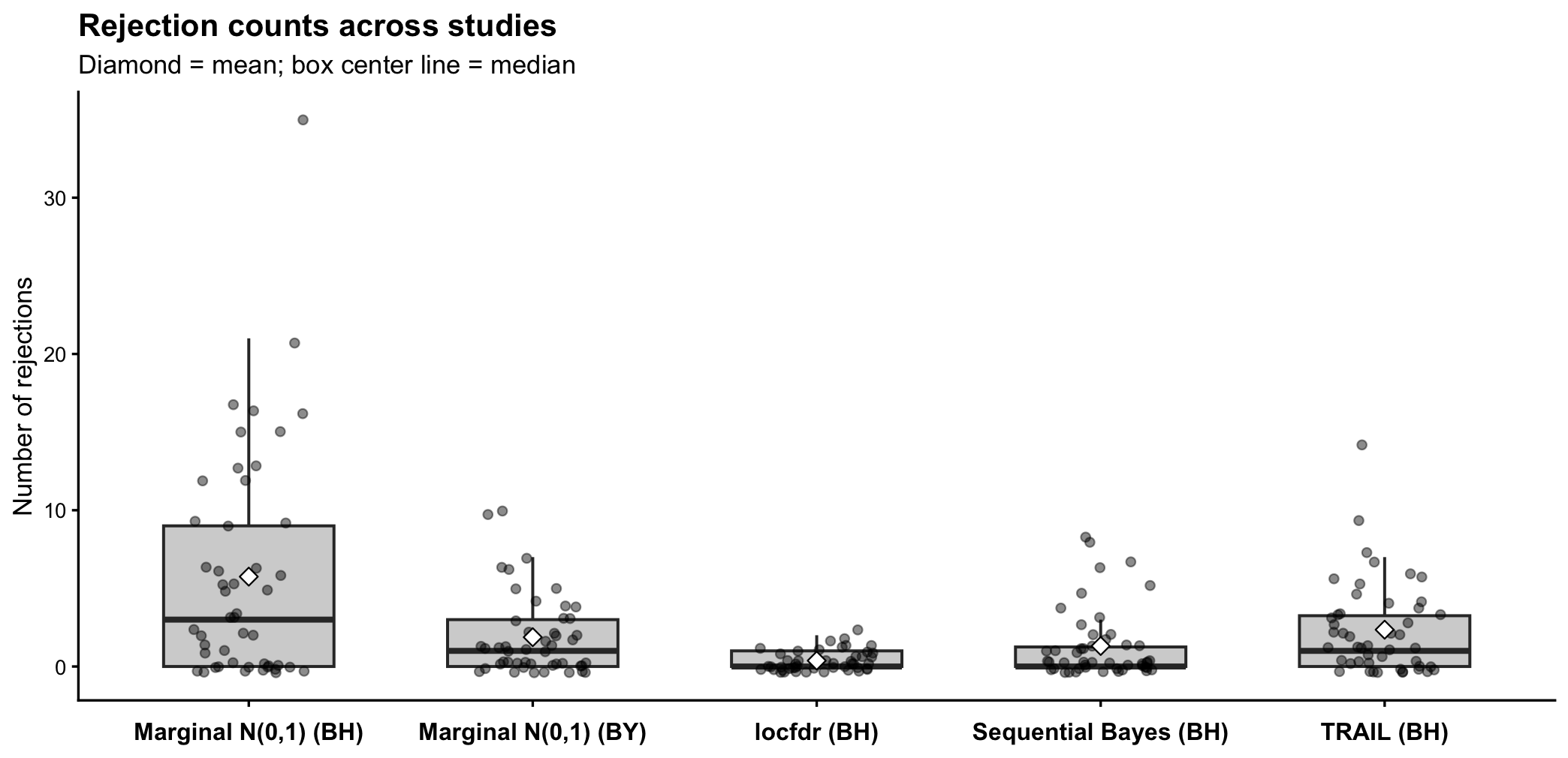}
    \caption{Boxplot of rejection counts across 48 studies}
    \label{boxplot_realdata}
\end{figure}

Figure \ref{boxplot_realdata} and Table \ref{rejection_summary_realdata} summarize the rejection counts across the 48 studies. The procedures based on \texttt{locfdr} and \texttt{Sequential Bayesian} yielded relatively few discoveries, with mean rejection counts of 0.38 and 1.31, respectively, and a median of zero for both methods. Both procedures produced fewer rejections on average than $N(0,1)$ with Benjamini--Yekutieli (BY) \citep{BenjaminiYekutieli2001} adjustment, whose mean and median rejection counts were 1.88 and 1, respectively. Thus, despite using BH adjustment, these two empirical null procedures exhibited more restrictive rejection behavior than the marginal BY benchmark.

In contrast, $N(0,1)$ with BH adjustment produced the largest mean rejection count, 5.75, and the largest between-study standard deviation, 7.28. \texttt{TRAIL\_BIC} with BH adjustment occupied an intermediate position between the two marginal procedures in terms of both the mean and median rejection counts, yielding 2.35 and 1, respectively. The mean rejection rates followed the same ordering: 2.80\% for marginal BH, 1.16\% for \texttt{TRAIL\_BIC}, and 0.918\% for marginal BY. Accordingly, \texttt{TRAIL\_BIC} yielded more discoveries on average than the other empirical null procedures and marginal BY, while making fewer rejections than marginal BH. The larger discovery count of marginal BH could reflect additional true associations, additional false rejections, or both; these possibilities cannot be distinguished from the rejection counts alone. The intermediate rejection profile of \texttt{TRAIL\_BIC} should therefore be interpreted as a difference in discovery behavior rather than direct evidence of a particular FDR--power trade-off.

\begin{table}[t]
\centering
\resizebox{\textwidth}{!}{%
\begin{tabular}{lcrrrrrr}
\toprule
Method 
& Adjustment
& Studies Used 
& Mean Rejections 
& SD Rejections 
& Median Rejections 
& Mean Rejection Rate 
& SD Rejection Rate \\
\midrule
$N(0,1)$ & BH & 48 & 5.75  & 7.28  & 3 & 0.0280  & 0.0353  \\
\midrule
$N(0,1)$ & BY & 48 & 1.88  & 2.57  & 1 & 0.0092 & 0.0125  \\
\texttt{locfdr} & BH             & 48 & 0.38 & 0.61 & 0 & 0.0018 & 0.0031 \\
\texttt{Sequential Bayesian} & BH   & 48 & 1.31  & 2.23  & 0 & 0.0064 & 0.0111  \\
\texttt{TRAIL\_BIC} & BH              & 48 & 2.35  & 2.91  & 1 & 0.0116  & 0.0143  \\
\bottomrule
\end{tabular}%
}
\caption{Rejection summary across 48 studies}
\label{rejection_summary_realdata}
\end{table}

\begin{table}[t]
\centering
\begin{threeparttable}
\begin{tabular}{lcrr}
\toprule
Method 
& Adjustment
& With $N(0,1)+\operatorname{BH}$ 
& Without $N(0,1)+\operatorname{BH}$ \\
\midrule
$N(0,1)$ & BH & 31 & -- \\
\midrule
$N(0,1)$ & BY & 1  & 16 \\
\texttt{locfdr} & BH             & 0  & 1  \\
\texttt{Sequential Bayesian} & BH   & 4  & 14  \\
\texttt{TRAIL\_BIC} & BH              & 8 & 24 \\
\midrule
$\#$ of all zero studies & -- & 14 & 14 \\
Total studies & -- & 48 & 48 \\
\bottomrule
\end{tabular}

\begin{tablenotes}[flushleft]
\footnotesize
\item \textit{Note:} Ties for the largest positive number of rejections are counted for each tied method. Studies with zero rejections across all methods are reported separately as "all-zero studies." Hence, first-place counts need not sum to the total number of studies.
\end{tablenotes}

\end{threeparttable}
\caption{Frequency of largest positive rejection counts, with and without the $N(0,1)+\operatorname{BH}$ method}
\label{firstplace_count_realdata}
\end{table}

Table \ref{firstplace_count_realdata} complements these aggregate summaries by reporting how frequently each procedure attained the largest positive rejection count within a study. All procedures made zero rejections in 14 of the 48 studies. These studies were reported separately and excluded from the first-place comparisons, leaving 34 studies with at least one rejection by at least one procedure. Ties for the largest positive count were credited to every tied procedure, so the first-place counts need not sum to the number of studies being compared. We report the comparison both with all five procedures included and with marginal BH omitted, the latter providing an additional comparison among the empirical null procedures and the marginal BY benchmark.

With all procedures included, marginal BH attained the largest positive rejection count in 31 studies. \texttt{TRAIL\_BIC} had the second-highest first-place frequency, attaining the largest count in 8 studies, followed by \texttt{Sequential Bayesian} in 4 studies and marginal BY in 1 study. When marginal BH was omitted, \texttt{TRAIL\_BIC} had the highest first-place frequency, attaining the largest positive rejection count, including ties, in 24 of the 34 studies (70.6\%). The corresponding counts were 16 for marginal BY, 14 for \texttt{Sequential Bayesian}, and 1 for \texttt{locfdr}. Together with the aggregate summaries, these results show that the relatively high discovery yield of \texttt{TRAIL\_BIC} was accompanied by frequent study-level maxima among the empirical null procedures and marginal BY.

\begin{figure}[t]
    \centering    
    \includegraphics[width=0.9\linewidth]{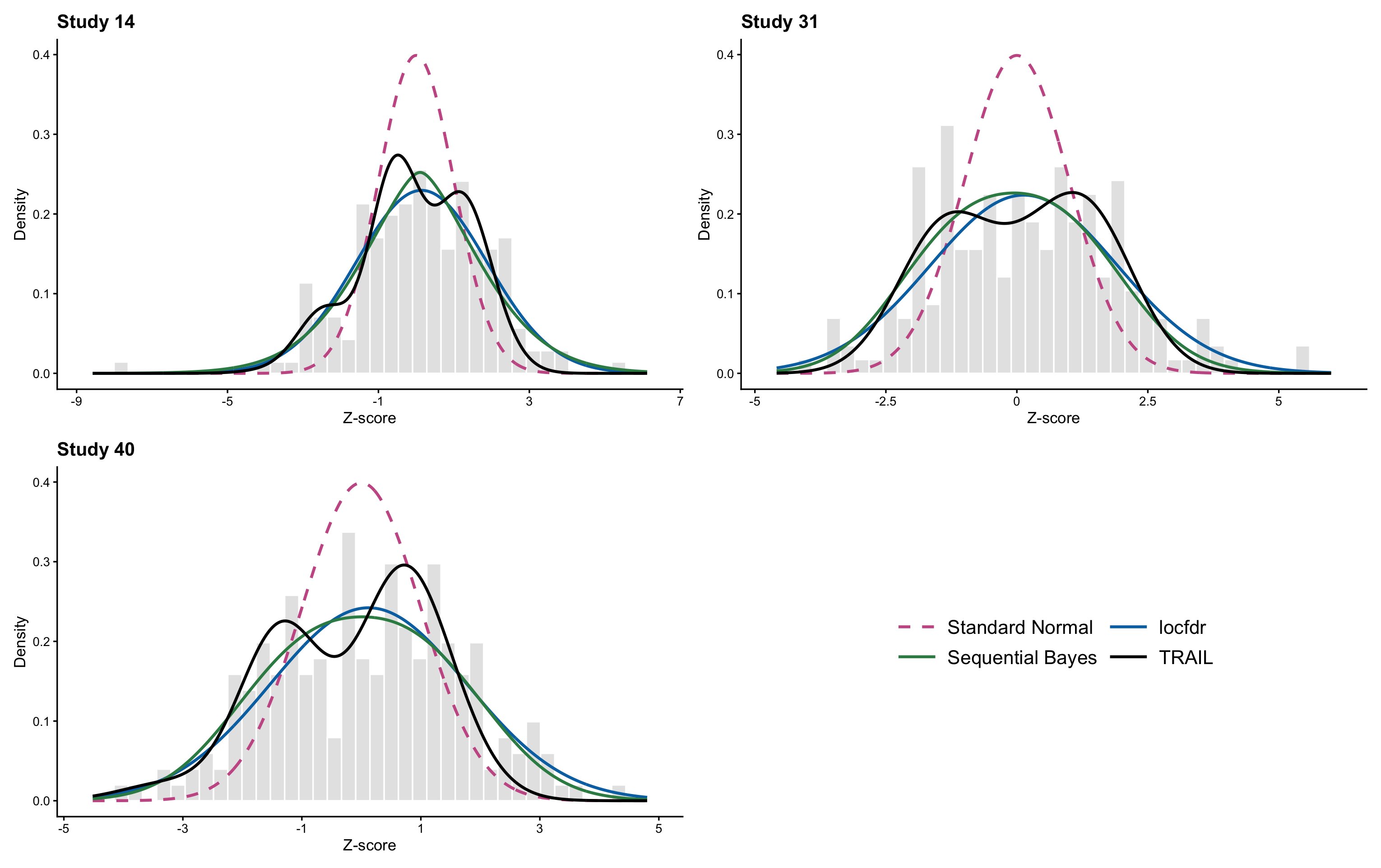}
    \caption{Estimated null densities of Studies 14, 31, 40}
    \label{fig:real_data_null_densities}
\end{figure}

To complement the aggregate comparisons, we examine Studies 14, 31, 40 as illustrative examples. These studies were selected since their empirical distributions of the test statistics exhibit complex bimodal patterns, providing an opportunity to examine how the different null-calibration methods accommodate such departures from a Gaussian shape and how these differences are reflected in the resulting $p$-values and rejection decisions.

Figure \ref{fig:real_data_null_densities} compares the estimated null densities with the empirical distributions in these studies. The standard normal reference and the null densities fitted by \texttt{locfdr} and \texttt{Sequential Bayesian} remain unimodal and do not reproduce the observed two-mode structure. In the implementations considered here, the Gaussian null specification of \texttt{locfdr} and the symmetric generalized normal specification of \texttt{Sequential Bayesian} restrict their ability to accommodate bimodality. In contrast, \texttt{TRAIL\_BIC} produces bimodal density estimates that more closely follow the observed empirical shapes. These examples illustrate the additional flexibility of \texttt{TRAIL\_BIC} relative to the unimodal null specifications used by the competing procedures.

\begin{figure}[t]
    \centering

    \begin{subfigure}{0.48\textwidth}
        \centering
        \includegraphics[width=\linewidth]{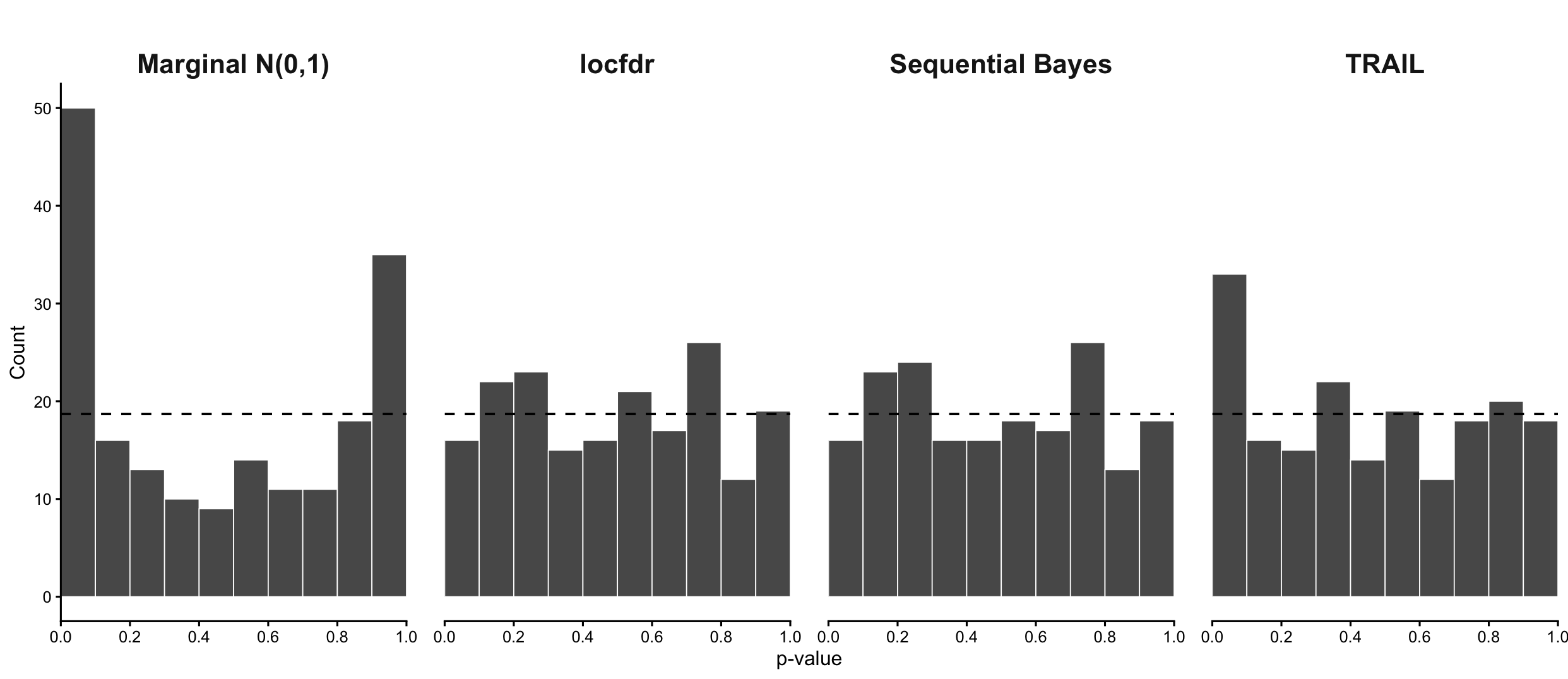}
        \caption{Study 14}
    \end{subfigure}
    \hfill
    \begin{subfigure}{0.48\textwidth}
        \centering
        \includegraphics[width=\linewidth]{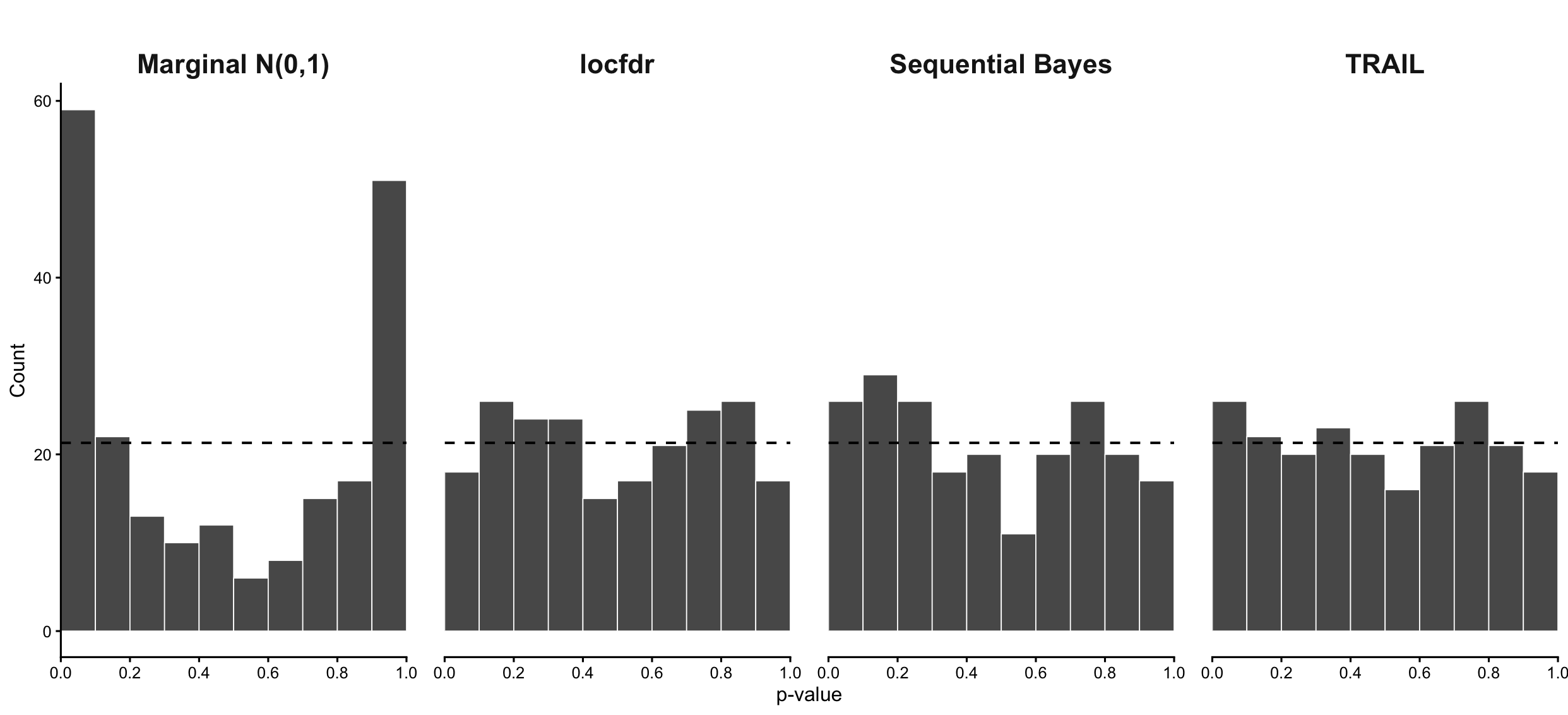}
        \caption{Study 31}
    \end{subfigure}

    \par\vspace{0.5em}

    \begin{subfigure}{0.48\textwidth}
        \centering
        \includegraphics[width=\linewidth]{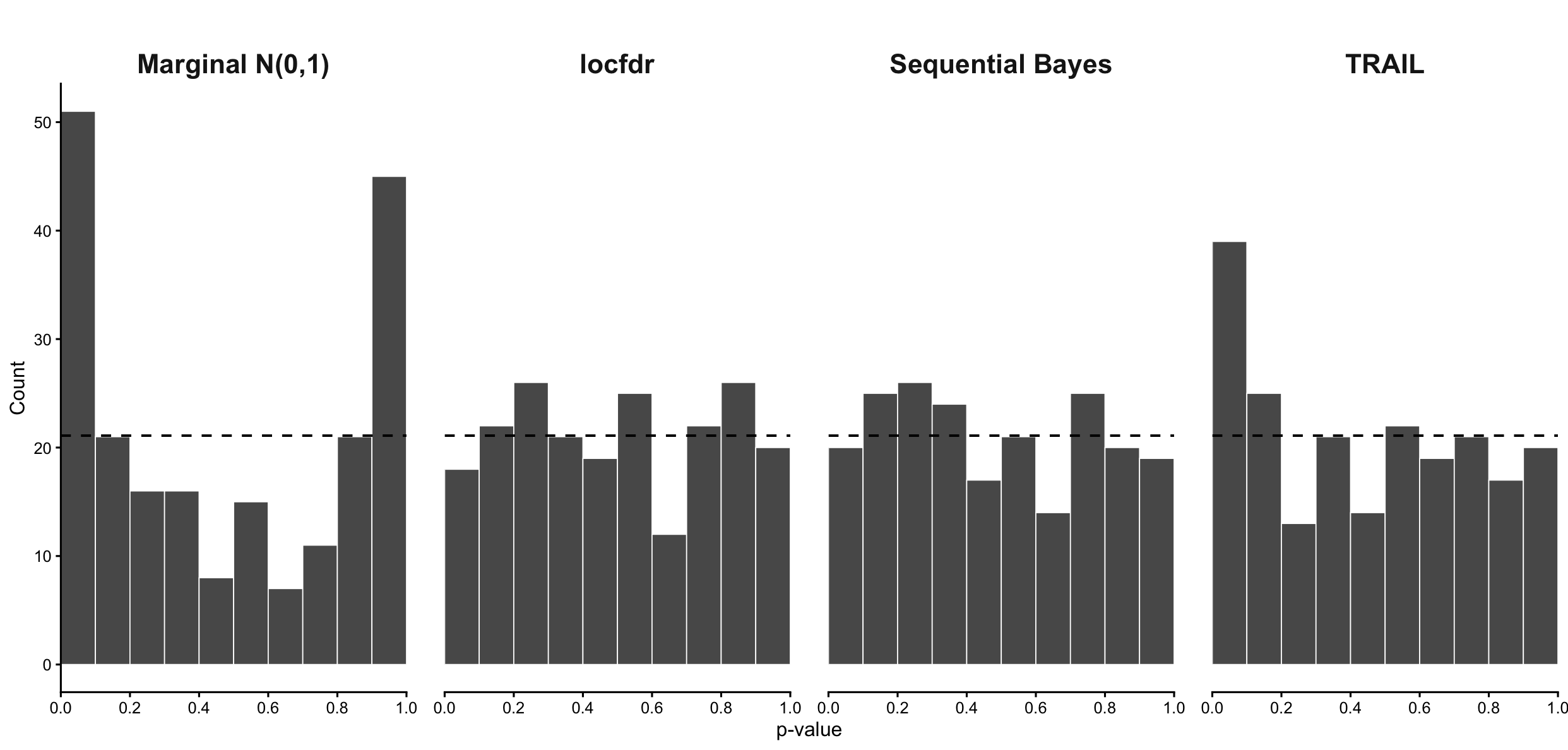}
        \caption{Study 40}
    \end{subfigure}

    \caption{Histograms of estimated $p$-values for Studies 14, 31, 40.}
    \label{fig:real_data_pvalues}
\end{figure}

The corresponding $p$-value histograms in Figure \ref{fig:real_data_pvalues} provide another perspective on these differences. The values computed from the theoretical $N(0,1)$ null exhibit a pronounced dip in the intermediate $p$-value range. Together with the bimodal patterns in Figure \ref{fig:real_data_null_densities}, this raises concern about the adequacy of the fixed standard normal reference as a representation of the pooled null realized within these studies. It therefore calls for caution in interpreting the larger rejection counts of marginal BH as evidence of greater detection power. The $p$-values obtained from \texttt{locfdr} and \texttt{Sequential Bayesian} also exhibit uneven frequencies, whereas those obtained from \texttt{TRAIL\_BIC} have less pronounced dips over the moderate-to-large $p$-value range. These differences are consistent with the greater flexibility of \texttt{TRAIL\_BIC} in accommodating the observed distributional shapes.

\begin{table}[t]
\centering
\begin{tabular}{cccccc}
\toprule
\diagbox{Study}{Methods}
& $N(0,1)$ & $N(0,1)$ & \texttt{locfdr} & \texttt{Seq. Bayes.} & \texttt{TRAIL\_BIC} \\
\midrule
Adjustment & BH & BY & BH & BH & BH \\
\midrule
Study 14  & 16  & 5  & 0 & 0 & 6  \\
Study 31 & 21 & 10  & 0 & 2 & 9 \\
Study 40 & 15 & 2 & 0 & 0 & 14 \\
\bottomrule
\end{tabular}
\caption{Rejection counts for selected studies}
\label{tab:real_data_rejections}
\end{table}

Table \ref{tab:real_data_rejections} summarizes the rejection counts for Studies 14, 31, and 40. Marginal BH attains the largest rejection count in all three studies, with 16, 21, and 15 rejections, respectively. Among the empirical null methods, \texttt{TRAIL\_BIC} consistently yields the largest rejection count, with 6, 9, and 14 rejections across the three studies. In contrast, \texttt{locfdr} makes no rejections in any of the selected studies, while \texttt{Sequential Bayesian} makes only 2 rejections in Study 31 and none in Studies 14 and 40.

Compared with marginal BY, \texttt{TRAIL\_BIC} yields more rejections in Studies 14 and 40. In Study 31, \texttt{TRAIL\_BIC} makes 9 rejections, slightly fewer than the 10 obtained by marginal BY. Particularly notable is Study 40, where \texttt{TRAIL\_BIC} nearly matches marginal BH, with 14 versus 15 rejections, while the other empirical null methods make no rejections. Taken together, these results illustrate that \texttt{TRAIL\_BIC} can yield substantially more discoveries than the competing empirical null methods, although it does not uniformly outperform marginal BY in terms of rejection counts.

A possible explanation for the higher rejection counts of \texttt{TRAIL\_BIC} concerns how the competing empirical null models represent unequal left and right flanks. When a pooled null has a broadly dispersed left flank and a more rapidly declining right flank over the range relevant for testing, a symmetric fit can carry the left-side dispersion into the estimated right tail. In Example~1 of Figure \ref{fig:examples}, this behavior produces excess right-tail mass for \texttt{Sequential Bayesian} and, to a lesser extent, \texttt{locfdr}. Such excess mass inflates the estimated right-sided $p$-values at potential signals and can reduce the number of BH rejections. The Gaussian--Hermite representation used by \texttt{TRAIL\_BIC} accommodates unequal flank shapes and can consequently avoid this source of overly large right-tail probabilities. This interpretation is consistent with its higher power in the heteroscedastic bimodal settings with $\kappa<0$ in Figure \ref{fig:power}. The negative-skewness simulations also show higher power than \texttt{Sequential Bayesian}, although \texttt{locfdr} attains the highest power in those settings. A similar left--right imbalance in the study-specific pooled nulls could therefore contribute to the CAD rejection differences, although its prevalence across studies remains to be assessed.

\section{Discussion}
\label{sec:discussion}

We have proposed \texttt{TRAIL} for estimating a flexible pooled conditional null in dependent multiple-testing problems. Building on the empirical-null perspective
\citep{Efron2007,AzrielSchwartzman2015}, the method combines a Gaussian--Hermite representation with progressive tail reconstruction. The factor model defines the null distribution to be estimated within the realized experiment, while the Hermite coefficients allow this distribution to be fitted without requiring the null correlation matrix or individual factor loadings. Joint selection of the Gaussian reference location and truncation order in \texttt{TRAIL\_BIC} further adapts the finite-dimensional representation to the observed data.

The numerical results suggest that this flexibility is particularly useful when the realized null exhibits sharp peaks, unequal tail decay, or multiple modes. Across the settings considered, \texttt{TRAIL\_BIC} generally provides a favorable balance between estimated FDR and power, with the relative performance varying across configurations. The shifted and unshifted comparisons also illustrate the importance of the Gaussian reference location for finite-order approximation. Since a common location shift preserves the ordering of the test statistics, its effect is expressed through the fitted tail probabilities and the resulting rejection threshold. The CAD analysis shows that differences in null calibration can translate into substantially different discovery patterns, with \texttt{TRAIL\_BIC} producing more rejections on average than the other empirical null procedures. The signal-enriched nature of the selected variants and the unknown study-specific null labels leave the error rates of these discoveries undetermined.

While density estimation from the full observed sample is well established, reconstructing the contribution in regions where two distributions overlap presents additional methodological challenges. Assessing the accuracy of such reconstruction is particularly difficult because the underlying null density in the contaminated region is not directly observed. In \texttt{TRAIL}, the center fraction $q$ determines the extent of the sample subjected to reconstruction. A larger $q$ retains more original observations and confines reconstruction to a smaller tail region, favoring greater preservation of the observed distributional structure. Conversely, a smaller $q$ replaces a larger portion of the sample with pseudo-observations, giving the procedure greater scope to reshape the estimated null through model based reconstruction. The choice of $q$ therefore governs the balance between preserving observed structure and reconstructing potentially contaminated regions. Developing principled criteria for selecting this extent of reconstruction, together with diagnostics for assessing the plausibility and stability of the reconstructed tail, remains an important direction for future research.

The theoretical analysis connects reconstruction tracking, pooled-null consistency, and testing with the same observations. The results apply to a regularized reconstruction with $c=0$ and $M=1$, using prescribed Hermite orders and iteration counts under the stated dependence, center-purity, and crossing conditions. An important next step is to extend this analysis to the data-adaptive implementation, including joint shift--order selection and averaging across pseudo-sample replications. The BIC-type score is evaluated on candidate-specific reconstructed data, so understanding its selection behavior requires accounting for both density estimation and pseudo-sample generation. Testing regimes in which the BH threshold approaches zero also motivate more refined control of tail-estimation error.

Further developments include adaptive retention schedules and uncertainty quantification for the tails of the estimated null density. Such uncertainty measures could help allocate reconstruction effort and assess the stability of rejection decisions. Extension to two-sided testing would additionally require a reconstruction scheme for both potentially contaminated tails around a retained central region. Overall, the results emphasize the role of flexible pooled-null estimation in balancing discovery
and error control under dependence.

\section*{Online Supplementary Note}
The online supplementary note contains the conditions, the proofs of Theorems 1–3 and Corollaries 1–2, and the supporting technical lemmas for the regularized reconstruction analyzed in Section 4.

\section*{Acknowledgement}
Research of J. Park and K. Kim  was supported by the National Research Foundation of Korea (NRF) grant funded by the Korea government (MSIT) (No. RS-2025-00556575). Research of J. Kim was supported by the National Research Foundation of Korea (NRF) grant funded by the Korea government (MSIT) (RS-2026-25489986).

\bibliographystyle{apalike}
\bibliography{refs}

\newpage

\end{document}


\begin{center}
\LARGE\textbf{Supplementary Note}
\end{center}
\vspace{0.5em}
\begingroup
\small
\tableofcontents
\endgroup
\clearpage

\section{Reconstruction in proof notation}
\label{supp:sec:algorithm}

Throughout this note, sample-size indices are retained on the observed-data quantities, empirical fits, and reconstruction states. For the class counts denoted by $n_j$ in the main text, we write $n_{j,n}=|\mathcal H_{j,n}|$ for $j=0,1$. The main text retains sample-size indices selectively in the algorithm description and explicitly in Section~4. Estimates of the null density and its CDF omit the null label $0$; true null targets and class labels retain it. Iteration indices on density, CDF, and coefficient iterates are written as parenthesized superscripts, and the population label $\mathrm{pop}$ is a subscript. Generic inputs such as $(\mathsf P,b)$, trial coefficients, and fixed population limits retain their generic notation.

Use the $c=0$, $M=1$ reconstruction and the family $\mathcal F_k$
defined in the main text.  Retain the main-text coefficient vector
$\boldsymbol w=(w_1,\ldots,w_k)^{\mathsf T}$ and set
\begin{equation}
 \begin{aligned}
 \delta_k&=e^{-k^2},\qquad
 \mathcal P_k=\{R_{\boldsymbol w}:\boldsymbol w\in\mathcal K_k\},\\
 \mathcal K_k&=\{\boldsymbol w\in\mathbb R^k:
 R_{\boldsymbol w}(z)\ge\delta_k\ (z\in\mathbb R),\quad
 \sqrt2w_2\le\log k\}.
 \end{aligned}
 \label{supp:eq:model-sieve}
\end{equation}
The second-moment identity is $\int z^2f(z;\boldsymbol w)\,dz=1+\sqrt2w_2$.
Choose a fixed normalized finite-order Hermite initializer $g$ with
$\inf_zg(z)/\phi(z)>0$, and use orders for which $g\in\mathcal F_k$.

Let $\mathcal I_{0,n}$ be the indices of the $N_n=\lfloor qn\rfloor$
smallest original statistics, breaking ties by index.  Thus
\[
 \tau_n=n-N_n,\qquad b_n=Z_{n,(N_n)},\qquad
 \mathsf P_n=\frac1{N_n}\sum_{i\in\mathcal I_{0,n}}\delta_{Z_{n,i}}.
\]
For a finite measure $P$, write $Ph=\int h\,dP$.  Define the normalized
restrictions
\[
 \mathcal C_bf=\frac{f\mathbf1_{(-\infty,b]}}{F_f(b)},\qquad
 \mathcal T_af=\frac{f\mathbf1_{(a,\infty)}}{1-F_f(a)}.
\]
For $T\ge2$ and fixed $\eta\in(0,1)$, use the main-text retention counts
and fresh-batch sizes with the schedule
\begin{equation}
 \begin{aligned}
 M_{n,t}
 &=\left\lfloor\frac{(1-\eta)\tau_n t}{T-1}\right\rfloor
   -\left\lfloor\frac{(1-\eta)\tau_n(t-1)}{T-1}\right\rfloor,
       &&1\le t<T,\\
 M_{n,T}&=\tau_n-\lfloor(1-\eta)\tau_n\rfloor,\\
 m_{n,t}&=\tau_n-\sum_{s=1}^{t-1}M_{n,s},
       &&1\le t\le T.
 \end{aligned}
 \label{supp:eq:algorithm-counts}
\end{equation}
The stated asymptotic schedules give positive nonterminal block counts
eventually.

Let $\mathcal X_{0,n}^{(s)}$ denote the block of pseudo-observations retained at stage $s$, corresponding to $\mathcal X_{0}^{(s)}$ in the main text. For the proofs, represent the cumulative collection of pseudo-observations retained through stage $t$ by the finite measure
\begin{equation}
 B_{n,t}
:= \frac{1}{\tau_n}
    \sum_{s=1}^{t}
    \sum_{y\in\mathcal X_{0,n}^{(s)}} \delta_y,
\qquad B_{n,0}=0,
\qquad
B_{n,t}(\mathbb R)
= \frac{1}{\tau_n}\sum_{s=1}^{t}M_{n,s}.
 \label{supp:eq:frozen-measure-definition}
\end{equation}
Here $\delta_y$ is the unit point mass at $y$, and the summation preserves
multiplicities.  Each retained pseudo-observation contributes mass
$1/\tau_n$, so $\mathsf B_{n,t}$ records both the retained values and the
fraction of the pseudo-tail already retained.  The refit below applies
the current pseudo-tail weight $1-c_{n,t}$ to this measure.

Initialize $\widehat f_n^{(0)}=g$ and $a_{n,0}=b_n$.
At iteration $t$, retain the main-text center weight
\[
 c_{n,t}=F_{\widehat f_n^{(t-1)}}(b_n)
\]
and generate $Y_{n,t,1}^{(1)},\ldots,Y_{n,t,m_{n,t}}^{(1)}$ conditionally iid
from $\mathcal T_{a_{n,t-1}}\widehat f_n^{(t-1)}$.
The incoming history contains $U$, the original observations, and all
preceding simulation variables.  Refit by
\begin{equation}
 \begin{aligned}
 \widehat{\boldsymbol w}_n^{(t,1)}
 &\in\argmax_{\boldsymbol w\in\mathcal K_k}\bigg[
 c_{n,t}\mathsf P_n\log R_{\boldsymbol w}\\
 &\qquad +(1-c_{n,t})\left\{
       \mathsf B_{n,t-1}\log R_{\boldsymbol w}
       +\frac1{\tau_n}\sum_{j=1}^{m_{n,t}}
          \log R_{\boldsymbol w}(Y_{n,t,j}^{(1)})\right\}\bigg],\\
 \widehat f_n^{(t)}&=f(\cdot;\widehat{\boldsymbol w}_n^{(t,1)}).
 \end{aligned}
 \label{supp:eq:algorithm-refit}
\end{equation}
With $M=1$, the main-text averages satisfy
$\overline{\widehat{\boldsymbol w}}_n^{(t)}
 =\widehat{\boldsymbol w}_n^{(t,1)}$ and
$\overline Y_{n,t,(j)}=Y_{n,t,(j)}^{(1)}$.
For empirical observations, adding the finite Gaussian term
$\log\phi$ leaves the maximizer unchanged. Population refits use
the displayed log-density-ratio criterion directly.
Using the same fresh batch, for $t<T$ set
\begin{equation}
 \begin{aligned}
 \mathsf B_{n,t}&=\mathsf B_{n,t-1}
       +\frac1{\tau_n}\sum_{j=1}^{M_{n,t}}
                          \delta_{\overline Y_{n,t,(j)}},\\
 a_{n,t}&=\overline Y_{n,t,(M_{n,t}+1)}.
 \end{aligned}
 \label{supp:eq:algorithm-freeze}
\end{equation}
After the final refit retain the entire remaining batch. Then
$\mathsf B_{n,T}(\mathbb R)=1$ and
$\widehat f_{n;k}^{(T)}=\widehat f_n^{(T)}$. Let $\widehat F_{n;k}^{(T)}$ denote its CDF.

For the population reconstruction at a center $(\mathsf P,b)$, initialize
$(f,\mathsf B,a)=(g,0,b)$ and use cumulative retained fractions $\xi_0=0$,
$\xi_t=(1-\eta)t/(T-1)$ for $t<T$, and $\xi_T=1$.
For empirical reconstruction put
$\xi_{n,t}=\tau_n^{-1}\sum_{s=1}^t M_{n,s}$.
In the population recursion, $\mathsf B$ is the finite measure of
previously retained tail blocks, the counterpart of \eqref{supp:eq:frozen-measure-definition}.
At iteration $t$, the incoming state $(f_\beta,\mathsf B,a)$ has live mass
$\lambda=1-\xi_{t-1}$.  Use $\mathsf P,b$ in place of $\mathsf P_n,b_n$ in
\eqref{supp:eq:algorithm-refit} and replace the fresh empirical measure
by $\lambda \mathcal T_af_\beta$. The center weight is $F_{f_\beta}(b)$.
A nonterminal block of mass $\omega=\xi_t-\xi_{t-1}$ retains
$\lambda(\mathcal T_af_\beta)\mathbf1_{(a,a^\dagger]}$, where
$a^\dagger=F_{\mathcal T_af_\beta}^{-1}(\omega/\lambda)$, and advances the boundary to $a^\dagger$.
Here $F^{-1}(t)=\inf\{x:F(x)\ge t\}$.
The terminal refit uses all remaining mass.  Write its density as
$f^{(T)}_{k,\mathrm{pop}}$.

For the proofs, use center-normalized coordinates
\begin{equation}
 \begin{aligned}
 \Psi_k&=(\psi_0,\ldots,\psi_k)^{\mathsf T},&
 \boldsymbol h_b&=\int_{-\infty}^b\phi(x)\Psi_k(x)\,dx,\\
 r_\beta&=\beta^{\mathsf T}\Psi_k,&
 A_\beta&=\beta_0,\\
 f_\beta&=\phi r_\beta/A_\beta,&
 s_\beta(a)&=\int_a^\infty\phi(x)r_\beta(x)\,dx.
 \end{aligned}
 \label{supp:eq:beta-coordinates}
\end{equation}
The domain $\mathcal D_{k,b}$ consists of the images of
$R_{\boldsymbol w}/F_{f(\cdot;\boldsymbol w)}(b)$ for $\boldsymbol w\in\mathcal K_k$.
Under this change of coordinates, $f_\beta=f(\cdot;\boldsymbol w)$. Equivalently,
\[
 \boldsymbol h_b^{\mathsf T}\beta=1,\qquad
 r_\beta(x)\ge\delta_kA_\beta\ (x\in\mathbb R),\qquad
 \sqrt2\beta_2\le(\log k)A_\beta.
\]
Integration gives $1<A_\beta\le\{\delta_k\Phi(b)\}^{-1}$ and
$A_\beta-1\ge\delta_k\{1-\Phi(b)\}$.  The conditional criterion becomes
\begin{equation}
 \begin{aligned}
 L_{\mathsf P,b}(f_\beta)&=\mathsf P\log r_\beta=:L_{\mathsf P}(\beta),\qquad \mathcal V_{\mathsf P}=-L_{\mathsf P},\\
 \beta_*&\in\argmin_{\beta\in\mathcal D_{k,b}}\mathcal V_{\mathsf P}(\beta),\qquad
 p_*=f_{\beta_*}=p_{k,\mathsf P,b}.
 \end{aligned}
 \label{supp:eq:conditional-target}
\end{equation}
For the empirical iterate, let $\widehat\beta_n^{(t)}$ be its
center-normalized coefficient vector, so
$f_{\widehat\beta_n^{(t)}}=\widehat f_n^{(t)}$.
For the population recursion, write $\beta_{\mathrm{pop}}^{(t)}$ for the
corresponding coefficient vector; in particular,
$f_{\beta_{\mathrm{pop}}^{(T)}}=f_{k,\mathrm{pop}}^{(T)}$.
Write $F_*=F_{p_*}$ and $e_0=(1,0,\ldots,0)^{\mathsf T}$.
The local sampling calculations also use $\tau_n$ for the total pseudo-tail size.

\section{Conditions}
\label{supp:sec:conditions}
\begin{condition}[Null model and center fraction]
\label{supp:ass:factor}
The null statistics follow the Gaussian factor model in the main
text with a common $U\sim N_d(0,I_d)$, fixed finite $d\ge0$,
deterministic loadings, and Gaussian residual vectors independent
of $U$. Let $L_n$ be the matrix with rows
$\ell_{n,i}^{\mathsf T}$ for $i\in\mathcal H_{0,n}$.
The loading matrix and residual covariance matrix satisfy
\[
 R_{0,n}=L_nL_n^{\mathsf T}+B_{0,n},
 \qquad B_{0,n}\succeq0.
\]
For every $n$ and $i\in\mathcal H_{0,n}$,
\[
 \sigma_{n,i}^2:=(B_{0,n})_{ii}
 =1-\|\ell_{n,i}\|^2\ge\sigma_{0,*}^2>0,
\]
where $\sigma_{n,i}>0$ is the null residual standard deviation
and $\sigma_{0,*}>0$ is a uniform lower bound.
Any factor decomposition satisfying these requirements is allowed.
The index sets are deterministic,
$\pi_{0,n}=n_{0,n}/n\to\pi_0\in(0,1)$, and $0<q<\pi_0$.
\end{condition}

\begin{condition}[Quantitative residual dependence]
\label{supp:ass:residual}
For some $\rho_{\mathrm{dep}}\in(0,1/2]$ and $C<\infty$,
\begin{equation}
 \vartheta_n
 :=\frac{\|B_{0,n}\|_F}{n_{0,n}}
 \le Cn^{-\rho_{\mathrm{dep}}}.
 \label{supp:eq:spectral-rate}
\end{equation}
\end{condition}

\begin{remark}
For the residual vectors in Condition~\ref{supp:ass:factor}, define
\[
 D_n
 :=\frac1{n_{0,n}^2}
   \sum_{i,j\in\mathcal H_{0,n}}
   \left|
     \operatorname{Corr}(\varepsilon_{n,i},\varepsilon_{n,j})
   \right|.
\]
The residual-variance lower bound and Cauchy--Schwarz give
\begin{equation}
 D_n
 =\frac1{n_{0,n}^2}
   \sum_{i,j\in\mathcal H_{0,n}}
   \frac{|(B_{0,n})_{ij}|}{\sigma_{n,i}\sigma_{n,j}}
 \le
 \frac1{\sigma_{0,*}^2n_{0,n}^2}
 \sum_{i,j\in\mathcal H_{0,n}}|(B_{0,n})_{ij}|
 \le
 \frac{\|B_{0,n}\|_F}{\sigma_{0,*}^2n_{0,n}}
 =\sigma_{0,*}^{-2}\vartheta_n.
 \label{supp:eq:general-residual-correlation-bound}
\end{equation}
Consequently, Condition~\ref{supp:ass:residual} implies
$D_n=O(n^{-\rho_{\mathrm{dep}}})$ for the chosen factor
decomposition.
\end{remark}

\begin{condition}[A quantitative zero condition on the retained center]
\label{supp:ass:center}
For some $\zeta>0$ and almost every $U$,
\begin{equation}
 \chi_n:=\frac1n
 \sum_{i\in\mathcal H_{1,n}\cap\mathcal I_{0,n}}
       \{1+\log(1+|Z_{n,i}|)\}
 =O_{\Pp (\cdot \mid U)}(n^{-\zeta}).
 \label{supp:eq:zero-assumption}
\end{equation}
All original statistics are finite almost surely.
\end{condition}

\begin{condition}[Reconstruction operations]
\label{supp:ass:operations}
Use the reconstruction in Section~\ref{supp:sec:algorithm} with fixed
$\eta\in(0,1)$ and fixed initializer $g$.
Every refit attains the global maximum over $\mathcal K_k$.
The uniforms used for fresh simulation are mutually independent and
independent of $U$ and all original observations.
Use a measurable maximizer selection throughout.
\end{condition}

\begin{condition}[Signal margins and a weak Gaussian copula spectrum]
\label{supp:ass:signal}
For $i\in\mathcal H_{1,n}$,
\begin{equation}
 Z_{n,i}=\mu_{n,i}+\ell_{n,i}^{\mathsf T}U
                  +\sigma_{n,i}K^{-1}\{\Phi(V_{n,i})\},
 \label{supp:eq:signal-model}
\end{equation}
where $K$ is a fixed continuous proper CDF on $\R$. The parameters
are deterministic and satisfy
\begin{equation}
 0\le\mu_{n,i}<\infty,\qquad \|\ell_{n,i}\|\le C_\ell,\qquad
 0<\sigma_*\le\sigma_{n,i}\le\sigma^*<\infty.
 \label{supp:eq:signal-bounds}
\end{equation}
The vector $(V_{n,i}:i\in\mathcal H_{1,n})$ is jointly
Gaussian with standard normal margins and is independent of $U$.
For its correlation matrix $R_{1,n}^{V}$,
\begin{equation}
 \lambda_{\max}(R_{1,n}^{V})/n_{1,n}\longrightarrow0.
 \label{supp:eq:signal-spectrum}
\end{equation}
The coupling of null and signal residuals is otherwise unrestricted.
\end{condition}

\begin{condition}[Uniformly separated positive moving crossing]
\label{supp:ass:crossing}
For almost every $U$ there is $\underline t_U>0$ such that
$t_{n,U}^{\star}\ge\underline t_U$ eventually. For every
$0<\epsilon<\underline t_U/2$, there is $m_{U,\epsilon}>0$ such that eventually
\begin{align}
 \sup_{t\in[t_{n,U}^{\star}-\epsilon,\,t_{n,U}^{\star}]}
       \{\mathsf H_{n,U}(t)-t/\alpha\}&\ge m_{U,\epsilon},
 \label{supp:eq:moving-left-margin}\\
 \sup_{t\in[t_{n,U}^{\star}+\epsilon,\,\alpha]}
       \{\mathsf H_{n,U}(t)-t/\alpha\}&\le-m_{U,\epsilon}.
 \label{supp:eq:moving-right-margin}
\end{align}
The second requirement is ignored for an empty interval.
\end{condition}

\section{Proof of Theorem 1}
\label{supp:sec:main-proofs}
\label{supp:sec:proof-theorem-1}

\begin{proof}
Let $\mathsf P$ be the population center in the theorem. Choose a compact
interval $I\subset(-\infty,b]$ containing $k+1$ distinct points of
$\operatorname{supp}(\mathsf P)$. For a nonzero polynomial $P$ of degree at
most $k$, at least one of these points has $P\ne0$. A neighborhood
of that point has positive $\mathsf P$-mass, so $\int_I P^2\,d\mathsf P>0$ after
choosing the left endpoint of $I$ below the smallest selected point.
Compactness of the coefficient unit sphere gives a positive lower
bound for this Gram form. Explicitly, for
$G_I=\int_I\Psi_k\Psi_k^{\mathsf T}\,d\mathsf P$,
\[
 \lambda_{\min}(G_I)
 =\min_{\|h\|_2=1}\int_I(h^{\mathsf T}\Psi_k)^2\,d\mathsf P>0.
\]
The integrand is continuous in $h$, and no unit $h$ can give zero
because a nonzero degree-at-most-$k$ polynomial has at most $k$
distinct roots. At fixed $k$, choose $C_0$ large enough that this
minimum exceeds $e^{-C_0(k+1)^2}$ and that the assumed logarithmic
moment is at most $C_0$. The assumed logarithmic moment and this
bound verify \eqref{supp:eq:alg-center-event} for a finite constant
at the fixed order $k$. Lemma~\ref{supp:lem:alg-primitive-bounds}
gives a continuous strongly convex conditional objective on a compact
convex domain, hence a unique target.
In population reconstruction every simulation discrepancy is
zero. Lemma~\ref{supp:lem:alg-quantitative-freezing} gives
\[
 \|f^{(T)}_{k,\mathrm{pop}}-p_{k,\mathsf P,b}\|_1\le 2\Gamma_k/T
\]
for sufficiently large $T$.

For the empirical assertion, fix a factor realization $U=u$ on which
Conditions~\ref{supp:ass:factor}--\ref{supp:ass:operations} hold.
Lemma~\ref{supp:lem:cdf-stat} gives, with conditional probability tending
to one, $|b_n|\le b_{\max}(u)$ and a bounded logarithmic moment of
$\mathsf P_n$. Consider the interval
\[
 [-b_{\max}(u)-2,-b_{\max}(u)-1].
\]
This interval is fixed conditional on $u$ and its right endpoint is
strictly below every $b_n\in[-b_{\max}(u),b_{\max}(u)]$.
Lemma~\ref{supp:lem:curvature-stat} bounds its empirical Gram matrix
below by $\exp\{-C_u k\log(k+1)\}/2$ on events with conditional
probability tending to one. These bounds verify all parts of
\eqref{supp:eq:alg-center-event} with constants depending on $u,k$
and fixed reconstruction settings, uniformly over the original-data
realizations in those events. In particular $p_{n,k}$ is unique there.

Condition on such an original-data realization. Stop before the first
incoming state outside \eqref{supp:eq:bootstrap-region}.
The indicator that stage $t$ is reached before exit is measurable
before that stage's fresh batch is drawn. Conditional application of
Lemma~\ref{supp:lem:mc-rank-freezing} gives
\[
 \E\!\left[\sum_{t=1}^{T_n}\nu_{n,t}
       \ind\{t\text{ is reached before exit}\}
       \,\middle|\,U,\text{original data}\right]
 \le \mathfrak A_kT_n/\sqrt{\tau_n}.
\]
This follows from the tower property: condition first on the full
incoming history at each stage, multiply its conditional batch bound
by the measurable reached-stage indicator, and then sum. In
particular, it is not a bound obtained by conditioning on a future
refit or retained block.
Here $\tau_n\asymp n$ and $T_n=n^{\upsilon+o(1)}$ with $\upsilon<1/2$.
At fixed $k$, $T_n\ge\Gamma_k^2$ eventually. Markov's inequality
makes the stopped total discrepancy at most $\Gamma_k^{-2}$ with
conditional probability tending to one. If the stopped sum is
$S_n^{\mathrm{stop}}$, the explicit bound is
\[
 \Pp\{S_n^{\mathrm{stop}}>\Gamma_k^{-2}\mid U,\text{original data}\}
 \le\Gamma_k^2\mathfrak A_kT_n/\sqrt{\tau_n}\longrightarrow0.
\]
The constants are uniform over the localized original-data event. The first-exit argument in
Lemma~\ref{supp:lem:alg-quantitative-freezing} then keeps the entire
trajectory inside its localization region. The same expectation bound
shows that its total discrepancy is $O_{\Pp (\cdot \mid U)}(T_n/\sqrt n)$.
The terminal estimate and density conversion in that lemma prove
\eqref{supp:eq:main-fixed-tracking} below, which is the empirical
assertion of the theorem:
\begin{equation}
 \|\widehat f_{n;k}^{(T_n)}-p_{n,k}\|_1
 =O_{\Pp (\cdot \mid U)}(T_n^{-1}+T_n n^{-1/2}).
 \label{supp:eq:main-fixed-tracking}
\end{equation}
To remove the conditioning on the original data, add the probability
of the complement of their localization event, which tends to zero,
and average the uniform conditional bound over that event. Since
$T_n^{-1}=n^{-\upsilon+o(1)}$ and
$T_n/\sqrt n=n^{\upsilon-1/2+o(1)}$, both terms vanish under the
stated schedule.
\end{proof}

\section{Proof of Theorem 2}
\label{supp:sec:proof-theorem-2}

\begin{proof}
Put $\rho_{\mathrm{stat}}=\min\{\rho_{\mathrm{dep}}/3,\zeta\}$ and
$s_{n,k}=k^2\{k+\log\log(en)\}n^{-\rho_{\mathrm{stat}}}$.
The following argument gives, for finite $C_U,\rho_U>0$ uniform over the null loading laws,
\begin{align}
 \|\widehat f_{n;k}^{(T)}-\overline f_{0,n,U}\|_1
 =O_{\Pp (\cdot \mid U)}\bigg(&
 e^{C_U k^{9/10}}\{k e^{-\rho_Uk}+\sqrt{s_{n,k}}\}
 +\frac{1+\log k}{\sqrt k}
 \notag\\[-2pt]
 &+\exp\!\{\exp(C_U(k+1)^3)\}
       \{T^{-1}+Tn^{-1/2}\}\bigg).
 \label{supp:eq:consistency-bound}
\end{align}
The bound holds along the schedule of Theorem~\MainConsistencyTheorem{}.  The additional
fixed-target error under \eqref{supp:eq:optional-loading-limit} is at most
$C_{\sigma_{0,*}}(1+\|U\|)\mathcal W_1(H_{0,n},H_0)$.

Fix an environment $U=u$ for which Condition~\ref{supp:ass:center} holds.
Condition~\ref{supp:ass:residual} and
\eqref{supp:eq:general-residual-correlation-bound} give
$D_n=O(n^{-\rho_{\mathrm{dep}}})$. All constants may depend on $u$, but the
statistical bounds below are uniform over loading laws supported on $\{\ell:\|\ell\|^2\le1-\sigma_{0,*}^2\}$.
In this proof $f_{0,n}=\overline f_{0,n,u}$ and
$F_{0,n}=\overline F_{0,n,u}$.
Lemmas~\ref{supp:lem:cdf-stat} and \ref{supp:lem:curvature-stat}, together with
\eqref{supp:eq:center-log-event}, give the bounded-cutoff, Gram, and
logarithmic-moment events in \eqref{supp:eq:alg-center-event} with conditional
probability tending to one.  Lemma~\ref{supp:lem:curvature-stat} uses the single interval
$[-b_{\max}(u)-2,-b_{\max}(u)-1]$ throughout. Its right
endpoint is below every admissible $b_n$, and its Gram bound directly
verifies \eqref{supp:eq:alg-center-event}.
The condition $k_n^3=o(\log\log n)$ makes the CDF error smaller than
$e^{-C_uk_n^3}$ with conditional probability tending to one.
Indeed $e^{-C_uk_n^3}/n^{-\rho_{\mathrm{dep}}/3}\to\infty$, while
Lemma~\ref{supp:lem:cdf-stat} bounds the CDF error by
$O_{\Pp (\cdot \mid U=u)}(n^{-\rho_{\mathrm{dep}}/3})$. The Gram lower bound
$e^{-C_uk_n\log(k_n+1)}/2$ is then at least
$e^{-C'_u(k_n+1)^2}$, verifying the scale in
\eqref{supp:eq:alg-center-event}.
The positive center Gram matrix makes the refit strictly concave in
$\boldsymbol w$, and the conditional criterion is strongly concave in its normalized
coordinates.  Hence the refits and $p_{n,k}$ are well defined and unique
on these events.  A fixed measurable maximizer selection completes the definition on
all other data realizations.

Condition on such an original-data realization.  Apply
Lemma~\ref{supp:lem:mc-rank-freezing} with the entire incoming history and
stop immediately before the first state outside the region
\eqref{supp:eq:bootstrap-region} of
Lemma~\ref{supp:lem:alg-quantitative-freezing}.
The local input-radius requirement is enforced after its $J_k$ initial stages. Those stages use the global regret estimate.
The stopping indicator is measurable before the fresh sample is drawn.
The conditional expectation bound gives
\begin{equation}
 \E\!\left[\sum_{t=1}^T\nu_{n,t}\mathbf1\{t\text{ is reached before exit}\}
          \,\middle|\,\text{original data},U\right]
 \le \mathfrak A_k T/\sqrt{\tau_n}.
 \label{supp:eq:main-mc-sum}
\end{equation}
Since $\tau_n\asymp n$ and
\[
 \Gamma_k=\exp\{\exp(C_u(k+1)^3)\}=n^{o(1)},
 \quad \mathfrak A_k=n^{o(1)},
\]
we have $T\ge\Gamma_k^2$ eventually and
$\Gamma_k^2\mathfrak A_kT/\sqrt{\tau_n}\to0$.
For these growth comparisons, the two logarithmic scales are
\[
 \log\Gamma_k=\exp\{C_u(k+1)^3\}
       =(\log n)^{o(1)}=o(\log n),\qquad
 \log\mathfrak A_k=O_u((k+1)^3)=o(\log n).
\]
Consequently $T/\Gamma_k^2=n^{\upsilon-o(1)}\to\infty$ and
$\Gamma_k^2\mathfrak A_kT/\sqrt{\tau_n}
=n^{\upsilon-1/2+o(1)}\to0$.
Markov's inequality makes the stopped total noise at most
$\Gamma_k^{-2}$ with conditional probability tending to one.
Writing its sum as $S_n^{\mathrm{stop}}$, equation
\eqref{supp:eq:main-mc-sum} gives
\[
 \Pp\{S_n^{\mathrm{stop}}>\Gamma_k^{-2}
                    \mid U,\text{original data}\}
 \le\Gamma_k^2\mathfrak A_kT/\sqrt{\tau_n}=o(1).
\]
This is uniform over the original-data localization event, so it
continues to hold after averaging over those data.
The one-step margin and bootstrap in the reconstruction lemma then
exclude an exit, so the stopped and actual paths agree on that event.
Increasing $C_u$ absorbs $\mathfrak A_k$ and all coordinate constants into
$\Gamma_k$.  Consequently
\begin{equation}
 \|\widehat f_{n;k}^{(T)}-p_{n,k}\|_1
 =O_{\Pp (\cdot \mid U)}\!\left(
 \Gamma_k\{T^{-1}+Tn^{-1/2}\}\right).
 \label{supp:eq:main-alg-density}
\end{equation}
This argument conditions on the complete observed center and controls
the trajectory from the fixed initializer $g$.

Lemma~\ref{supp:lem:empirical-stat} gives
$\mathcal S_{n,k}=O_{\Pp (\cdot \mid U)}(s_{n,k})$.
Lemma~\ref{supp:lem:projection-stat} therefore bounds
$\|p_{n,k}-\overline f_{0,n,u}\|_1$ by the first two terms of
\eqref{supp:eq:consistency-bound}, after decreasing $\rho_U$ if necessary.
The two errors are combined by the density triangle
\[
 \|\widehat f_{n;k}^{(T)}-\overline f_{0,n,u}\|_1
 \le\underbrace{\|\widehat f_{n;k}^{(T)}-p_{n,k}\|_1}
                  _{\text{reconstruction error}}
    +\underbrace{\|p_{n,k}-\overline f_{0,n,u}\|_1}
                  _{\text{conditional-target error}}.
\]
Use \eqref{supp:eq:main-alg-density} for the first term and
Lemma~\ref{supp:lem:projection-stat} for the second. Splitting its
square root by $\sqrt{a+b}\le\sqrt a+\sqrt b$ and decreasing
$\rho_U$ absorbs $\sqrt{k}e^{-\rho_{\mathrm{app}}k/2}$ into
$ke^{-\rho_Uk}$. This gives \eqref{supp:eq:consistency-bound}.
The approximation term tends to zero because
$C_uk^{9/10}-\rho_Uk\to-\infty$. The statistical term tends to zero
because $\rho_{\mathrm{stat}}>0$ and $k$ is sublogarithmic.
In particular, the logarithm of its deterministic scale is
\[
 C_uk^{9/10}+\log k+\tfrac12\log\{k+\log\log(en)\}
            -\tfrac12\rho_{\mathrm{stat}}\log n\longrightarrow-\infty.
\]  The moment remainder
tends to zero because $k\to\infty$.  Finally, the algorithm terms
are $n^{-\upsilon+o(1)}$ and $n^{\upsilon-1/2+o(1)}$.
This proves conditional moving-target $L^1$ consistency.
For probability densities,
$\|\widehat F_{n;k}^{(T)}-\overline F_{0,n,u}\|_\infty\le\frac12\|\widehat f_{n;k}^{(T)}-\overline f_{0,n,u}\|_1$.
Conditional convergence for almost every $U$, followed by bounded
convergence of the corresponding conditional error probabilities,
proves the unconditional moving-target statements.
Explicitly, if $X_n$ denotes either nonnegative estimation error,
then for each $\epsilon>0$
\[
 \Pp(X_n>\epsilon)
 =\E\{\Pp (X_n>\epsilon \mid U)\}\longrightarrow0.
\]
The conditional probabilities are bounded by one, so this step
requires no integrability of the environment-dependent constants
$C_U$ or $\Gamma_k$.

For the further assertion only, suppose
$\mathcal W_1(H_{0,n},H_0)\to0$.  Lemma~\ref{supp:lem:target-stability}
gives
\[
 \|\widehat f_{n;k}^{(T)}-f_{0,u}\|_1
 \le\|\widehat f_{n;k}^{(T)}-\overline f_{0,n,u}\|_1
       +C_{\sigma_{0,*}}(1+\|u\|)\mathcal W_1(H_{0,n},H_0).
\]
The second term tends to zero for each fixed $u$.  This proves the fixed-target density assertion and the
stated error bound.  The CDF assertion follows by total variation,
and bounded convergence of conditional error probabilities gives its
unconditional version.  In both cases the unconditional target retains
the realized factor $U$.
\end{proof}

\section{Proof of Theorem 3}
\label{supp:sec:proof-theorem-3}

\begin{proof}
Fix $U=u$ on the full-probability set of the conditions and use the main-text shorthand
$\widehat F_n=\widehat F_{n;k_n}^{(T_n)}$. Put
$\epsilon_{\mathrm{fit},n}=\|\widehat F_n-\overline F_{0,n,u}\|_\infty$.
Theorem~\MainConsistencyTheorem{} gives
$\epsilon_{\mathrm{fit},n}\to0$ in conditional probability.
The mixture bound $\mathsf H_{n,u}(t)\le\pi_{0,n}t+1-\pi_{0,n}$ gives
\begin{equation}
 t^\star_{n,u}\le
 \frac{\alpha(1-\pi_{0,n})}{1-\alpha\pi_{0,n}}<\alpha
 \label{supp:eq:crossing-upper-bound}
\end{equation}
for all sufficiently large $n$. Since $\pi_{0,n}\to\pi_0\in(0,1)$,
this upper bound stays strictly below $\alpha$. This follows
quantitatively from
\[
 \alpha-\frac{\alpha(1-\pi_{0,n})}{1-\alpha\pi_{0,n}}
 =\frac{\alpha\pi_{0,n}(1-\alpha)}{1-\alpha\pi_{0,n}},
\]
whose limit is strictly positive.
Condition~\ref{supp:ass:crossing} gives a positive lower bound
$\underline t_u$ and both separation margins.

For $j=0,1$, let $\mathbb G^o_{j,n}$ and
$\widehat{\mathbb G}_{j,n}$ be the oracle and fitted empirical
$p$-value CDFs. Lemmas~\ref{supp:lem:signal-law}--\ref{supp:lem:testing-transfer}
show that the two all-test empirical CDFs differ from $\mathsf H_{n,u}$
by $o_{\Pp (\cdot \mid U=u)}(1)$ on every compact subinterval of $(0,1)$.
For example, the all-test oracle CDF is
$\mathbb G_n^o=\pi_{0,n}\mathbb G_{0,n}^o+
(1-\pi_{0,n})\mathbb G_{1,n}^o$, so on such an interval $J$,
\[
 \|\mathbb G_n^o-\mathsf H_{n,u}\|_{\infty,J}
 \le\pi_{0,n}\epsilon_{0,n}
      +(1-\pi_{0,n})\|\mathbb G_{1,n}^o-
                                    \overline G_{1,n,u}\|_{\infty,J}.
\]
The fitted version uses the analogous mixture identity and the
pathwise sandwich in Lemma~\ref{supp:lem:testing-transfer}.
The null error is bounded by $\epsilon_{0,n}$ for the oracle and
by $\epsilon_{0,n}+\epsilon_{\mathrm{fit},n}$ for the fit.
The signal error uses the fluctuation bound and the common local
modulus of continuity in Lemma~\ref{supp:lem:signal-law}.
Lemma~\ref{supp:lem:local-bh} now gives
$\widehat t_n-t_{n,u}^{\star}\to0$ and
$t_n^o-t_{n,u}^{\star}\to0$ in conditional probability.

For any BH vector, exactly $R$ coordinates are at most $\alpha R/n$.
If $r>R$ such coordinates existed, then
$p_{(r)}\le\alpha R/n\le\alpha r/n$, contradicting maximality of $R$.
Thus, on $\{\widehat R_n>0\}$,
\[
 \operatorname{FDP}(\widehat{\mathcal R}_n)
 =\alpha\pi_{0,n}
       \frac{\widehat{\mathbb G}_{0,n}(\widehat t_n)}{\widehat t_n}.
\]
Here the numerator count is
$\widehat V_n=n_{0,n}\widehat{\mathbb G}_{0,n}(\widehat t_n)$,
and BH gives $\widehat R_n=n\widehat t_n/\alpha$.
The positive lower bound for the limiting crossing ensures that
this ratio is evaluated with a denominator bounded away from zero
with conditional probability tending to one.
With conditional probability tending to one,
$\widehat t_n\ge\underline t_u/2$. On that event,
\[
 |\operatorname{FDP}(\widehat{\mathcal R}_n)-\alpha\pi_0|
 \le\alpha|\pi_{0,n}-\pi_0|
   +\frac{2\alpha\pi_{0,n}}{\underline t_u}
          (\epsilon_{0,n}+\epsilon_{\mathrm{fit},n})
   \xrightarrow{\Pp (\cdot \mid U=u)}0.
\]
The oracle argument sets $\epsilon_{\mathrm{fit},n}=0$.

For either procedure, its empirical signal CDF $\mathbb G_{1,n}$ satisfies
$\operatorname{TDP}(\mathcal R_n)=\mathbb G_{1,n}(t_n)$.
The threshold bounds place $t_n$ and $t_{n,u}^{\star}$ in a common
compact interval $J\subset(0,1)$ with conditional probability tending
to one. On that event,
\[
 \left|\mathbb G_{1,n}(t_n)-\overline G_{1,n,u}(t_{n,u}^{\star})\right|
 \le\sup_{t\in J}|\mathbb G_{1,n}(t)-\overline G_{1,n,u}(t)|
      +\omega_{u,J}(|t_n-t_{n,u}^{\star}|).
\]
Both terms tend to zero in conditional probability.
The rank identity \eqref{supp:eq:rank-set-identity} proves the
rejection-set assertion. More explicitly,
\[
 \frac{|\widehat{\mathcal R}_n\triangle\mathcal R_n^o|}{n}
 =\frac{|\widehat t_n-t_n^o|}{\alpha}
 \le\frac{|\widehat t_n-t_{n,u}^\star|
                 +|t_n^o-t_{n,u}^\star|}{\alpha}
 \xrightarrow{\Pp (\cdot \mid U=u)}0.
\]
All proportions, threshold errors, and power-target errors are bounded.
Conditional convergence in probability therefore gives conditional
mean absolute convergence. Bounded convergence over $U$ proves the
unconditional probability, FDR, and power-target conclusions.
For completeness, if $X_n$ is any of the bounded absolute errors,
with $0\le X_n\le B$, then
\[
 \E[X_n \mid U=u]\le\epsilon+B\Pp (X_n>\epsilon \mid U)\longrightarrow\epsilon.
\]
Let $\epsilon\downarrow0$, then average over $U$ by dominated
convergence. This also applies to the moving power-target error
$|\operatorname{TDP}(\mathcal R_n)-
       \overline G_{1,n,U}(t_{n,U}^\star)|$.
\end{proof}

\section{Proof of Corollary 1}
\label{supp:sec:covariance-proof}

\begin{proof}
Let $L_n$ have columns $\sqrt{\lambda_{r,n}}e_{r,n}$, $1\le r\le d$.
The spectral theorem gives
$R_{0,n}=L_nL_n^{\mathsf T}+B_{0,n}$ with both summands positive
semidefinite. Indeed,
\[
 B_{0,n}=\sum_{r>d}\lambda_{r,n}e_{r,n}e_{r,n}^{\mathsf T},
 \qquad \|B_{0,n}\|_F^2=\sum_{r>d}\lambda_{r,n}^2.
\]
Orthogonality of the eigenvectors proves the second identity.  On one probability space choose a common
$U\sim N_d(0,I_d)$ and centered Gaussian residual vectors
$\varepsilon_{0,n}$ with covariance $B_{0,n}$, independent of $U$.
Then $Z_{n,i}=\ell_{n,i}^{\mathsf T}U+\varepsilon_{n,i}$ has the
specified joint Gaussian law for every $n$. Its covariance is
\[
 \Cov(L_nU+\varepsilon_{0,n})
 =L_n\Cov(U)L_n^{\mathsf T}+\Cov(\varepsilon_{0,n})
 =L_nL_n^{\mathsf T}+B_{0,n}=R_{0,n},
\]
where independence removes the cross terms.
The unit diagonal of $R_{0,n}$ gives
$(B_{0,n})_{ii}=1-\|\ell_{n,i}\|^2\ge \sigma_{0,*}^2$.

Cauchy--Schwarz gives, with $n_{0,n}^2$ matrix entries,
\[
 \sum_{i,j}|(B_{0,n})_{ij}|
 \le\left(\sum_{i,j}1\right)^{1/2}
          \left(\sum_{i,j}(B_{0,n})_{ij}^2\right)^{1/2}
 =n_{0,n}\|B_{0,n}\|_F.
\]
Consequently
\[
 D_n=\frac1{n_{0,n}^2}\sum_{i,j}\frac{|(B_{0,n})_{ij}|}{\sqrt{(B_{0,n})_{ii}(B_{0,n})_{jj}}}
 \le\frac1{\sigma_{0,*}^2n_{0,n}^2}\sum_{i,j}|(B_{0,n})_{ij}|
 \le\frac{\|B_{0,n}\|_F}{\sigma_{0,*}^2n_{0,n}}=\sigma_{0,*}^{-2}\vartheta_n.
\]
Also
\[
 \sum_{r>d}\lambda_{r,n}^2
 \le\lambda_{d+1,n}\sum_{r>d}\lambda_{r,n}
 \le\lambda_{d+1,n}\operatorname{tr}(R_{0,n})
 =\lambda_{d+1,n}n_{0,n}.
\]
Dividing by $n_{0,n}^2$ and taking square roots yields
\[
 \vartheta_n\le\sqrt{\lambda_{d+1,n}/n_{0,n}},\qquad
 D_n\le\sigma_{0,*}^{-2}\vartheta_n.
\]
Thus $\vartheta_n=O(n^{-\rho_{\mathrm{dep}}})$ directly gives the
required correlation rate, and
$\lambda_{d+1,n}/n_{0,n}=O(n^{-2\rho_{\mathrm{dep}}})$
is a sufficient condition for that rate.
The lower bound $\|B_{0,n}\|_F^2\ge\sum_i(B_{0,n})_{ii}^2\ge n_{0,n}\sigma_{0,*}^4$
is consistent with $\rho_{\mathrm{dep}}\le1/2$ when $n_{0,n}\asymp n$.
\end{proof}

\section{Proof of Corollary 2}
\label{supp:sec:full-left-proof}

\begin{proof}
Assume Conditions~\ref{supp:ass:factor}--\ref{supp:ass:operations}
and the schedule of Theorem~\MainConsistencyTheorem{}.
Only the marginal part of Condition~\ref{supp:ass:signal} is needed:
assume a fixed continuous proper CDF $K$ and conditional signal margins
\[
 \Pp (Z_{n,i}\le z \mid U)
 =K\!\left(\frac{z-\mu_{n,i}-\ell_{n,i}^{\mathsf T}U}{\sigma_{n,i}}\right),
 \qquad i\in\mathcal H_{1,n},
\]
with deterministic parameters satisfying \eqref{supp:eq:signal-bounds}.
No Gaussian-copula or spectral condition on their joint conditional
law is imposed in this proof. Suppose also that
\begin{equation}
 K(x)>0\qquad\text{for every }x\in\mathbb R.
 \label{supp:eq:signal-full-left}
\end{equation}
Fix $U=u$ where these conditions and the moving-center bounds hold.
Lemma~\ref{supp:lem:compatibility}, which uses
Condition~\ref{supp:ass:center}, gives
\[
 \frac1{n_{1,n}}\sum_{i\in\mathcal H_{1,n}}
            \Pp (Z_{n,i}\le x \mid U=u)\longrightarrow0
 \quad\text{for every fixed }x<-b_{\max}(u).
\]
For finite $M\ge0$, choose
$x<\min\{-b_{\max}(u),M+C_\ell\|u\|\}$. For every $i$ with
$\mu_{n,i}\le M$, the signal parameter bounds give
\[
 \Pp (Z_{n,i}\le x \mid U=u)
 \ge K\!\left(\frac{x-M-C_\ell\|u\|}{\sigma_*}\right)>0.
\]
The lower bound is uniform in $n$ and $i$. To check the scale
choice, the actual numerator is at least
$x-M-C_\ell\|u\|<0$. If the numerator is negative, division by
$\sigma\ge\sigma_*$ gives a standardized value at least
$(x-M-C_\ell\|u\|)/\sigma_*$; if it is nonnegative the same bound
is immediate. Monotonicity of $K$ therefore proves the displayed
lower bound in both cases. Writing this probability lower bound as
$\underline p_{u,M,x}>0$, we obtain
\[
 \underline p_{u,M,x}\frac{\#\{i\in\mathcal H_{1,n}:\mu_{n,i}\le M\}}{n_{1,n}}
 \le\frac1{n_{1,n}}\sum_{i\in\mathcal H_{1,n}}
                    \Pp (Z_{n,i}\le x \mid U=u)\longrightarrow0.
\]
Hence
\begin{equation}
 \frac1{n_{1,n}}\#\{i\in\mathcal H_{1,n}:\mu_{n,i}\le M\}\longrightarrow0
 \qquad\text{for every finite }M.
 \label{supp:eq:full-left-escape}
\end{equation}
The fraction is deterministic, so this conclusion is independent of
the chosen factor realization. Negative $M$ give an empty set.

For any finite $z$ and $M>z+C_\ell\|u\|$,
\[
 \frac1{n_{1,n}}\sum_{i\in\mathcal H_{1,n}}\Pp (Z_{n,i}\le z \mid U=u)
 \le\frac1{n_{1,n}}\#\{i\in\mathcal H_{1,n}:\mu_{n,i}\le M\}
      +K\!\left(\frac{z-M+C_\ell\|u\|}{\sigma^*}\right).
\]
Let $n\to\infty$ and then $M\to\infty$. Properness of $K$ shows that
the average signal left-tail probability tends to zero at every finite $z$.
The order of limits is explicit: for each fixed $M$ the bounded-location
fraction vanishes, leaving a limit superior bounded by
$K((z-M+C_\ell\|u\|)/\sigma^*)$; this tends to zero as
$M\to\infty$.
Markov's inequality gives $\mathbb F_{1,n}(z)\to0$ in conditional
probability. For fixed $\epsilon\in(0,1)$,
Lemma~\ref{supp:lem:target-stability} bounds the null quantiles
$z_{n,u}(\epsilon)$ above uniformly in $n$. Consequently,
\begin{equation}
 \begin{aligned}
 \sup_{t\in[\epsilon,1]}|\overline G_{1,n,u}(t)-1|&\longrightarrow0,\\
 \sup_{t\in[\epsilon,1]}|\mathbb G^o_{1,n}(t)-1|
                  &\xrightarrow{\Pp (\cdot \mid U=u)}0.
 \end{aligned}
 \label{supp:eq:escaping-signal-cdfs}
\end{equation}
Monotonicity gives each supremum from its value at $\epsilon$.
These steps use only the signal margins and the retained-center property.

For $t>0$, the mixtures $\mathsf H_{n,u}$ converge locally uniformly to
$\mathsf H_\infty(t)=\pi_0t+1-\pi_0$. Define
\begin{equation}
 t_\star=\frac{\alpha(1-\pi_0)}{1-\alpha\pi_0}\in(0,\alpha).
 \label{supp:eq:explicit-crossing}
\end{equation}
Then
\[
 \mathsf H_\infty(t)-t/\alpha
       =(\alpha^{-1}-\pi_0)(t_\star-t).
\]
Every fixed $s\in(0,t_\star)$ is eventually feasible for the
population BH crossing. The elementary mixture bound
$\mathsf H_{n,u}(t)\le\pi_{0,n}t+1-\pi_{0,n}$ also gives
\[
 t_{n,u}^{\star}\le
 \frac{\alpha(1-\pi_{0,n})}{1-\alpha\pi_{0,n}}\longrightarrow t_\star.
\]
It follows that $t_{n,u}^{\star}\to t_\star$.

To verify Condition~\ref{supp:ass:crossing}, take
$\underline t_u=t_\star/2$. For $0<\epsilon<t_\star/4$, eventually
$|t_{n,u}^{\star}-t_\star|<\epsilon/4$, and
$s=t_\star-\epsilon/2$ belongs to
$[t_{n,u}^{\star}-\epsilon,t_{n,u}^{\star}]$.
Local uniform convergence on $[t_\star/2,\alpha]$ gives a positive
margin at $s$ and a negative margin for every
$t\ge t_{n,u}^{\star}+\epsilon$. Both inequalities in
Condition~\ref{supp:ass:crossing} hold eventually with
the claimed margin. In detail, put $\kappa_{\mathrm{BH}}=\alpha^{-1}-\pi_0>0$ and
make the local uniform error smaller than $\kappa_{\mathrm{BH}}\epsilon/8$. At
$s=t_\star-\epsilon/2$,
\[
 \mathsf H_{n,u}(s)-s/\alpha
 \ge \kappa_{\mathrm{BH}}\epsilon/2-\kappa_{\mathrm{BH}}\epsilon/8
 =3\kappa_{\mathrm{BH}}\epsilon/8.
\]
For $t\ge t_{n,u}^\star+\epsilon$, the bound
$|t_{n,u}^\star-t_\star|<\epsilon/4$ gives
\[
 \mathsf H_{n,u}(t)-t/\alpha
 \le-3\kappa_{\mathrm{BH}}\epsilon/4+\kappa_{\mathrm{BH}}\epsilon/8
 =-5\kappa_{\mathrm{BH}}\epsilon/8.
\]
In particular both required inequalities hold with
\[
 m_{u,\epsilon}=(\alpha^{-1}-\pi_0)\epsilon/8.
\]
The condition has therefore been verified from the stated hypotheses.

Use the empirical-law argument in the proof of
Theorem~\MainTestingTheorem{}, rather than invoking that theorem
with its full Condition~\ref{supp:ass:signal}. Specifically,
Theorem~\MainConsistencyTheorem{} gives fitted-null CDF consistency;
Lemma~\ref{supp:lem:null-calibration} gives the null empirical law;
and \eqref{supp:eq:escaping-signal-cdfs} supplies the signal empirical
law directly, without a copula fluctuation bound. The pathwise
sandwich in Lemma~\ref{supp:lem:testing-transfer} transfers this
local signal law to the fitted values. Thus both all-test empirical
CDFs converge locally uniformly in conditional probability to
$\mathsf H_\infty$ on $(0,1)$.
The separated-crossing argument gives $t_n\to t_\star>0$ for both
procedures. For either procedure let $\mathbb G_{1,n}$ denote its
signal empirical CDF. These CDFs converge to one uniformly near
$t_\star$, so
$\operatorname{TDP}(\mathcal R_n)=\mathbb G_{1,n}(t_n)\to1$.
Boundedness yields the conditional mean limits and, after averaging
over $U$, the unconditional probability and power limits. This also
proves the corollary under the stronger full Condition~\ref{supp:ass:signal},
but that additional joint-dependence assumption is not used here.
\end{proof}

\section{Technical lemmas of Theorems 1--2}
\label{supp:sec:lemmas}
\label{supp:sec:lemmas-theorems-1-2}

\begin{lemma}[Empirical information and a moving center]
\label{supp:lem:cdf-stat}
Put $f_{0,n}(\cdot\mid u)=\overline f_{0,n,u}$,
$F_{0,n}(\cdot\mid u)=\overline F_{0,n,u}$, and
\[
 \mathbb F_{0,n}(z)=\frac1{n_{0,n}}\sum_{i\in\mathcal H_{0,n}}
       \mathbf1\{Z_{n,i}\le z\},\qquad
 \epsilon_{0,n}=\|\mathbb F_{0,n}-\overline F_{0,n,U}\|_\infty.
\]
Under Conditions~\ref{supp:ass:factor}--\ref{supp:ass:residual},
\begin{equation}
 \E [\epsilon_{0,n} \mid U]\le C D_n^{1/3}\le C \sigma_{0,*}^{-2/3}\vartheta_n^{1/3},
 \label{supp:eq:cdf-mean-stat}
\end{equation}
and hence
\begin{equation}
 \epsilon_{0,n}=O_{\Pp (\cdot \mid U)}(n^{-\rho_{\mathrm{dep}}/3}).
 \label{supp:eq:cdf-rate-stat}
\end{equation}
If Condition~\ref{supp:ass:center} also holds, there are deterministic
$p_q\in(0,1/2)$ and $C_q<\infty$ such that, with conditional
probability tending to one,
\begin{equation}
 p_q\le F_{0,n}(b_n\mid U)\le1-p_q,\qquad
 |b_n|\le b_{\max}(U):=\|U\|+C_q.
 \label{supp:eq:cutoff-stat}
\end{equation}
Moreover, for a finite constant $C_U$,
\begin{equation}
 \mathsf P_n\log(2+|x|)\le C_U
 \label{supp:eq:center-log-event}
\end{equation}
with conditional probability tending to one.  For all sufficiently
large $n$, let
\[
 b_{0,n}(U)=\overline F_{0,n,U}^{-1}(N_n/n_{0,n}).
\]
Then
\begin{equation}
 |b_n-b_{0,n}(U)|=O_{\Pp (\cdot \mid U)}(n^{-\rho_{\mathrm{dep}}/3}+n^{-\zeta}).
 \label{supp:eq:moving-cutoff-rate}
\end{equation}
\end{lemma}

\begin{lemma}[Uniform target stability and compact loading topology]
\label{supp:lem:target-stability}
Let $K_*=\{\ell\in\mathbb R^d:\|\ell\|^2\le1-\sigma_{0,*}^2\}$.
For a probability $H$ on $K_*$ and $u\in\mathbb R^d$, define
\begin{equation}
 f_{H,u}(z)=\int_{K_*}
 \frac{\phi((z-\ell^{\mathsf T}u)/\sqrt{1-\|\ell\|^2})}
      {\sqrt{1-\|\ell\|^2}}\,H(d\ell),
 \qquad F_{H,u}(z)=\int_{-\infty}^zf_{H,u}(x)\,dx.
 \label{supp:eq:generic-pool}
\end{equation}
Let $\mathcal W_1$ denote first-order Wasserstein distance on $K_*$
with Euclidean cost.
For probabilities $H,H'$ on $K_*$ and every $u\in\R^d$,
\begin{align}
 \|f_{H,u}-f_{H',u}\|_1
 &\le C_{\sigma_{0,*}}(1+\|u\|)\mathcal W_1(H,H'),
 \label{supp:eq:pool-l1-lipschitz}\\
 \|F_{H,u}-F_{H',u}\|_\infty
 &\le\tfrac12 C_{\sigma_{0,*}}(1+\|u\|)\mathcal W_1(H,H').
 \label{supp:eq:pool-cdf-lipschitz}
\end{align}
All densities in this class are uniformly bounded below on every fixed
compact interval, for fixed $u$.  Their quantiles on
$[\epsilon,1-\epsilon]$ lie in a common compact interval and are
uniformly Lipschitz in the probability argument.

On $K_*$, $\mathcal W_1(H_n,H)\to0$ if and only if $H_n\Rightarrow H$.
In particular, if
\begin{equation}
 \mathcal W_1(H_{0,n},H_0)\longrightarrow0,
 \label{supp:eq:optional-loading-limit}
\end{equation}
then the pooled densities and CDFs converge to
\begin{equation}
 f_{0,U}=f_{H_0,U},\qquad F_{0,U}=F_{H_0,U}.
 \label{supp:eq:model-limit-pool}
\end{equation}
\end{lemma}

\begin{lemma}[Compactness and a uniform logarithmic envelope]
\label{supp:lem:compact-stat}
For $R\in\mathcal P_k$, write
$F_{\phi R}(b)=\int_{-\infty}^b\phi(x)R(x)\,dx$.
The set $\mathcal P_k$ is nonempty, compact and convex in coefficient space.
There is an absolute constant $C$ such that every $R\in\mathcal P_k$ satisfies
\begin{align}
 R(x)&\leq C(k+1)^2[2(1+|x|)]^k,\label{supp:eq:poly-envelope-stat}\\
 |\log R(x)|&\leq C\{k^2+k\log(2+|x|)\}.\label{supp:eq:log-envelope-stat}
\end{align}
For $L\geq1$, the total variation of $\log R$ on $[-L,L]$ is at most
\begin{equation}
 C\{k^3+k^2\log(2+L)\}.\label{supp:eq:variation-stat}
\end{equation}
For a fixed $b$, the conditional parameter set
\[
 \left\{S=R/F_{\phi R}(b):R\in\mathcal P_k\right\}
\]
is also compact and convex.  In this parameterization the conditional
log-likelihood is concave.
\end{lemma}

\begin{lemma}[Coefficient bounds and a fixed-interval empirical Gram bound]
\label{supp:lem:curvature-stat}
Fix $u\in\mathbb R^d$ and a cutoff bound $0 \le \mathfrak{b}<\infty$, and set
$I_{\mathfrak b}=[-\mathfrak b-2,-\mathfrak b-1]$.
There is $C_{u,\mathfrak b}<\infty$ such that the following assertions
hold for every $k\ge2$ and $|b|\le\mathfrak b$.
\begin{enumerate}
\item Every $\beta\in\mathcal D_{k,b}$ satisfies
\[
 \|\beta\|_2\le e^{C_{u,\mathfrak b}k^2},\qquad
 \sup_{x\in I_{\mathfrak b}}r_\beta(x)
       \le e^{C_{u,\mathfrak b}k^2}.
\]
The domain diameter is at most $2e^{C_{u,\mathfrak b}k^2}$.
\item Uniformly in $n$,
\[
 \lambda_{\min}\!\left(\int_{I_{\mathfrak b}}
       \Psi_k(x)\Psi_k(x)^{\mathsf T} f_{0,n}(x\mid u)\,dx\right)
 \ge e^{-C_{u,\mathfrak b}k\log(k+1)}.
\]
\item For the retained empirical center $\mathsf P_n$, on
$|b_n|\le\mathfrak b$ and for all sufficiently large $n$,
\begin{equation}
 \lambda_{\min}\{\mathsf P_n[\ind_{I_{\mathfrak b}}\Psi_k\Psi_k^{\mathsf T}]\}
 \ge e^{-C_{u,\mathfrak b}k\log(k+1)}
       -e^{C_{u,\mathfrak b}k\log(k+1)}\epsilon_{0,n}.
 \label{supp:eq:empirical-gram-stat}
\end{equation}
On the event
\begin{equation}
 |b_n|\le\mathfrak b,\qquad
 \epsilon_{0,n}\le e^{-C'_{u,\mathfrak b}k^3},
 \label{supp:eq:curvature-event-stat}
\end{equation}
with a sufficiently large $C'_{u,\mathfrak b}$, the empirical
criterion $L_{\mathsf P_n}(\beta)=\mathsf P_n\log r_\beta$ satisfies
\begin{equation}
 -h^{\mathsf T}\nabla^2L_{\mathsf P_n}(\beta)h
 =\mathsf P_n\!\left[\frac{(h^{\mathsf T}\Psi_k)^2}{r_\beta^2}\right]
 \ge e^{-C''_{u,\mathfrak b}k^3}\|h\|_2^2.
 \label{supp:eq:empirical-curvature-stat}
\end{equation}
This proves strong concavity on the feasible domain, including its
boundary, by integration along feasible segments.
\end{enumerate}
For $k_n^3=o(\log n)$ and $\mathfrak b=b_{\max}(U)$ from
Lemma~\ref{supp:lem:cdf-stat}, event
\eqref{supp:eq:curvature-event-stat} has conditional probability tending
to one under Conditions~\ref{supp:ass:factor}--\ref{supp:ass:center}.
\end{lemma}

\begin{lemma}[Analytic pooled target and positive KL comparator]
\label{supp:lem:comparator-stat}
Fix $u\in\R^d$.  For every $n$ such that $n_{0,n} \ge 1$, the density
$f_{0,n}=f_{H_{0,n},u}$ is positive and real analytic, and its ratio
$f_{0,n}/\phi$ satisfies
\begin{equation}
 \frac{f_{0,n}(x)}{\phi(x)}\leq M_u:=\sigma_{0,*}^{-1}\exp(\norm u^2/2),\qquad x\in\R.
 \label{supp:eq:true-upper-stat}
\end{equation}
Let $\KL$ denote Kullback--Leibler divergence.
Let $q_*=(1+\sigma_{0,*}^2)^{-1}\in[1/2,1)$ and
$\rho_{\mathrm{app}}:=\frac14\log(1+\sigma_{0,*}^2)>0$.  For every sufficiently large $k$ there
exists $f_{n,k}^\circ\in\mathcal F_k$ such that
\begin{align}
 \norm{f_{n,k}^\circ-f_{0,n}}_1&\leq C_u e^{-\rho_{\mathrm{app}}k},\label{supp:eq:positive-l1-stat}\\
 \KL(f_{0,n}\Vert f_{n,k}^\circ)&\leq C_u k e^{-\rho_{\mathrm{app}}k},\label{supp:eq:positive-kl-stat}\\
 \left|\int x^2\{f_{n,k}^\circ(x)-f_{0,n}(x)\}\,dx\right|
 &\le C_u\sqrt{k+1}\,e^{-\rho_{\mathrm{app}}k}.
 \label{supp:eq:positive-moment-stat}
\end{align}
For any $b$ with $F_{0,n}(b)\geq p_{\min}>0$,
\begin{equation}
 \KL(\mathcal C_b f_{0,n}\Vert \mathcal C_b f_{n,k}^\circ)\leq p_{\min}^{-1}C_u k e^{-\rho_{\mathrm{app}}k}.
 \label{supp:eq:positive-central-kl-stat}
\end{equation}
The constants and the threshold for sufficiently large $k$ are uniform
over $n$, over all loading laws in $\mathcal P(K_*)$, and over $b$ in the stated range,
with dependence on $u$ and $\sigma_{0,*}^2$.
\end{lemma}

\begin{lemma}[Polynomial continuation from a left half-line]
\label{supp:lem:extension-stat}
Fix $0\le \mathfrak{b} < \infty$.  For every sufficiently large $k$, every $|b|\leq \mathfrak b$,
and every real polynomial $\mathfrak p$ of degree at most $k$,
\begin{equation}
 \int_{-M_k}^{M_k}\phi(x)|\mathfrak p(x)|\,dx
 \leq E_k(\mathfrak b)\int_{-\infty}^b\phi(x)|\mathfrak p(x)|\,dx,
 \qquad M_k=k^{1/4},\label{supp:eq:extension-stat}
\end{equation}
where, with constants depending only on $\mathfrak b$,
\begin{equation}
 E_k(\mathfrak b)\leq C_\mathfrak b k^{31/20}\exp(C_\mathfrak b k^{9/10})
 \leq\exp(C_\mathfrak b'k^{9/10}).\label{supp:eq:extension-factor-stat}
\end{equation}
Consequently, let $f=\phi R$ and $g=\phi S$ be two normalized
nonnegative polynomial densities of degree at most $k$, with second
moments $V_f,V_g$. Write $F=F_f$ and $G = F_g$ for their CDFs, and suppose $G(b)\geq p_{\min}>0$.  Then
\begin{equation}
 \norm{f-g}_1\leq2E_k(\mathfrak b)\norm{\mathcal C_b f-\mathcal C_b g}_1
      +\frac{2(V_f+p_{\min}^{-1}V_g)}{M_k^2}.
 \label{supp:eq:normalized-extension-stat}
\end{equation}
The conclusion applies to every $F(b)>0$.
\end{lemma}

\begin{lemma}[Uniform retained-center criterion bound]
\label{supp:lem:empirical-stat}
For a density $f$ with $F_f(b_n)>0$, define the following criteria
whenever the displayed integrals are finite:
\[
 \begin{aligned}
 \widehat L_{n,b_n}(f)
 &=L_{\mathsf P_n,b_n}(f)
 =\mathsf P_n\log(f/\phi)-\log F_f(b_n),\\
 L_{0,n,b_n}(f)
 &=\frac{\int_{-\infty}^{b_n}f_{0,n}(x)\log\{f(x)/\phi(x)\}\,dx}{F_{0,n}(b_n)}
 -\log F_f(b_n).
 \end{aligned}
\]
Under Conditions~\ref{supp:ass:factor}--\ref{supp:ass:center}, set
\[
 \mathcal S_{n,k}
 :=\sup_{f\in\mathcal F_k}|\widehat L_{n,b_n}(f)-L_{0,n,b_n}(f)|.
\]
For any deterministic sequence $k=k_n\geq2$,
\begin{equation}
 \mathcal S_{n,k}
 =O_{\Pp (\cdot \mid U)}\!\left(
    [k^3+k^2\log\log(en)](n^{-\rho_{\mathrm{dep}}/3})
       +k^2n^{-\zeta}+k^2n^{-2}\right).\label{supp:eq:empirical-stat}
\end{equation}
In particular, for $\rho_{\mathrm{stat}}=\min(\rho_{\mathrm{dep}}/3,\zeta)>0$,
\begin{equation}
 \mathcal S_{n,k}=O_{\Pp (\cdot \mid U)}\!\left(k^2[k+\log\log(en)]n^{-\rho_{\mathrm{stat}}}\right).
 \label{supp:eq:empirical-simple-stat}
\end{equation}
\end{lemma}

\begin{lemma}[Consistency of the empirical conditional target]
\label{supp:lem:projection-stat}
Let $p_{n,k}=p_{k,\mathsf P_n,b_n}$ be the maximizer of
\eqref{supp:eq:conditional-target}.  For the degree sequences used in
Theorem~\MainConsistencyTheorem{}, this maximizer exists and is unique on
events with conditional probability tending to one.  On such events,
\begin{align}
 \KL(\mathcal C_{b_n} f_{0,n}\|\mathcal C_{b_n} p_{n,k})
 &\le C_U k e^{-\rho_{\mathrm{app}}k}+2\mathcal S_{n,k},
 \label{supp:eq:projection-kl}\\
 \|p_{n,k}-f_{0,n}\|_1
 &\le C_U\left[
 e^{C_Uk^{9/10}}\sqrt{\mathcal S_{n,k}+ke^{-\rho_{\mathrm{app}}k}}
       +\frac{1+\log k}{\sqrt k}\right].
 \label{supp:eq:projection-l1}
\end{align}
The bound depends on the retained center and the sieve.
\end{lemma}

\paragraph{Fixed-envelope convention for Lemmas~\ref{supp:lem:mc-envelopes}--\ref{supp:lem:alg-quantitative-freezing}.}
Choose the interval-envelope constants first, so that the cutoff
$\mathfrak a_k$ in Lemma~\ref{supp:lem:alg-primitive-bounds} belongs to
$I_k$ in Lemma~\ref{supp:lem:mc-envelopes}. For each fixed collection of
cutoff, center, initializer, and $\eta$ bounds, fix $C_{\mathrm{prim}}>0$
and an integer $p_{\mathrm{alg}}\ge5$ once and set
\[
 \mathfrak A_{k,0}=\exp\{C_{\mathrm{prim}}(k+1)^3\}\ge2,\qquad
 \mathfrak A_k=\mathfrak A_{k,0}^{p_{\mathrm{alg}}}
             =\exp\{C_1(k+1)^3\},\qquad C_1=p_{\mathrm{alg}}C_{\mathrm{prim}}.
\]
Choose $C_{\mathrm{prim}}$ to cover the primitive bounds in
Lemmas~\ref{supp:lem:mc-envelopes}--\ref{supp:lem:alg-primitive-bounds},
including the scaled tail-Hessian and perturbation-Lipschitz bounds
used in Lemma~\ref{supp:lem:alg-global-regret}. Retain a fixed strong-convexity
modulus $m_k\in[\mathfrak A_{k,0}^{-1},1]$.
Choose $p_{\mathrm{alg}}$ to dominate the finitely many fixed powers of
$\mathfrak A_{k,0}$ needed for the metric conversions, inverse gaps,
localization, and rank-retention bounds below, and to ensure
$\mathfrak A_k\ge2/\eta^2$. These choices precede the localization
radii and stopping margins; neither envelope is enlarged later.
Further accumulated bounds are written as fixed powers
$\mathfrak A_k^p$, with the exponent fixed before $\Gamma_k$ is chosen.

\begin{lemma}[Envelopes and weak score control]
\label{supp:lem:mc-envelopes}
Fix a compact range $|b|\le \mathfrak b$, and put $r_k=4k+4$.
For a mass-zero signed measure $D$ with finite $(r_k+1)$st absolute
moment, define
\begin{equation}
 \mathcal N_k(D)=\sup_x|D((-\infty,x])|
       +\int_{\mathbb R}(1+|x|)^{r_k}|D((-\infty,x])|\,dx.
 \label{supp:eq:mc-weak-norm}
\end{equation}
Use the fixed envelopes specified above. For the fixed sufficiently large
interval constant $C_{\mathrm{env}}$, set
$I_k=[-\mathfrak b,C_{\mathrm{env}}(k+1)]$ and
$I_k^+=[-\mathfrak b,C_{\mathrm{env}}(k+1)+1]$.
Uniformly over the fitted sieve, $a\in I_k^+$, and
$0\le s\le 2r_k+4$,
\begin{equation}
 \{1-F_{f_\beta}(a)\}\ge e^{-Ck^2},
 \qquad
 \int(1+|x|)^s\,\mathcal T_af_\beta(dx)\le e^{Ck^2}.
 \label{supp:eq:mc-tail-envelopes}
\end{equation}
For the log-density score
\[
 S_\beta(x)
 =\nabla_\beta\log f_\beta(x)
 =\frac{\Psi_k(x)}{r_\beta(x)}-\frac{e_0}{A_\beta},
\]
one has
\begin{equation}
 \sup_\beta\|\partial_x S_\beta(x)\|_2
 \le \mathfrak A_k(1+|x|)^{r_k},\qquad
 \sup_\beta\left\|\int S_\beta\,dD\right\|_2
 \le \mathfrak A_k\mathcal N_k(D).
 \label{supp:eq:mc-score-control}
\end{equation}
The same bound holds for the spatial derivative and pairing of the
parameter-dependent log-density
$\log r_\beta-\log A_\beta$ and its parameter Hessian, with the
absolute value and operator norm, respectively.
\end{lemma}

\begin{lemma}[Primitive bounds on the polynomial sieve]
\label{supp:lem:alg-primitive-bounds}
Let $\mathsf P$ be a probability supported on $(-\infty,b]$, where
$b$ ranges over a fixed compact set. Suppose that a fixed compact
interval $I$ and a finite constant $C_0$ satisfy
\begin{equation}
 \int_I(h^{\mathsf T}\Psi_k)^2\,d\mathsf P
 \ge e^{-C_0(k+1)^2}\|h\|_2^2\quad(h\in\mathbb R^{k+1}),\qquad
 \int\log(2+|x|)\,\mathsf P(dx)\le C_0.
 \label{supp:eq:alg-center-event}
\end{equation}
The interval $I$ is fixed across the centers under consideration.
For a single population center it may contain the endpoint $b$.
With the fixed-envelope convention above, the following bounds hold.
They also hold with the sharper primitive envelope $\mathfrak A_{k,0}$
in place of $\mathfrak A_k$. The fixed constants depend only on the
cutoff range, $I,C_0$, the initializer, and $\eta$.
\begin{enumerate}
\item The domain $\mathcal D_{k,b}$ has diameter at most $\mathfrak A_k$.
The continuous function $\mathcal V_{\mathsf P}=-\mathsf P\log r_\beta = -\int \log r_\beta (x) \mathsf{P}(dx)$ is
strongly convex on this domain with modulus $m_k\ge \mathfrak A_k^{-1}$.
The regret at the fixed initializer is at most $\mathfrak A_k$.
\item There is $\mathfrak a_k\le C_2(k+1)$ such that
\[
 \frac{1-F_*(\mathfrak a_k)}{1-F_*(b)}\le\frac\eta4
 \qquad(p_*\in\mathcal F_k).
\]
On $[b,\mathfrak a_k+1]$, all fitted densities and survival
normalizers are at least $\mathfrak A_k^{-1}$. The polynomial numerators and
their derivatives used below are at most $\mathfrak A_k$.
\item The log-ratio, score and Hessian pairings with mass-zero
signed measures are bounded by $\mathfrak A_k\mathcal N_k(\nu)$. At each fixed cutoff $a\in[b,\mathfrak a_k+1]$,
the parameter derivatives through total order three
of the fitted-tail expectations and the reference
tail refit criterion used below are bounded by
$\mathfrak A_k$ on feasible parameter segments.
The same envelope bounds the scaled full-refresh
tail Hessian.

\end{enumerate}
\end{lemma}

\begin{lemma}[Global learning under a small completion perturbation]
\label{supp:lem:alg-global-regret}
Under \eqref{supp:eq:alg-center-event}, let
$\beta\in\mathcal D_{k,b}$ and let the refit use the probability measure
\[
 F_{f_\beta}(b)\mathsf P+\{1-F_{f_\beta}(b)\}\mathsf Q^{\mathrm{comp}},
 \qquad F_{f_\beta}(b)=A_\beta^{-1},
\]
where $\mathsf Q^{\mathrm{comp}}$ is a tail probability measure supported on $(b,\infty)$ with finite
$(r_k+1)$st absolute moment. If $\theta$ is an
exact global refit and
$\epsilon=\mathcal N_k(\mathsf Q^{\mathrm{comp}}-\mathcal T_bf_\beta)$, then
\[
 \sqrt{\mathcal V_{\mathsf P}(\theta)-\mathcal V_{\mathsf P}(\beta_*)}
 \le(1-\mathfrak A_k^{-1})
        \sqrt{\mathcal V_{\mathsf P}(\beta)-\mathcal V_{\mathsf P}(\beta_*)}
       +\mathfrak A_k\epsilon,
\]
with the single fixed choice of $\mathfrak A_k$ specified above.
\end{lemma}

\begin{lemma}[Quantified local contraction around a retained reference]
\label{supp:lem:alg-local-reference}
Under \eqref{supp:eq:alg-center-event}, let
\[
 \mathsf B_a^*=(\mathcal T_bp_*)\mathbf1_{(b,a]},\qquad
 \lambda_a^*=\frac{1-F_*(a)}{1-F_*(b)},
 \qquad b\le a\le\mathfrak a_k+1.
\]
For the fixed envelope $\mathfrak A_k$, there are a positive definite
metric $\mathsf M_*$ with eigenvalues in $[\mathfrak A_k^{-1},\mathfrak A_k]$,
and a gap $\gamma_{0,k}\in[\mathfrak A_k^{-1},1]$ with the following
property. Set
\begin{equation}
 \begin{aligned}
 R_k&=\min\{1,\gamma_{0,k}/(32\mathfrak A_k)\},&
 r_k^{\mathrm{in}}&=R_k/(4\mathfrak A_k),\\
 \gamma_k&=\gamma_{0,k}/2,&
 \varepsilon_k^{\mathrm{loc}}
   &=\min\{r_k^{\mathrm{in}},
                 \gamma_k r_k^{\mathrm{in}}/(2\mathfrak A_k)\}.
 \end{aligned}
 \label{supp:eq:local-radii}
\end{equation}
These quantities are bounded below by inverse fixed powers of $\mathfrak A_k$.
Consider a state $(\mathsf B,a,\lambda)$ where $\mathsf B$ is
supported on $(b,a]$ and has mass $1-\lambda$. Define
\[
 D=\mathsf B+\lambda \mathcal T_ap_*-\mathcal T_bp_*,\qquad
 \mathfrak d=\mathcal N_k(D).
\]
A fresh-batch perturbation is a mass-zero signed measure $E$ for
which the perturbed completed tail remains a probability, and put
$\nu=\mathcal N_k(E)$. If
\begin{equation}
 \|\beta-\beta_*\|_{\mathsf M_*}\le r_k^{\mathrm{in}},\qquad
 \mathfrak d+\nu\le\varepsilon_k^{\mathrm{loc}},
 \label{supp:eq:local-input-domain}
\end{equation}
then every exact global refit $\theta$ satisfies
\begin{equation}
 \|\theta-\beta_*\|_{\mathsf M_*}
 \le(1-\gamma_k)\|\beta-\beta_*\|_{\mathsf M_*}
                 +\mathfrak A_k(\mathfrak d+\nu)
 \le r_k^{\mathrm{in}}.
 \label{supp:eq:local-contraction}
\end{equation}
The bounds use the logarithmic center moment in
\eqref{supp:eq:alg-center-event}.
\end{lemma}

\begin{lemma}[Fresh batches and empirical rank retention]
\label{supp:lem:mc-rank-freezing}
Condition on an incoming sigma-field $\mathcal F_n^-$ containing the original data,
the environment and the complete pseudo-sampling history.
Let $\beta,a,\mathsf B,\lambda$ be measurable with respect to it, where
$\beta\in\mathcal D_{k,b}$, $a\in I_k$, $\lambda=m/\tau_n$, and
$\eta \tau_n\le m\le \tau_n$.  The measure $\mathsf B$ is a finite nonnegative
measure of mass $1-\lambda$ with finite $(r_k+1)$st absolute moment.
Conditionally draw $Y_{n,1},\ldots,Y_{n,m}$ independently from
$\mathsf Q=\mathcal T_af_\beta$, and define
\[
 \mathsf L_n=\frac1{\tau_n}\sum_{j=1}^m\delta_{Y_{n,j}},\qquad
 \nu_n=\mathcal N_k(\mathsf L_n-\lambda \mathsf Q)+\frac1{\tau_n}.
\]
Then
\begin{equation}
 \mathbb E(\nu_n\mid\mathcal F_n^-)\le \frac{\mathfrak A_k}{\sqrt{\tau_n}}.
 \label{supp:eq:mc-noise-expectation}
\end{equation}

For a nonterminal block, let $0\le h<m$,
$\omega=h/\tau_n$, and $\lambda_+=\lambda-\omega\ge\eta$.
Set $a^\dagger=F_{\mathsf Q}^{-1}(h/m)$, with $a^\dagger=a$ if $h=0$, and suppose
$a^\dagger\in I_k$.  Define
\[
 \mathsf E_n=\frac1{\tau_n}\sum_{j=1}^h\delta_{Y_{n,(j)}},\qquad
 \widehat a_n=Y_{n,(h+1)},\qquad
 \mathsf E_n^\circ=\lambda \mathsf Q\mathbf1_{(a,a^\dagger]}.
\]
There is a deterministic $\underline q_k\ge e^{-Ck^2}$ such that
$(\mathcal T_a f_\beta)(x)\ge\underline q_k$ for $a<x\in I_k^+$.
On the event $\nu_n\le\eta\underline q_k/2$,
\begin{align}
 \widehat a_n&\in I_k^+,
 &
 |\widehat a_n-a^\dagger|&\le\frac{\nu_n}{\eta\underline q_k},
 \label{supp:eq:mc-quantile}\\
 \mathcal N_k(\mathsf E_n-\mathsf E_n^\circ)&\le\nu_n.
 \label{supp:eq:mc-block}
\end{align}

Let $p^*=f_{\beta^*}$, $\beta^*\in\mathcal D_{k,b}$, be any fitted
reference density measurable with respect to the incoming sigma-field,
and put
\[
 D=\mathsf B+\lambda \mathcal T_ap^*-\mathcal T_bp^*,\qquad
 \widehat D_{n,+}=\mathsf B+\mathsf E_n+\lambda_+\mathcal T_{\widehat a_n}p^*-\mathcal T_bp^*.
\]
On the same event,
\begin{equation}
 \mathcal N_k(\widehat D_{n,+})
 \le \mathcal N_k(D)
       +\mathfrak A_k\omega\|\beta-\beta^*\|_2+\mathfrak A_k\nu_n.
 \label{supp:eq:mc-debt}
\end{equation}
All these statements hold for a reference $\beta^*$ measurable
with respect to the incoming sigma-field, in particular the
conditional-center MLE.

At the terminal block $h=m$, no new cutoff is defined and
$\widehat D_{n,+}=\mathsf B+\mathsf L_n-\mathcal T_bp^*$ satisfies
\begin{equation}
 \mathcal N_k(\widehat D_{n,+})
 \le\mathcal N_k(D)
       +\mathfrak A_k\lambda\|\beta-\beta^*\|_2+\mathfrak A_k\nu_n.
 \label{supp:eq:mc-terminal}
\end{equation}
This terminal statement requires only the incoming $a\in I_k$.
\end{lemma}

\begin{lemma}[One refit per stage: a quantitative deterministic bound]
\label{supp:lem:alg-quantitative-freezing}
Fix $\eta\in(0,1)$, and retain a total fraction $1-\eta$ in $T-1$
blocks with maximum mass at most $C/T$, followed by the final refit
and full terminal retention.  In the empirical algorithm, require
$T\le \tau_n$ and use the integer rounded block masses, so that the
same bound holds after increasing $C$.  Suppose the center satisfies
\eqref{supp:eq:alg-center-event}.  There is
\[
 \Gamma_k=\exp\{\exp(C_3(k+1)^3)\}
\]
such that, for all sufficiently large $T$ and all sample paths with
\[
 T\ge\Gamma_k^2,\qquad
 \sum_{t=1}^T\nu_{n,t}\le\Gamma_k^{-2},
\]
the actual one-refit-per-stage output satisfies
\begin{equation}
 \|\widehat\beta_n^{(T)}-\beta_*\|_2
 \le\Gamma_k\left\{T^{-1}+\sum_{t=1}^T\nu_{n,t}\right\}.
 \label{supp:eq:alg-terminal-bound}
\end{equation}
For the population recursion there is no simulation discrepancy, and under
the same lower bound on $T$ the corresponding coefficient bound is
\[
 \|\beta_{\mathrm{pop}}^{(T)}-\beta_*\|_2\le\Gamma_k T^{-1}.
\]
The density conversions are
\[
 \begin{aligned}
 \|\widehat f_{n;k}^{(T)}-p_*\|_1
 &\le2\|\widehat\beta_n^{(T)}-\beta_*\|_2,\\
 \|f_{k,\mathrm{pop}}^{(T)}-p_*\|_1
 &\le2\|\beta_{\mathrm{pop}}^{(T)}-\beta_*\|_2.
 \end{aligned}
\]
This bound holds for every fixed admissible initializer and every
fixed $\eta\in(0,1)$.
\end{lemma}

\section{Proof of Lemmas S1--S14}
\label{supp:sec:proofs-lemmas-theorems-1-2}

\subsection[Proof of Lemma S1]{Proof of Lemma~\ref{supp:lem:cdf-stat}}
\label{supp:sec:proof-lemma-s1}

\begin{proof}
Fix $U=u$ and write $F_{0,n}=\overline F_{0,n,u}$.
For standardized jointly Gaussian variables $X,Y$ with correlation
$\rho$, orthonormal Hermite expansions of centered square-integrable
functions give
\[
 |\operatorname{Cov}(g(X),h(Y))|
 \le |\rho|\sqrt{\operatorname{Var}g(X)\operatorname{Var}h(Y)}.
\]
Indeed, the Gaussian--Hermite covariance identity and Parseval's
identity give
\[
 \begin{aligned}
 |\Cov(g(X),h(Y))|
 &=\left|\sum_{j\ge1}\rho^ja_jb_j\right|\\
 &\le |\rho|\Big(\sum_{j\ge1}a_j^2\Big)^{1/2}
                 \Big(\sum_{j\ge1}b_j^2\Big)^{1/2},
 \end{aligned}
\]
where $\{a_j\}, \{b_j\}$ are coefficients satisfying 
\[
g(x) - \mathbb E g(X) = \sum_{j=1}^\infty a_j \psi_j(x),
\quad
h(y) - \mathbb E h(Y) = \sum_{j=1}^\infty b_j \psi_j(y),
\]
for $\psi_j(\cdot) = \frac{1}{\sqrt{j!}} \operatorname{He}_j(\cdot)$. One may first use finite Hermite sums and then pass to the limit in
$L^2$; the identity also covers $\rho=\pm1$.
For $I_i(x)=\ind\{Z_{n,i}\le x\}$, each variance is at most $1/4$.
Since $\E[\mathbb F_{0,n}(x) \mid U=u]=F_{0,n}(x)$, it follows that
\[
 \begin{aligned}
 \Var_u\{\mathbb F_{0,n}(x)\}
 &=\frac1{n_{0,n}^2}\sum_{i,j\in\mathcal H_{0,n}}
                            \Cov_u(I_i(x),I_j(x))\\
 &\le\frac1{4n_{0,n}^2}\sum_{i,j\in\mathcal H_{0,n}}
          |\operatorname{Corr}(\varepsilon_{n,i},\varepsilon_{n,j})|
 =D_n/4.
 \end{aligned}
\]

The function $F_{0,n}$ is continuous and strictly increasing.
For $0<\epsilon<1$, choose a quantile grid with probability
spacing at most $\epsilon/2$ and at most $3/\epsilon$ interior points.
On an interval between consecutive grid points, monotonicity of both
CDFs bounds the error by the larger endpoint error plus $\epsilon/2$;
the same argument handles the two unbounded end intervals. Thus an
error larger than $2\epsilon$ forces an error larger than $\epsilon$
at a grid point. Chebyshev's inequality at each grid point gives
$D_n/(4\epsilon^2)$, and a union bound gives
\[
 \Pp\{\|\mathbb F_{0,n}-F_{0,n}\|_\infty>2\epsilon \mid U=u\}
 \le C D_n\epsilon^{-3}.
\]
Because $0\le\epsilon_{0,n}\le1$, the tail-integral formula, with
$a=D_n^{1/3}$, yields
\[
 \E[\epsilon_{0,n} \mid U=u]
 \le a+CD_n\int_a^1 t^{-3}\,dt
 \le a+CD_na^{-2}\le C'D_n^{1/3}.
\]
Combining this bound with
\eqref{supp:eq:general-residual-correlation-bound} and
Condition~\ref{supp:ass:residual} proves
\eqref{supp:eq:cdf-mean-stat}--\eqref{supp:eq:cdf-rate-stat}.

Let $S_n=|\mathcal H_{1,n}\cap\mathcal I_{0,n}|$ and let $E_n$
be the number of null observations equal to $b_n$ that were not
retained.  Since $F_{0,n}$ is continuous, every jump of
$\mathbb F_{0,n}$ is at most $2\epsilon_{0,n}$. In fact,
\[
 \mathbb F_{0,n}(x)-\mathbb F_{0,n}(x-)
 \le |\mathbb F_{0,n}(x)-F_{0,n}(x)|
       +|F_{0,n}(x)-\mathbb F_{0,n}(x-)|
 \le2\epsilon_{0,n}.
\]
The left-limit error obeys the same supremum bound by continuity of
$F_{0,n}$; moreover $S_n/n\le\chi_n$.  Therefore
\[
 0\le E_n/n_{0,n}\le2\epsilon_{0,n},\qquad
 N_n=n_{0,n}\mathbb F_{0,n}(b_n)-E_n+S_n,
\]
and
\begin{equation}
 \left|F_{0,n}(b_n)-N_n/n_{0,n}\right|
 \le3\epsilon_{0,n}+(n/n_{0,n})\chi_n.
 \label{supp:eq:center-mass-moving}
\end{equation}
To see the last inequality explicitly, subtract the count identity
from $n_{0,n}F_{0,n}(b_n)$ and use
\[
 \left|F_{0,n}(b_n)-\frac{N_n}{n_{0,n}}\right|
 \le\epsilon_{0,n}+\frac{E_n}{n_{0,n}}+\frac{S_n}{n_{0,n}}.
\]
Here $N_n/n_{0,n}\to p=q/\pi_0\in(0,1)$.
For $p_q=\min(p,1-p)/4$, the first inequality in
\eqref{supp:eq:cutoff-stat} follows.  Every component mean has magnitude
at most $\|u\|$ and standard deviation at most one.  Gaussian tail
bounds put all the corresponding mixture quantiles between
$-\|u\|-C_q$ and $\|u\|+C_q$, uniformly over the loading law.
This proves the cutoff bound for a fixed suitable $C_q$.
The deterministic quantile $b_{0,n}(u)$ lies in the same
interval for large $n$.  On this interval all mixture densities have
a common strictly positive lower bound, by
Lemma~\ref{supp:lem:target-stability}.  Denote that lower bound by $m_u>0$. On the stated event,
\[
 \begin{aligned}
 m_u|b_n-b_{0,n}(u)|
 &\le|F_{0,n}(b_n)-F_{0,n}(b_{0,n}(u))|\\
 &=\left|F_{0,n}(b_n)-\frac{N_n}{n_{0,n}}\right|\\
 &\le3\epsilon_{0,n}+(n/n_{0,n})\chi_n.
 \end{aligned}
\]
The two error rates prove \eqref{supp:eq:moving-cutoff-rate}.

For \eqref{supp:eq:center-log-event}, put $g(x)=\log(2+|x|)$.
The bounded conditional Gaussian means and variances give
$\sup_{n,i\in\mathcal H_{0,n}}\E[g(Z_{n,i})^2 \mid U=u] <\infty$.
The covariance inequality above yields
\[
 \operatorname{Var}_u\left\{\frac1{n_{0,n}}
                 \sum_{i\in\mathcal H_{0,n}}g(Z_{n,i})\right\}
 \le C_uD_n\to0.
\]
The null average thus lies below a fixed finite constant with
conditional probability tending to one.  Its contribution to the
retained average is at most $n_{0,n}/N_n$ times that full average.
The nonnull contribution is bounded using
$\log(2+|x|)\le1+\log(1+|x|)$, so altogether
\[
 \mathsf P_ng
 \le\frac{n_{0,n}}{N_n}
       \left\{\frac1{n_{0,n}}\sum_{i\in\mathcal H_{0,n}}g(Z_{n,i})\right\}
       +\frac n{N_n}\chi_n.
\]
Both deterministic ratios are bounded, and the last term is
$o_{\Pp (\cdot \mid U=u)}(1)$. This proves the logarithmic bound.
\end{proof}

\subsection[Proof of Lemma S2]{Proof of Lemma~\ref{supp:lem:target-stability}}
\label{supp:sec:proof-lemma-s2}

\begin{proof}
Write $f_{m,\sigma^2}(x)=\sigma^{-1}\phi((x-m)/\sigma)$, where
$m=\ell^{\mathsf T}u$ and $\sigma=(1-\|\ell\|^2)^{1/2}\ge \sigma_{0,*}>0$.
For $G\sim N(0,1)$, differentiation and a change of variables give
\[
 \|\partial_mf_{m,\sigma^2}\|_1=\frac{\E|G|}{\sigma},\qquad
 \|\partial_{\sigma^2}f_{m,\sigma^2}\|_1=\frac{\E|G^2-1|}{2\sigma^2}.
\]
Indeed,
\[
 \partial_mf_{m,\sigma^2}
   =\frac{x-m}{\sigma^2}f_{m,\sigma^2},\qquad
 \partial_{\sigma^2}f_{m,\sigma^2}
   =\frac{(x-m)^2-\sigma^2}{2\sigma^4}f_{m,\sigma^2}.
\]
The chain rule, $\nabla_\ell m=u$, and
$\nabla_\ell\sigma^2=-2\ell$ therefore show that
\[
 \int\|\nabla_\ell f_{m,\sigma^2}(x)\|_2\,dx
 \le\|u\|\|\partial_mf_{m,\sigma^2}\|_1
       +2\|\ell\|\|\partial_{\sigma^2}f_{m,\sigma^2}\|_1.
\]
Consequently $\|\nabla_\ell f_{m,\sigma^2}\|_{L^1(\R;\R^d)}
\le C_{\sigma_{0,*}}(1+\|u\|)$.
The segment $\ell_t=(1-t)\ell+t\ell'$ stays in $K_*$. The
fundamental theorem of calculus and the preceding derivative bound give
\[
 \|f_{\ell^{\mathsf T}u,1-\|\ell\|^2}
       -f_{\ell'^{\mathsf T}u,1-\|\ell'\|^2}\|_1
 \le C_{\sigma_{0,*}}(1+\|u\|)\|\ell-\ell'\|.
\]
For any coupling $\Pi$ of $H,H'$, Tonelli's theorem and the triangle
inequality then give
\[
 \|f_{H,u}-f_{H',u}\|_1
 \le C_{\sigma_{0,*}}(1+\|u\|)
                \int\|\ell-\ell'\|\,\Pi(d\ell,d\ell').
\]
Taking the infimum over couplings proves
\eqref{supp:eq:pool-l1-lipschitz}.  The CDF bound follows because the
supremum CDF distance between densities is at most half their $L^1$
distance.

For $|x|\le \mathfrak b$, every component density is at least
$(2\pi)^{-1/2}\exp\{-(\mathfrak b+\|u\|)^2/(2\sigma_{0,*}^2)\}$.
Every component mean has magnitude at most $\|u\|$ and standard
deviation at most one.  Gaussian tail bounds therefore place all
mixture quantiles with probabilities in $[\epsilon,1-\epsilon]$
in an interval depending only on $u,\epsilon$.  Integrating the
uniform density lower bound on this interval gives the asserted
inverse-CDF Lipschitz bound. More explicitly, if the lower bound
is $\underline f_{u,\epsilon}^{\mathrm{qnt}}>0$ and $\epsilon\le s<t\le1-\epsilon$, then
\[
 t-s=\int_{F_{H,u}^{-1}(s)}^{F_{H,u}^{-1}(t)}f_{H,u}(x)\,dx
 \ge \underline f_{u,\epsilon}^{\mathrm{qnt}}
       \{F_{H,u}^{-1}(t)-F_{H,u}^{-1}(s)\}.
\]

The compact-support topology statement is also discussed by \citet{Santambrogio2015}.
If $H_n\Rightarrow H$, partition $K_*$ into finitely many Borel
sets of diameter at most $\epsilon$ whose boundaries have $H$-mass
zero.  Such partitions follow from a finite ball cover with radii
chosen outside the countably many boundary atoms.  Match the common
mass of $H_n,H$ within each cell and couple the remaining mass
arbitrarily.  The cost is at most
\[
 \epsilon+\frac{\operatorname{diam}(K_*)}{2}
                 \sum_A|H_n(A)-H(A)|.
\]
The second term tends to zero, and then $\epsilon\downarrow0$ proves
Wasserstein convergence.  Conversely, a coupling with mean distance
approaching zero makes integrals of every uniformly continuous bounded
function converge, by splitting at a distance $\epsilon$ and using
Markov's inequality.  All continuous functions on $K_*$ are uniformly
continuous. In particular, for a continuous $v$ and a coupling
$(\widetilde\ell_n,\widetilde\ell)$ with laws $H_n,H$, respectively,
\[
 |\E v(\widetilde\ell_n)-\E v(\widetilde\ell)|
 \le\omega_v(\epsilon)
       +2\|v\|_\infty\epsilon^{-1}\E\|\widetilde\ell_n-\widetilde\ell\|,
\]
where $\omega_v$ is its modulus of continuity. Letting first $n\to\infty$
and then $\epsilon\downarrow0$ proves the converse.  The singleton case is immediate.
\end{proof}

\subsection[Proof of Lemma S3]{Proof of Lemma~\ref{supp:lem:compact-stat}}
\label{supp:sec:proof-lemma-s3}

\begin{proof}
The polynomial $R=1$ is feasible. For each fixed $x$, the condition
$R(x)\ge\delta_k$ is a closed affine inequality in the coefficients;
the moment constraint has the same form. Their intersection is
therefore closed and convex. Since $R\ge0$ and $\int\phi R=1$,
\[
 1\ge\int_{-1}^1\phi(x) R(x) dx\ge\phi(1)\int_{-1}^1R(x) dx,
 \qquad \int_{-1}^1R(x) dx\le1/\phi(1).
\]
Let $M=\|R\|_{\infty,[-1,1]}$ and choose $x_0$ attaining it.
Markov's polynomial inequality (see \citealp{Shadrin2004}) gives
$\|R'\|_{\infty,[-1,1]}\le k^2M$. The mean value theorem then implies
\[
 R(x)\ge M-k^2M|x-x_0|\ge M/2
 \quad\text{if }|x-x_0|\le(2k^2)^{-1},\quad x\in[-1,1].
\]
At least a one-sided interval of this length is available, including
when $x_0$ is an endpoint. Hence
\[
 \frac{M}{4k^2}\le\int_{-1}^1R(x) dx\le\frac1{\phi(1)},
 \qquad M\le\frac{4k^2}{\phi(1)}.
\]
Evaluate $R$ at any fixed $k+1$ distinct points of $[-1,1]$.
Its monomial coefficient vector is the inverse Vandermonde matrix
applied to these bounded values. For fixed $k$ this matrix is fixed
and invertible, so all coefficients are bounded. The change to the
Hermite basis is also a fixed invertible linear map. Heine--Borel
therefore gives compactness in coefficient space.

For $|x|\ge1$, the Chebyshev extremal inequality gives
\[
 |R(x)|\le M|\mathrm T_k(x)|
 \le M\exp\{k\operatorname{arcosh}|x|\}
 \le M(2|x|)^k.
\]
Combine this with the bound on $[-1,1]$ to obtain
\eqref{supp:eq:poly-envelope-stat}. Taking logarithms and using the
floor gives the two-sided estimates
\[
 -k^2\le\log R(x)
 \le\log C+2\log(k+1)+k\log2+k\log(1+|x|),
\]
which imply \eqref{supp:eq:log-envelope-stat} after increasing $C$.
Unless $R$ is constant, $R'$ has at most $k-1$ real zeros. Since
$(\log R)'=R'/R$ and $R>0$, there are at most $k$ monotonicity
intervals. On each one the variation is no more than the full range
of $\log R$ on $[-L,L]$. Consequently
\[
 \TV_{[-L,L]}(\log R)
 \le2k\|\log R\|_{\infty,[-L,L]}
 \le C\{k^3+k^2\log(2+L)\},
\]
proving \eqref{supp:eq:variation-stat}; the constant case is immediate.

For the conditional parameterization,
$F_{\phi R}(b)\ge\delta_k\Phi(b)>0$. Thus
$R\mapsto R/F_{\phi R}(b)$ is continuous on the compact coefficient
set and has compact image. To check convexity, put
$S=R/F_{\phi R}(b)$ and $A=\int\phi S=1/F_{\phi R}(b)$. Its image is
exactly the set of degree-at-most-$k$ polynomials satisfying
\[
 \int_{-\infty}^b\phi(x) S(x) dx =1,\qquad
 S(x)\ge\delta_k\int_{-\infty}^{\infty} \phi(t) S(t) dt\quad(x\in\R),
 \]
 \[
 \int_{-\infty}^{\infty} x^2\phi(x) S(x) dx\le(1+\log k)\int_{-\infty}^{\infty} \phi(x) S(x) dx.
\]
The forward implication follows by dividing each original constraint
by $F_{\phi R}(b)$. Conversely, these constraints give
$A\ge\delta_kA$, hence $A\ge0$. If $A=0$, then $S\ge0$ and
$\int\phi S=0$, contradicting its center integral being one.
Thus $A>0$, and $R=S/A$ satisfies the original normalization, floor,
and moment cap. Also $F_{\phi R}(b)=1/A$, so it maps back to $S$.
This proves the equivalence. All displayed constraints are affine,
and therefore the image is convex.

On the center, the conditional density is $\phi S$. Its
parameter-dependent log-likelihood is $\mathsf P\log S$ (or the
corresponding finite sum). Concavity of $\log$ gives, wherever the
criteria are finite,
\[
 \mathsf P\log\{tS_1+(1-t)S_2\}
 \ge t\mathsf P\log S_1+(1-t)\mathsf P\log S_2,
 \qquad 0\le t\le1.
\]
The Gaussian term is independent of $S$, completing the claim.
\end{proof}

\subsection[Proof of Lemma S4]{Proof of Lemma~\ref{supp:lem:curvature-stat}}
\label{supp:sec:proof-lemma-s4}

\begin{proof}
\textbf{Coefficient and pointwise bounds.}
By Lemma~\ref{supp:lem:compact-stat},
$\|R\|_{\infty,[-1,1]}\le C(k+1)^2$ for $R\in\mathcal P_k$.
Write its Chebyshev expansion as $R=\sum_{j=0}^kc_j\mathrm T_j$.
The cosine coefficient formula implies $|c_j|\le2C(k+1)^2$.
If $s_j$ is the sum of the absolute monomial coefficients of
$\mathrm T_j$, the recurrence
$\mathrm T_{j+1}=2x\mathrm T_j-\mathrm T_{j-1}$ gives
$s_{j+1}\le2s_j+s_{j-1}$ and hence $s_j\le3^j$.
Thus the monomial coefficients $a_m$ of $R$ satisfy
\[
 \sum_{m=0}^k|a_m|
 \le\sum_{j=0}^k|c_j|s_j\le C(k+1)^2 3^{k+1}.
\]
The Hermite inversion identity is
\[
 x^m=m!\sum_{r=0}^{\lfloor m/2\rfloor}
       \frac{\psi_{m-2r}(x)}{2^r r!\sqrt{(m-2r)!}}.
\]
The sum of its coefficients is at most
$m!\sum_{r\ge0}(2^rr!)^{-1}=e^{1/2}m!$. Therefore the Euclidean
Hermite coefficient norm of $R$ is no more than
$C(k+1)^2 3^{k+1}e^{1/2}k!$. Dividing by
$F_{\phi R}(b)\ge e^{-k^2}\Phi(-\mathfrak b)$ gives
\[
 \|\beta\|_2
 \le\frac{C(k+1)^2 3^{k+1}e^{1/2}k!e^{k^2}}
          {\Phi(-\mathfrak b)}
 \le e^{C_{u,\mathfrak b}k^2},
\]
where $\log(k!)\le k\log k$ is absorbed into the last exponent.
For $x\in I_{\mathfrak b}$ the polynomial envelope similarly gives
\[
 r_\beta(x)
 \le\frac{C(k+1)^2[2(\mathfrak b+3)]^ke^{k^2}}
          {\Phi(-\mathfrak b)}
 \le e^{C_{u,\mathfrak b}k^2}.
\]
The diameter bound follows by the triangle inequality.

\textbf{Population Gram bound.}
For $\|h\|_2=1$, put $P=h^{\mathsf T}\Psi_k$ and
$M=\sup_{I_{\mathfrak b}}|P|$. Hermite orthonormality gives
$\int P^2\phi=\sum_{j=0}^kh_j^2=1$.
Map the length-one interval $I_{\mathfrak b}$ affinely to $[-1,1]$.
The Chebyshev extremal inequality, including the trivial bound inside
the interval, yields
\[
 |P(x)|\le M[C_{\mathfrak b}(1+|x|)]^k,\qquad x\in\R.
\]
For $G\sim N(0,1)$, use
$\E|G|^{2k}=(2k-1)!!\le(2k)^k$ to obtain
\[
 \begin{aligned}
 1=\int P^2\phi
 &\le M^2 C_{\mathfrak b}^{2k}\E(1+|G|)^{2k}\\
 &\le M^2 C_{\mathfrak b}^{2k}2^{2k-1}\{1+(2k)^k\}.
 \end{aligned}
\]
Taking logarithms gives $M^2\ge e^{-C_{\mathfrak b}k\log(k+1)}$.
Markov's polynomial inequality on an interval of length one gives
$\|P'\|_\infty\le2k^2M$. At a maximizer of $|P|$, the mean value
theorem therefore supplies a one-sided interval of length
$1/(4k^2)$ on which $|P|\ge M/2$. Hence
\[
 \int_{I_{\mathfrak b}}P^2\,dx
 \ge\frac{M^2}{16k^2}
 \ge e^{-C_{\mathfrak b}k\log(k+1)}.
\]
Every Gaussian component of $f_{0,n}(\cdot\mid u)$ has mean bounded
by $\|u\|$ and variance in $[\sigma_{0,*}^2,1]$. On $I_{\mathfrak b}$
it is bounded below by
\[
 m_{u,\mathfrak b}:=(2\pi)^{-1/2}
 \exp\{-(\mathfrak b+2+\|u\|)^2/(2\sigma_{0,*}^2)\}>0.
\]
The mixture has the same lower bound. Thus
\[
 h^{\mathsf T}\left(\int_{I_{\mathfrak b}}
        \Psi_k\Psi_k^{\mathsf T}f_{0,n}\right)h
 =\int_{I_{\mathfrak b}}P^2f_{0,n}
 \ge m_{u,\mathfrak b}\int_{I_{\mathfrak b}}P^2.
\]
Taking the infimum over unit $h$ by the Rayleigh--Ritz formula proves
(ii), with the fixed factor $m_{u,\mathfrak b}$ absorbed into $C_{u,\mathfrak b}$.

\textbf{Empirical transfer and curvature.}
The explicit Hermite coefficient formula and
$\psi_j'=\sqrt j\,\psi_{j-1}$ give
\[
 \sup_{|x|\le\mathfrak b+2}
 \left\{\sum_{j=0}^k\psi_j(x)^2+
              \sum_{j=0}^k\psi_j'(x)^2\right\}
 \le e^{C_{\mathfrak b}k\log(k+1)}.
\]
Writing $I_{\mathfrak b}=[a,c]$, Cauchy--Schwarz and $\|h\|_2=1$
therefore imply
\[
 \begin{aligned}
 \TV(P^2\ind_{I_{\mathfrak b}})
 &\le P(a)^2+P(c)^2+2\int_a^c|PP'|\,dx\\
 &\le e^{C_{\mathfrak b}k\log(k+1)},
 \end{aligned}
\]
uniformly over $h$; this variation includes both endpoint jumps.
Lebesgue--Stieltjes integration by parts consequently yields
\[
 \left|\int_{I_{\mathfrak b}}P^2\,d(\mathbb F_{0,n}-F_{0,n})\right|
 \le e^{C_{\mathfrak b}k\log(k+1)}\epsilon_{0,n}.
\]
On $|b_n|\le\mathfrak b$, every null observation in $I_{\mathfrak b}$
is retained because $\sup I_{\mathfrak b}<-\mathfrak b\le b_n$.
The squared polynomial is nonnegative, so dropping the retained
nonnull terms only decreases the empirical quadratic form. Thus
\[
 \begin{aligned}
 \mathsf P_n[\ind_{I_{\mathfrak b}}P^2]
 &\ge\frac{n_{0,n}}{N_n}\int_{I_{\mathfrak b}}P^2\,d\mathbb F_{0,n}\\
 &\ge\frac{n_{0,n}}{N_n}\left\{
        \int_{I_{\mathfrak b}}P^2 f_{0,n}
       -e^{C_{\mathfrak b}k\log(k+1)}\epsilon_{0,n}\right\}.
 \end{aligned}
\]
The ratio $n_{0,n}/N_n$ is eventually bounded above and away from
zero. Taking the infimum over unit $h$ proves
\eqref{supp:eq:empirical-gram-stat}, without any correction at $b_n$.
On \eqref{supp:eq:curvature-event-stat}, choose $C'_{u,\mathfrak b}$
large enough that the error term is at most half the positive term.
Twice differentiating the finite empirical sum gives
\[
 -\nabla^2 L_{\mathsf P_n}(\beta)
 =\mathsf P_n\left[\frac{\Psi_k\Psi_k^{\mathsf T}}{r_\beta^2}\right].
\]
Restricting its quadratic form to $I_{\mathfrak b}$ and using (i),
\[
 \begin{aligned}
 -h^{\mathsf T}\nabla^2 L_{\mathsf P_n}(\beta)h
 &\ge e^{-2C_{u,\mathfrak b}k^2}
             \mathsf P_n[\ind_{I_{\mathfrak b}}(h^{\mathsf T}\Psi_k)^2]\\
 &\ge\tfrac12e^{-2C_{u,\mathfrak b}k^2-C_{u,\mathfrak b}k\log(k+1)}
                  \|h\|_2^2
 \ge e^{-C''_{u,\mathfrak b}k^3}\|h\|_2^2.
 \end{aligned}
\]
This is \eqref{supp:eq:empirical-curvature-stat}. All observed values
are finite and all feasible polynomial values are positive, so the
calculation is valid along every feasible segment. Integrating the
Hessian bound along such segments proves strong concavity even when
the endpoints lie on the boundary.

Finally put $\rho_{\mathrm{cdf}}=\rho_{\mathrm{dep}}/3>0$.
If $k_n^3=o(\log n)$, then
\[
 \frac{e^{-C'_{u,\mathfrak b}k_n^3}}{n^{-\rho_{\mathrm{cdf}}}}
 =\exp\{\rho_{\mathrm{cdf}}\log n-C'_{u,\mathfrak b}k_n^3\}
 \longrightarrow\infty.
\]
Together with $\epsilon_{0,n}=O_{\Pp (\cdot \mid U=u)}(n^{-\rho_{\mathrm{cdf}}})$ and the cutoff
bound in Lemma~\ref{supp:lem:cdf-stat}, this proves the final
high-probability assertion.
\end{proof}

\subsection[Proof of Lemma S5]{Proof of Lemma~\ref{supp:lem:comparator-stat}}
\label{supp:sec:proof-lemma-s5}

\begin{proof}
For one component write $m=\ell^{\mathsf T}u$, $\sigma^2=1-\norm\ell^2$ with $\sigma>0$, and
$f_{m,\sigma^2}(x)=\sigma^{-1}\phi((x-m)/\sigma)$. For
$\sigma^2<1$, the exact log-ratio is
\[
 \log\frac{f_{m,\sigma^2}(x)}{\phi(x)}
 =-\log\sigma
  -\frac{1-\sigma^2}{2\sigma^2}
       \left(x-\frac{m}{1-\sigma^2}\right)^2
  +\frac{m^2}{2(1-\sigma^2)}.
\]
Since $m^2=(\ell^{\mathsf T}u)^2\le(1-\sigma^2)\|u\|^2$ by Cauchy--Schwarz inequality, taking the supremum in this expression gives
\[
 \sup_x\frac{f_{m,\sigma^2}(x)}{\phi(x)}
 =\sigma^{-1}\exp\!\left\{\frac{m^2}{2(1-\sigma^2)}\right\}
 \leq \sigma_{0,*}^{-1}e^{\norm u^2/2}.
\]
If $\sigma^2=1$, then $\ell=0$, $m=0$, and the ratio is one.  Mixing proves
\eqref{supp:eq:true-upper-stat}.  The same Gaussian formula, considered at
a complex argument, is locally uniformly bounded over $\|\ell\|^2\le1-\sigma_{0,*}^2$
because $\sigma^2\geq\sigma_{0,*}^2>0$ and $|m|\leq\|u\|$.
The holomorphic parameter-integral theorem therefore makes the
mixture an entire function.  Positivity on the real line is immediate.

Set $g_{m,\sigma^2}=\sqrt{f_{m,\sigma^2}/\phi}$ and expand
$g_{m,\sigma^2}=\sum_{j\geq0}c_j(m,\sigma^2)\psi_j$ in $L^2(\phi)$.
Completing the square once more yields
\[
 c_j(m,\sigma^2)=A_{m,\sigma^2}\E[\psi_j(X)],\qquad
 X\sim N\!\left(\frac{m}{1+\sigma^2},\frac{2\sigma^2}{1+\sigma^2}\right),
\]
where
\[
 A_{m,\sigma^2}=\left(\frac{2\sigma}{1+\sigma^2}\right)^{1/2}
                \exp\!\left\{-\frac{m^2}{4(1+\sigma^2)}\right\}\leq1.
\]
Here $\phi g_{m,\sigma^2}=\sqrt{\phi f_{m,\sigma^2}}$ is
$A_{m,\sigma^2}$ times the displayed Gaussian density; integrating
it against $\psi_j$ gives the coefficient formula. In particular,
put $a=m/(1+\sigma^2)$, $v=2\sigma^2/(1+\sigma^2)$,
$\chi_\sigma=(1-\sigma^2)/(1+\sigma^2)$, and
$y=m/\sqrt{1-\sigma^4}$. For $\sigma^2<1$,
\[
 \E e^{tX-t^2/2}
 =\exp\{at-(1-v)t^2/2\}
 =\exp\{y\sqrt{\chi_\sigma}\,t-\chi_\sigma t^2/2\}.
\]
Comparison with the Hermite generating function identifies the
coefficients as
\[
 c_j=A_{m,\sigma^2}\chi_{\sigma}^{j/2}\psi_j(y).
\]
The formula is interpreted by continuity at $\sigma^2=1,m=0$.
For $\sigma^2<1$,
$|y|=|\ell^{\mathsf T}u|/(\|\ell\|\sqrt{1+\sigma^2})\le\|u\|$;
the degenerate expression at $\ell=0$ is not needed separately.  We have
$\chi_{\sigma}\leq(1-\sigma_{0,*}^2)/(1+\sigma_{0,*}^2)$ and $|y|\leq\norm u$.
The diagonal Mehler identity
\[
 \sum_{j\geq0}t^j\psi_j(y)^2
 =(1-t^2)^{-1/2}\exp\!\left(\frac{ty^2}{1+t}\right),\qquad0\leq t<1,
\]
with $t=\chi_{\sigma}/q_*\leq1-\sigma_{0,*}^2$ shows uniformly over the components that
\begin{equation}
 \sum_{j\geq0}q_*^{-j}c_j(m,\sigma^2)^2\leq
 C_u':=[\sigma_{0,*}^2(2-\sigma_{0,*}^2)]^{-1/2}e^{\norm u^2/2}.
 \label{supp:eq:sqrt-coeff-stat}
\end{equation}
Indeed $1-t^2\ge\sigma_{0,*}^2(2-\sigma_{0,*}^2)$ and
$t/(1+t)\le1/2$, which give exactly the constant displayed above.

Put $J=\lfloor k/2\rfloor$ and
$p_{J,m,\sigma^2}=\sum_{j=0}^Jc_j(m,\sigma^2)\psi_j$.
Its squared error $d_{J,m,\sigma^2}$ is at most $C_u'q_*^{J+1}$, since Parseval and
\eqref{supp:eq:sqrt-coeff-stat} imply
\[
 d_{J,m,\sigma^2}=\sum_{j>J}c_j^2
 \le q_*^{J+1}\sum_{j>J}q_*^{-j}c_j^2
 \le C_u'q_*^{J+1}.
\]
Also $\int p_{J,m,\sigma^2}^2\phi=1-d_{J,m,\sigma^2}$ because
$\norm{g_{m,\sigma^2}}_2=1$.
Let
\[
 P_{n,J}(x)=\int p_{J,\ell^{\mathsf T}u,1-\norm\ell^2}(x)^2\,dH_{0,n}(\ell),
 \quad D_{n,J}=\int d_{J,\ell^{\mathsf T}u,1-\norm\ell^2}\,dH_{0,n}(\ell).
\]
Then $P_{n,J}$ is a nonnegative polynomial of degree at most $2J$,
$\int P_{n,J}\phi=1-D_{n,J}$, and $D_{n,J}\leq C_u'q_*^{J+1}$.
Cauchy--Schwarz gives
\[
 \int\phi|p_{J,m,\sigma^2}^2-g_{m,\sigma^2}^2|
 \leq\norm{p_{J,m,\sigma^2}-g_{m,\sigma^2}}_2
\left(\norm{p_{J,m,\sigma^2}}_2+\norm{g_{m,\sigma^2}}_2\right)
 \leq2\sqrt{d_{J,m,\sigma^2}}.
\]
After integrating over the mixing law, Tonelli's theorem and
Jensen's inequality for the square root give
\[
 \|\phi P_{n,J}-f_{0,n}\|_1
 \le2\int\sqrt{d_{J,\ell^{\mathsf T}u,1-\|\ell\|^2}}\,dH_{0,n}(\ell)
 \le2\sqrt{D_{n,J}}.
\]
For large $J$, $D_{n,J}<1/2$. Normalize to
$\widetilde f_{n,k}=\phi P_{n,J}/(1-D_{n,J})$, so
\[
 \norm{\widetilde f_{n,k}-f_{0,n}}_1\leq2\sqrt{D_{n,J}}+D_{n,J}.
\]
The normalization costs exactly $D_{n,J}$ in $L^1$, because
\[
 \left\|\frac{\phi P_{n,J}}{1-D_{n,J}}-\phi P_{n,J}\right\|_1
 =\frac{D_{n,J}}{1-D_{n,J}}\int\phi P_{n,J}=D_{n,J}.
\]
Set $\epsilon_k=q_*^{(J+1)/2}$ and
$f_{n,k}^\circ=(1-\epsilon_k)\widetilde f_{n,k}+\epsilon_k\phi$.
Then $f_{n,k}^\circ/\phi\geq\epsilon_k\geq e^{-k^2}$ for all
sufficiently large $k$, and
\[
 \norm{f_{n,k}^\circ-f_{0,n}}_1\leq C_u q_*^{(J+1)/2}
 \leq C_u e^{-\rho_{\mathrm{app}}k}.
\]
In the first inequality the added Gaussian component costs at most
$2\epsilon_k$ in $L^1$. In the second, $J+1\ge k/2$ and
$-\log q_*=4\rho_{\mathrm{app}}$. Thus the construction enforces
positivity without losing the stated geometric rate.

For densities satisfying $f/g\leq R$ with $R\geq1$, we have
\[
 \KL(f\Vert g)\leq
 \frac{R\log R}{R-1}\TV(f,g)\leq(1+\log R)\TV(f,g).
\]
Indeed, on $\{f>g\}$ write $r=f/g$ and use the increasing function
$r\log r/(r-1)$. Discard the nonpositive integrand on $\{f\leq g\}$.
The second inequality follows from $\log R\leq R-1$.
Equivalently, for $R>1$ the calculation is
\[
 \begin{aligned}
 \KL(f\Vert g)
 &\le\int_{\{f>g\}}(f-g)\frac{(f/g)\log(f/g)}{(f/g)-1}\\
 &\le\frac{R\log R}{R-1}\int_{\{f>g\}}(f-g)
 =\frac{R\log R}{R-1}\TV(f,g).
 \end{aligned}
\]
The case $R=1$ has $f=g$ almost everywhere.
Equation \eqref{supp:eq:true-upper-stat} implies
$f_{0,n}/f_{n,k}^\circ\leq M_u/\epsilon_k$, whose logarithm is $O_u(k)$.
This proves \eqref{supp:eq:positive-kl-stat}.

For the second-moment constraint, use
$x\psi_j=\sqrt{j+1}\psi_{j+1}+\sqrt j\psi_{j-1}$.  It implies
\[
 \norm{x(g_{m,\sigma^2}-p_{J,m,\sigma^2})}_{L^2(\phi)}^2
 \leq4\sum_{j>J}(j+1)c_j(m,\sigma^2)^2
 \leq C_u(J+1)q_*^J\longrightarrow0
\]
uniformly over the components, by \eqref{supp:eq:sqrt-coeff-stat}.
For the last tail sum, write $j=J+1+l$ and use
$(j+1)q_*^j\le C(J+1)q_*^J$ together with the weighted coefficient
bound. If $e_J=\|x(g_{m,\sigma^2}-p_{J,m,\sigma^2})\|_2$, then
$e_J\le C_u\sqrt{J+1}\,q_*^{J/2}$. Since
$\|xg_{m,\sigma^2}\|_2^2=m^2+\sigma^2\le\|u\|^2+1$,
\[
 \begin{aligned}
 \left|\int x^2\phi(p_{J,m,\sigma^2}^2-g_{m,\sigma^2}^2)\right|
 &\le e_J\{\|xp_{J,m,\sigma^2}\|_2+\|xg_{m,\sigma^2}\|_2\}\\
 &\le e_J\{2\sqrt{1+\|u\|^2}+e_J\}.
 \end{aligned}
\]
This proves uniform convergence of the component second moments.
The same bounds give a second-moment error at most
$C_u\sqrt{k+1}e^{-\rho_{\mathrm{app}}k}$ after mixing, normalization by $1-D_{n,J}$,
and addition of the $\epsilon_k\phi$ component.
This proves \eqref{supp:eq:positive-moment-stat} uniformly over $n$.
All target second moments are at most $1+\|u\|^2$, so the bound
$1+\log k$ holds beyond a threshold depending only on $u,\sigma_{0,*}^2$,
uniformly in $n$ and the loading law.  Thus the
constructed ratio belongs to $\mathcal P_k$.

Finally put $p_b=F_{0,n}(b)$ and $p_b^\circ=F_{f_{n,k}^\circ}(b)$.
The relative-entropy chain rule for the partition $(-\infty,b],(b,\infty)$
gives
\[
 \begin{aligned}
 \KL(f_{0,n}\Vert f_{n,k}^\circ)
 &=\KL\{\operatorname{Bernoulli}(p_b)\Vert\operatorname{Bernoulli}(p_b^\circ)\} +p_b\KL(\mathcal C_bf_{0,n}\Vert\mathcal C_bf_{n,k}^\circ)\\
 &\qquad+(1-p_b)\KL(\mathcal T_bf_{0,n}\Vert\mathcal T_bf_{n,k}^\circ)\\
 &\ge p_b\KL(\mathcal C_bf_{0,n}\Vert\mathcal C_bf_{n,k}^\circ).
 \end{aligned}
\]
Divide by $p_b\ge p_{\min}$ and use \eqref{supp:eq:positive-kl-stat}.  This proves the final assertion.
\end{proof}

\subsection[Proof of Lemma S6]{Proof of Lemma~\ref{supp:lem:extension-stat}}
\label{supp:sec:proof-lemma-s6}

\begin{proof}
Set $A_k=k^{9/20}$ and $I=[-A_k,b]$, of length $L=A_k+b$.
For large $k$, $L\geq A_k/2$ and $[-M_k,b]\subset I$.
The scaled Markov polynomial inequality is
\[
 \norm{\mathfrak p'}_{\infty,I}\leq\frac{2k^2}{L}\norm{\mathfrak p}_{\infty,I}.
\]
At a point maximizing $|\mathfrak p|$, a one-sided interval of length
$L/(4k^2)$ has $|\mathfrak p|\geq\norm{\mathfrak p}_{\infty,I}/2$.
Writing $M=\|\mathfrak p\|_{\infty,I}$, this follows from
$|\mathfrak p(x)|\ge M-(2k^2M/L)|x-x_0|$ at such a maximizer $x_0$ by mean value theorem and triangle inequality.
Integration over that one-sided interval gives
$\int_I|\mathfrak{p}(x)|dx\ge ML/(8k^2)$.
Consequently
\[
 \norm{\mathfrak p}_{\infty,I}\leq\frac{8k^2}{L}\int_I|\mathfrak p(x)|\,dx
 \leq\frac{Ck^2}{A_k}e^{A_k^2/2}
                    \int_{-\infty}^b\phi(x)|\mathfrak p(x)|\,dx.
\]
Here $A_k\geq \mathfrak b$ ensures that the minimum of $\phi$ on $I$ is
$\phi(A_k)$.

Map $I$ affinely to $[-1,1]$.  At $x\in[b,M_k]$ the transformed
coordinate is $y=1+2(x-b)/L\geq1$.  The Chebyshev extremal inequality
and $\mathrm T_k(y)=\cosh(k\operatorname{arcosh}y)$ give
\[
 |\mathfrak p(x)|\leq\norm{\mathfrak p}_{\infty,I}\,
 \exp\{k\operatorname{arcosh}(1+2(M_k-b)/L)\}.
\]
The elementary inequality
$\operatorname{arcosh}(1+s)\leq\sqrt{2s}$ follows from
$\cosh t\geq1+t^2/2$.  Since $|b|\leq \mathfrak b$, the last factor is at most
$\exp\{C_\mathfrak b k\sqrt{M_k/A_k}\}$ for large $k$.
Integrating over $[-M_k,M_k]$ against $\phi$, whose total mass is at
most one, proves \eqref{supp:eq:extension-stat}.  The two exponential
contributions are $A_k^2=k^{9/10}$ and
$k\sqrt{M_k/A_k}=k^{9/10}$, while $k^2/A_k=k^{31/20}$.
This proves \eqref{supp:eq:extension-factor-stat}.
In particular, the polynomial prefactor is absorbed into the last
exponential because $\log(C_\mathfrak b k^{31/20})=o(k^{9/10})$.
All choices above are uniform for $|b|\le\mathfrak b$.

For the normalized statement put $\vartheta=F(b)/G(b)\leq p_{\min}^{-1}$ and
$\mathfrak p=R-\vartheta S$.  Then
\[
 \int_{-\infty}^b\phi(x)|\mathfrak{p}(x)| dx
 =F(b)\norm{\mathcal C_b f-\mathcal C_b g}_1\leq\norm{\mathcal C_b f-\mathcal C_b g}_1.
\]
Outside $[-M_k,M_k]$, Markov's inequality gives
\[
 \int_{|x|>M_k} |f(x)-\vartheta g(x)|\,dx
 \leq \int_{|x|>M_k} |f(x)|\, dx + \vartheta \int_{|x|>M_k} |g(x)|\,dx
 \leq(V_f+\vartheta V_g)/M_k^2.
\]
Combining the two bounds gives
\[
 \norm{f-\vartheta g}_1\leq E_k(\mathfrak b)\norm{\mathcal C_b f-\mathcal C_b g}_1
                  +(V_f+p_{\min}^{-1}V_g)/M_k^2=:D_{\mathrm{ext}}.
\]
Normalization implies $|1-\vartheta|=|\int(f-\vartheta g)|\leq D_{\mathrm{ext}}$, and therefore
$\norm{f-g}_1\leq\norm{f-\vartheta g}_1+|\vartheta-1| \norm{g}_1 \leq2D_{\mathrm{ext}}$.
Only $G(b)$ is bounded below in this argument: the potentially small
$F(b)$ appears as a multiplier of the conditional $L^1$ error, not
as a denominator. This verifies the last sentence of the lemma.
\end{proof}

\subsection[Proof of Lemma S7]{Proof of Lemma~\ref{supp:lem:empirical-stat}}
\label{supp:sec:proof-lemma-s7}

\begin{proof}
Fix $u$ temporarily and localize to the event
\eqref{supp:eq:cutoff-stat}, with $|b_n|\leq \mathfrak b=b_{\max}(u)$ and
$F_{0,n}(b_n)\geq p_q$.  Put
$\mathcal L_n=\norm u+\sqrt{8\log n}+\mathfrak b+1$.  Gaussian tail bounds and a union
bound imply
\[
 \Pp\left\{\max_{i\in\mathcal H_{0,n}}|Z_{n,i}|>\mathcal L_n\mid U=u\right\}
 \leq2n^{-3}.
\]
Each conditional null variance is at most one. In fact, for a
component mean $m_i$ with $|m_i|\le\|u\|$,
\[
 \Pp(|Z_{n,i}|>\mathcal L_n \mid U=u)
 \le2\exp\{-(\mathcal L_n-\|u\|)^2/2\}\le2n^{-4},
\]
and summing this bound over at most $n$ indices gives the displayed
probability. No independence is used here.  On the complementary
event the empirical null measure is supported on $[-\mathcal L_n,\mathcal L_n]$.

Write $f=\phi R$ with $R\in\mathcal P_k$ and put $g_R=\log R$.
Integration by parts against
$\mathbb F_{0,n}-F_{0,n}$ and Lemma~\ref{supp:lem:compact-stat} give, uniformly
over $R\in\mathcal P_k$ and $b\in[-\mathfrak b,\mathfrak b]$,
\begin{align*}
 \left|\int_{[-\mathcal L_n,b]}g_R(x)\,d(\mathbb F_{0,n}-F_{0,n})(x)\right|
 &\leq\left(2\norm{g_R}_{\infty,[-\mathcal L_n,\mathcal L_n]}
                  +\operatorname{Var}_{[-\mathcal L_n,\mathcal L_n]}(g_R)\right)\epsilon_{0,n}\\
 &\leq C_u[k^3+k^2\log\log(en)]\epsilon_{0,n}.
\end{align*}
The omitted population tail is small uniformly over the candidates.
Indeed, \eqref{supp:eq:log-envelope-stat}, Cauchy--Schwarz, the bounded
second moments of the Gaussian components, and
$F_{0,n}(-\mathcal L_n)\leq n^{-4}$ imply
\[
 \int_{-\infty}^{-\mathcal L_n}|g_R(x)|f_{0,n}(x) dx
 \leq \left(\int g_R(x)^2 f_{0,n}(x) dx \right)^{1/2} F_{0,n}(-\mathcal L_n)^{1/2}
 \leq C_uk^2n^{-2}.
\]
The logarithmic envelope implies
$\int g_R^2f_{0,n}\le C_uk^4$, uniformly in $R$ and $n$. 
Thus
\begin{equation}
 \sup_{R\in\mathcal P_k}\sup_{|b|\leq \mathfrak b}
 \left|\int_{-\infty}^bg_R(x)\,d(\mathbb F_{0,n}-F_{0,n})(x)\right|
 \leq C_u\{[k^3+k^2\log\log(en)]\epsilon_{0,n}+k^2n^{-2}\}.
 \label{supp:eq:stieltjes-stat}
\end{equation}

It remains to account for the normalization, the nonnull observations,
and possible ties at $b_n$.  Write $\varrho_n=n_{0,n}/N_n$, and use the
notation $S_n,E_n$ from the proof of Lemma~\ref{supp:lem:cdf-stat}.
The summation of the initial center can first be written as
\[
 \mathsf P_ng_R
 =\varrho_n\int_{-\infty}^{b_n}g_R(x)\,d\mathbb F_{0,n}(x)
   -\frac{E_n}{N_n}g_R(b_n)
   +\frac1{N_n}\sum_{i\in\mathcal H_{1,n}\cap\mathcal I_{0,n}} g_R(Z_{n,i}).
\]
The second term removes precisely the unretained null ties. Add and
subtract $\varrho_n\int_{-\infty}^{b_n}g_R(x)f_{0,n}(x)dx$ to compare the
remaining expression with the normalized population criterion.
The criterion difference is exactly
\begin{align*}
 \widehat L_{n,b_n}(f)-L_{0,n,b_n}(f)
 &=\varrho_n\int_{-\infty}^{b_n}g_R(x)\,d(\mathbb F_{0,n}-F_{0,n})(x)\\
 &\quad+[\varrho_n F_{0,n}(b_n)-1]\,
       \frac1{F_{0,n}(b_n)}\int_{-\infty}^{b_n}g_R(x)f_{0,n}(x)dx\\
 &\quad-\frac{E_n}{N_n}g_R(b_n)
       +\frac1{N_n}\sum_{i\in\mathcal H_{1,n}\cap\mathcal I_{0,n}}g_R(Z_{n,i}).
\end{align*}
The terms $-\log F_{\phi R}(b_n)$ cancel exactly because both criteria use
probability measures.  The deterministic ratio $\varrho_n$ is bounded.
Moreover,
\[
 |\varrho_n F_{0,n}(b_n)-1|\leq3\varrho_n\epsilon_{0,n}+\frac{n}{N_n}\chi_n,
 \qquad E_n/N_n\leq2\varrho_n\epsilon_{0,n}.
\]
The logarithmic envelope, $F_{0,n}(b_n)\geq p_q$, and the bounded Gaussian
moments give uniformly
\[
 |g_R(b_n)|+
 \left|\frac1{F_{0,n}(b_n)}\int_{-\infty}^{b_n}g_R(x)f_{0,n}(x)dx\right|
 \leq C_uk^2.
\]
The nonnull contribution is at most $Ck^2\frac{n}{N_n}\chi_n$ by its defining weight.
Indeed, for $k\ge2$ the envelope gives
$|g_R(x)|\le Ck^2\{1+\log(1+|x|)\}$, so this bound holds even for
extreme negative retained signals.
Together with \eqref{supp:eq:stieltjes-stat}, this proves the pathwise bound
\[
 \mathcal S_{n,k}\leq C_u\{
 [k^3+k^2\log\log(en)]\epsilon_{0,n}+k^2\frac{n}{N_n}\chi_n+k^2n^{-2}\}
\]
on events with conditional probability tending to one.  Apply
\eqref{supp:eq:cdf-rate-stat} and Condition~\ref{supp:ass:center}
conditionally on the fixed realization $U=u$.  This proves
\eqref{supp:eq:empirical-stat}.  Since $\rho_{\mathrm{dep}}\leq1$, the $n^{-2}$ term is
absorbed in \eqref{supp:eq:empirical-simple-stat}.
\end{proof}

\subsection[Proof of Lemma S8]{Proof of Lemma~\ref{supp:lem:projection-stat}}
\label{supp:sec:proof-lemma-s8}

\begin{proof}
Work on the bounded-cutoff and empirical-curvature events. The
logarithmic envelope gives continuity of the finite empirical
criterion; compactness then gives existence, and strong concavity in
center-normalized coordinates gives uniqueness. Write
$f^\circ=f_{n,k}^\circ$ for the feasible comparator of
Lemma~\ref{supp:lem:comparator-stat}. Optimality and the definition of
$\mathcal S_{n,k}$ give the full comparison
\[
 \begin{aligned}
 L_{0,n,b_n}(p_{n,k})
 &\ge\widehat L_{n,b_n}(p_{n,k})-\mathcal S_{n,k}\\
 &\ge\widehat L_{n,b_n}(f^\circ)-\mathcal S_{n,k}\\
 &\ge L_{0,n,b_n}(f^\circ)-2\mathcal S_{n,k}.
 \end{aligned}
\]
For any candidate density $f$ with finite criterion,
\[
 \begin{aligned}
 L_{0,n,b_n}(f_{0,n})-L_{0,n,b_n}(f)
 &= \int_{-\infty}^{b_n} \frac{f_{0,n}(x)}{F_{0,n}(b_n)} \left[\log\frac{f_{0,n}(x)}{F_{0,n}(b_n)} - \log\frac{f(x)}{F_{f}(b_n)} \right] \,dx\\
 &=\KL(\mathcal C_{b_n}f_{0,n}\Vert\mathcal C_{b_n}f).
 \end{aligned}
\]
Thus the comparator's conditional KL bound proves
\eqref{supp:eq:projection-kl}; the parameter-independent Gaussian
term has canceled from the difference.

By Pinsker's inequality in the convention
$\TV(P,Q)=\|P-Q\|_1/2$, followed by the triangle inequality,
\[
 \begin{aligned}
 \|\mathcal C_{b_n}p_{n,k}-\mathcal C_{b_n}f^\circ\|_1
 &\le\sqrt{2\KL(\mathcal C_{b_n}f_{0,n}\Vert\mathcal C_{b_n}p_{n,k})}\\
 &\qquad+\sqrt{2\KL(\mathcal C_{b_n}f_{0,n}\Vert\mathcal C_{b_n}f^\circ)}\\
 &\le C_U\sqrt{\mathcal S_{n,k}+ke^{-\rho_{\mathrm{app}}k}}.
 \end{aligned}
\]
Both $p_{n,k}$ and $f^\circ$ are polynomial densities, so the
continuation lemma applies to this pair, rather than to the
nonpolynomial target $f_{0,n}$. Indeed,
\[
 F_{f^\circ}(b_n)
 \ge F_{0,n}(b_n)-\tfrac12\|f^\circ-f_{0,n}\|_1
 \ge p_q/2
\]
eventually, and $V_{p_{n,k}}\le1+\log k$ while $V_{f^\circ}\le C_U$.
Since $M_k^2=\sqrt k$, Lemma~\ref{supp:lem:extension-stat} gives
\[
 \begin{aligned}
 \|p_{n,k}-f^\circ\|_1
 &\le2E_k(b_{\max}(U))
           \|\mathcal C_{b_n}p_{n,k}-\mathcal C_{b_n}f^\circ\|_1\\
 &\qquad+\frac{2\{1+\log k+(2/p_q)C_U\}}{\sqrt k}.
 \end{aligned}
\]
Finally use $E_k(b_{\max}(U))\le e^{C_Uk^{9/10}}$ and add
$\|f^\circ-f_{0,n}\|_1\le C_Ue^{-\rho_{\mathrm{app}}k}$.
The latter is absorbed into the first term of
\eqref{supp:eq:projection-l1}, proving that bound.
\end{proof}

\subsection[Proof of Lemma S9]{Proof of Lemma~\ref{supp:lem:mc-envelopes}}
\label{supp:sec:proof-lemma-s9}

\begin{proof}
An elementary uniform polynomial envelope suffices.
If a nonnegative polynomial $R$ of degree at most $k$ satisfies
$\int_{\mathbb R} \phi(x)R(x) dx=1$, put $H_R=\sup_{[-1,1]}R$.
Markov's polynomial inequality gives
$\sup_{[-1,1]}|R'|\le k^2H_R$.
Around a maximizer, there is a one-sided interval of length at least
$1/(2k^2)$ on which $R\ge H_R/2$ by mean value theorem. Hence
$H_R\le Ck^2$.
The Chebyshev extremal bound outside $[-1,1]$ then gives
\begin{equation}
 0\le R(x)\le Ck^2[2(1+|x|)]^k.
 \label{supp:eq:mc-polynomial-envelope}
\end{equation}
Consequently, for $G\sim N(0,1)$, $s=O(k)$, and $f_\beta(x) = \phi(x)R(x), \; R \in \mathcal{P}_k$,
\[
 \int(1+|x|)^sf_\beta(x)\,dx
 \le Ck^2 2^k\E(1+|G|)^{s+k}
 \le\exp\{Ck\log(k+1)\}.
\]
The last inequality uses the Gaussian absolute-moment formula and,
if needed, the next larger even moment.
The lower envelope $f_\beta\ge e^{-k^2}\phi$ and the elementary
Gaussian tail bound
$\{1-\Phi(x)\}\ge\underline c e^{-x^2/2}/(1+x)$ for $x\ge0$
give the first part of \eqref{supp:eq:mc-tail-envelopes}. More precisely, with
$a\le C_{\mathrm{env}}(k+1)+1$
and $a_+=\max(a,0)$,
\[
 1-F_{f_\beta}(a)
 \ge e^{-k^2}\{1-\Phi(a)\}
 \ge\frac{\underline c\,e^{-k^2-a_+^2/2}}{1+a_+}
 \ge e^{-Ck^2}.
\]
Dividing the unconditional moment bound by this survival probability
gives its second part.

Also
$1\le A_\beta=1/F_{f_\beta}(b)
\le e^{k^2}/\Phi(-\mathfrak b)$.
Equation~\eqref{supp:eq:mc-polynomial-envelope} bounds
$\|r_\beta/A_\beta\|_{L^2(\phi)} = \|\beta\|_2 / A_\beta$ by
$\exp\{Ck\log(k+1)\}$, and therefore bounds $\|\beta\|_2$ by
$e^{Ck^2}$.
The explicit Hermite coefficients and
$\psi_j'=\sqrt j\,\psi_{j-1}$ now give
\[
 \|\Psi_k(x)\|_2+\|\Psi_k'(x)\|_2
       \le e^{Ck\log(k+1)}(1+|x|)^k,\qquad
 |r_\beta'(x)|\le e^{Ck^2}(1+|x|)^{k-1}.
 \]
Since $r_\beta\ge e^{-k^2}$,
\[
 \partial_x S_\beta
 =\frac{\Psi_k'}{r_\beta}
   -\frac{\Psi_kr_\beta'}{r_\beta^2}
 \]
has the stated envelope.  For the parameter Hessian, its spatial derivative is
\[
 \begin{aligned}
 \partial_x\left\{-\frac{\Psi_k\Psi_k^{\mathsf T}}{r_\beta^2}
                    +\frac{e_0e_0^{\mathsf T}}{A_\beta^2}\right\}
 &=-\frac{\Psi_k'\Psi_k^{\mathsf T}+\Psi_k(\Psi_k')^{\mathsf T}}
          {r_\beta^2}
    +\frac{2\Psi_k\Psi_k^{\mathsf T}r_\beta'}{r_\beta^3}.
 \end{aligned}
\]
The numerators have degrees at most $3k-1$. The fixed positive
floor controls every denominator, and $3k-1\le r_k=4k+4$.
The log-density term itself has spatial derivative $r_\beta'/r_\beta$.
These estimates prove all the required spatial envelopes.

If $g$ is absolutely continuous and
$\|g'(x)\|\le L(1+|x|)^{r_k}$, integration by parts gives
\[
 \int g(x)\,dD(x)=-\int g'(x)F_D(x)\,dx.
 \]
The boundary terms vanish because $D$ has finite
$(r_k+1)$st absolute moment and total mass zero. Indeed,
$\|g(x)\|\le C_g(1+|x|)^{r_k+1}$ and, on the positive tail,
\[
 (1+x)^{r_k+1}|F_D(x)|
 \le\int_{(x,\infty)}(1+|y|)^{r_k+1}|D|(dy)\longrightarrow0.
\]
Here $F_D(x)=-D((x,\infty))$ uses the zero total mass; the negative
tail is treated with $F_D(x)=D(( -\infty,x])$. Thus integration by
parts on finite intervals passes to the whole line.
Thus $\|\int g\,dD\|\le L\mathcal N_k(D)$.
Apply this to the displayed score, log-density term and Hessian.
\end{proof}

\subsection[Proof of Lemma S10]{Proof of Lemma~\ref{supp:lem:alg-primitive-bounds}}
\label{supp:sec:proof-lemma-s10}

\begin{proof}

Lemma~\ref{supp:lem:compact-stat} gives $R(x)\le C(k+1)^2[C(1+|x|)]^k$.
Consequently, for $G\sim N(0,1)$,
\[
\begin{aligned}
\|R\|_{L^2(\phi)}^2
&\le C(k+1)^4 C^{2k}
       \mathbb E(1+|G|)^{2k}\\
&\le \exp\{C(k+1)\log(k+2)\},
\end{aligned}
\]
where the last inequality follows from standard Gaussian
moment bounds, after enlarging $C$.
Since the normalized Hermite polynomials form an
orthonormal basis of $L^2(\phi)$, the center-normalized
coefficient vector satisfies
\[
\|\beta\|_2
=\|r_\beta\|_{L^2(\phi)}
=\frac{\|R\|_{L^2(\phi)}}{F_{\phi R}(b)}.
\]
The lower bound
$F_{\phi R}(b)\ge e^{-k^2}\Phi(b)$ therefore yields
the asserted coefficient and diameter bounds. On $I$, $r_\beta$ has a common upper
bound $e^{C(k+1)^2}$. Let $B_k=e^{C(k+1)^2}$ be that upper bound. For $0<u,v\le B_k$,
scalar strong convexity of $-\log$ gives
\[
 -\log\{(1-t)u+tv\}
 \le-(1-t)\log u-t\log v
       -\frac{t(1-t)}{2B_k^2}(u-v)^2.
\]
Use this inequality on $I$ and ordinary convexity outside $I$.
Since $r_\theta-r_\beta=(\theta-\beta)^{\mathsf T}\Psi_k$, integration
gives
\[
 \begin{aligned}
 \mathcal V_{\mathsf P}((1-t)\beta+t\theta)
 &\le(1-t)\mathcal V_{\mathsf P}(\beta)+t\mathcal V_{\mathsf P}(\theta)\\
 &\quad-\frac{t(1-t)}2 B_k^{-2}e^{-C_0(k+1)^2}
                                     \|\theta-\beta\|_2^2.
 \end{aligned}
\]
This proves strong convexity, with the weaker common modulus
$e^{-C(k+1)^3}$ sufficient here. The logarithmic envelope and dominated convergence
prove continuity on the feasible domain. In these coordinates
$\log r_\beta=\log R-\log F_{\phi R}(b)$ and
$|\log F_{\phi R}(b)|\le k^2-\log\Phi(b)$. Consequently all feasible
log-ratios are dominated by
$C\{(k+1)^2+k\log(2+|x|)\}$, integrable under the stated center
moment condition. The same envelope bounds
the initial regret. This proof uses only the displayed logarithmic
moment of the center.

The polynomial envelope and a Gaussian tail bound imply
\[
 \sup_{f\in\mathcal F_k}\int_{C_2(k+1)}^\infty f(x)\,dx
 \le\eta e^{-k^2}\{1-\Phi(b)\}/4
\]
for sufficiently large $C_2$, uniformly over the cutoff range.
Take $\mathfrak a_k=C_2(k+1)$. The Gaussian exponent dominates
the degree-$k$ numerator on this tail: for sufficiently large $C_2$
and $x\ge C_2(k+1)$,
$k\log\{2(1+x)\}\le x^2/4$. Thus its integral is at most a
constant times $k^2e^{-C_2^2(k+1)^2/4}$; increasing $C_2$ gives
the displayed bound, including the fixed cutoff and $\eta$ factors. Since
$1-F_f(b)\ge e^{-k^2}\{1-\Phi(b)\}$, this proves the tail-ratio
claim. The lower density and survival bounds follow from
$f\ge e^{-k^2}\phi$.
Lemma~\ref{supp:lem:mc-envelopes} proves the weak-pairing bounds.
Differentiation of the tail normalizers and polynomial numerators
at most three times introduces finitely many denominators bounded
below by $e^{-C(k+1)^2}$ and Gaussian moments of order $O(k)$.
These moments are at most $e^{C(k+1)\log(k+2)}$.
For example, put $v_a=\int_a^\infty\phi\Psi_k$, so
$s_\beta(a)=v_a^{\mathsf T}\beta$. Its log derivatives are
\[
 \begin{aligned}
 D\log s_\beta(a)[h]&=\frac{v_a^{\mathsf T}h}{s_\beta(a)},\\
 D^2\log s_\beta(a)[h_1,h_2]
 &=-\frac{(v_a^{\mathsf T}h_1)(v_a^{\mathsf T}h_2)}{s_\beta(a)^2},\\
 D^3\log s_\beta(a)[h_1,h_2,h_3]
 &=\frac{2\prod_{j=1}^3(v_a^{\mathsf T}h_j)}{s_\beta(a)^3}.
 \end{aligned}
\]

The same derivative formulas apply to $r_\beta(x)$
and $A_\beta$, with $v_a$ replaced by
$\Psi_k(x)$ and $e_0$, respectively.

We next verify the corresponding bounds for
fitted-tail expectations, which are used in the
subsequent refit analysis. For a fixed cutoff $a$,
write
\[
Q_{\beta,a}=\mathcal T_a f_\beta,\qquad
Q_{\beta,a}(dx)
=\frac{\phi(x)r_\beta(x)}{s_\beta(a)}
  \mathbf 1_{\{x>a\}}\,dx.
\]
For any function $g$ independent of $\beta$
and any direction $h$, differentiation gives
\[
D_\beta(Q_{\beta,a}g)[h]
=
Q_{\beta,a}\left[
g\left\{
\frac{h^{\mathsf T}\Psi_k}{r_\beta}
-\frac{h^{\mathsf T}v_a}{s_\beta(a)}
\right\}
\right].
\]
In particular, for
$g_\theta(x)=\log r_\theta(x)$,
\[
D_\theta(Q_{\beta,a}g_\theta)[h]
=
Q_{\beta,a}
\left[\frac{h^{\mathsf T}\Psi_k}{r_\theta}\right],
\]
and the mixed derivative is
\[
\begin{aligned}
&D_\beta D_\theta
 (Q_{\beta,a}g_\theta)[h_1,h_2]\\
&\quad=
Q_{\beta,a}\left[
\frac{h_2^{\mathsf T}\Psi_k}{r_\theta}
\left\{
\frac{h_1^{\mathsf T}\Psi_k}{r_\beta}
-\frac{h_1^{\mathsf T}v_a}{s_\beta(a)}
\right\}
\right].
\end{aligned}
\]
Higher mixed derivatives through total order three
are obtained by repeated differentiation of these
expressions and the preceding log-normalizer formulas.

To justify differentiation under the integral,
recall that, uniformly over feasible parameters,
\[
r_\beta(x)\ge e^{-k^2},\qquad
s_\beta(a)\ge e^{-C(k+1)^2},
\]
and
\[
\|\Psi_k(x)\|_2
\le e^{Ck\log(k+1)}(1+|x|)^k.
\]
Consequently, every integrand arising from at
most three parameter derivatives is bounded by
$e^{C(k+1)^2}$ times a polynomial weight of
degree $O(k)$, possibly multiplied by a logarithmic
factor. These envelopes are integrable uniformly
under $Q_{\beta,a}$ by
Lemma~\ref{supp:lem:mc-envelopes}, since the
required moment orders are covered by
$s\le 2r_k+4$.

In particular, for the reference tail criterion
$M_a(\beta,\theta)$ defined in
Lemma~\ref{supp:lem:alg-local-reference},
the resulting mixed derivative operators satisfy
\[
\sup_{\substack{\beta,\theta\in\mathcal D_{k,b}\\
                a\in[b,\mathfrak a_k+1]}}
\|D_\beta^iD_\theta^j M_a(\beta,\theta)\|_{\mathrm{op}}
\le e^{C(k+1)^2},
\qquad 1\le i+j\le3,
\]
where the cutoff $a$ and the reference retained
measure are held fixed during differentiation.
The retained reference measure has mass at most one
and is supported on $[b,a]$, so its contribution
obeys the same bound.

The scaled full-refresh tail Hessian used in
Lemma~\ref{supp:lem:alg-global-regret} is controlled
by the same argument. Indeed,
\[
\begin{aligned}
&D_\theta^2D_{\mathrm{prox}}(\beta,\theta)
   [h_1,h_2]\\
&=
\int_b^\infty
\frac{\phi(x)r_\beta(x)}
     {r_\theta(x)^2}
(h_1^{\mathsf T}\Psi_k(x))
(h_2^{\mathsf T}\Psi_k(x))\,dx\\
&\quad-
\frac{A_\beta}{A_\theta^2}
(h_1^{\mathsf T}e_0)(h_2^{\mathsf T}e_0),
\end{aligned}
\]
whose operator norm is at most
$e^{C(k+1)^2}$ by the positivity floor,
polynomial envelope, and Gaussian moment bounds.

All constants in these estimates depend only on
the fixed primitive bounds. They are therefore
covered by $\mathfrak A_{k,0}$ through the initial
choice of $C_{\mathrm{prim}}$, without enlarging
either fixed envelope.

\end{proof}

\subsection[Proof of Lemma S11]{Proof of Lemma~\ref{supp:lem:alg-global-regret}}
\label{supp:sec:proof-lemma-s11}

\begin{proof}
The completion identity has the EM and Kullback-proximal form
\citep{DempsterLairdRubin1977,ChretienHero2000}.
Multiplying the refit criterion by $A_\beta>0$ preserves its
maximizers. Up to terms independent of the candidate $\theta$,
the resulting criterion is
\[
 L_{\mathsf P}(\theta)
 +(A_\beta-1)\mathsf Q^{\mathrm{comp}}\log r_\theta
 -A_\beta\log A_\theta.
\]
For a full refresh,
$(A_\beta-1)\mathcal T_bf_\beta=\phi r_\beta\ind_{(b,\infty)}$.
Subtracting its value at $\theta=\beta$ gives
\[
 L_{\mathsf P}(\theta)-L_{\mathsf P}(\beta)
 +\int_b^\infty\phi(x) r_\beta(x)\log\{r_\theta(x)/r_\beta(x)\} \: dx
 -A_\beta\log(A_\theta/A_\beta).
\]
Thus the scaled full-refresh criterion equals
$L_{\mathsf P}(\theta)-D_{\mathrm{prox}}(\beta,\theta)$ up to an
input-dependent constant, where
\begin{align*}
 D_{\mathrm{prox}}(\beta,\theta)
 &=A_\beta\KL\{\operatorname{Bernoulli}(A_\beta^{-1})
                  \|\operatorname{Bernoulli}(A_\theta^{-1})\}\\
 &\quad +(A_\beta-1)\KL(\mathcal T_bf_\beta\|\mathcal T_bf_\theta)\\
 &=A_\beta\log(A_\theta/A_\beta)
       -\int_b^\infty\phi(x) r_\beta(x)\log\{r_\theta(x)/r_\beta(x)\} \: dx.
\end{align*}
Here $D_{\mathrm{prox}}\ge0$, $D_{\mathrm{prox}}(\beta,\beta)=0$, and its derivative in the second
argument vanishes on the affine plane $\boldsymbol h_b^{\mathsf T}\theta=1$
at $\theta=\beta$. In ambient coordinates this derivative is
\[
 \nabla_\theta D_{\mathrm{prox}}(\beta,\beta)
 =e_0-\int_b^\infty\phi(x)\Psi_k(x) \: dx=\boldsymbol h_b,
\]
whose inner product with every feasible direction is zero. Moreover, direct differentiation gives
\[
\nabla_\theta^2D_{\mathrm{prox}}(\beta,\theta)
=
-\frac{A_\beta}{A_\theta^2}e_0e_0^{\mathsf T}
+
\int_b^\infty
\phi(x)r_\beta(x)
\frac{\Psi_k(x)\Psi_k(x)^{\mathsf T}}
{r_\theta(x)^2}\,dx.
\]
The positivity floor, polynomial envelope,
and Gaussian moment bounds yield
\[
\sup_{\beta,\theta\in\mathcal D_{k,b}}
\|\nabla_\theta^2D_{\mathrm{prox}}(\beta,\theta)\|_{\mathrm{op}}
\le\mathfrak A_{k,0},
\]
as included in the primitive bounds of
Lemma~\ref{supp:lem:alg-primitive-bounds}. Taylor's theorem along the feasible segment and the uniform
tail-Hessian bound therefore give
\[
 D_{\mathrm{prox}}(\beta,\theta)
       \le\mathfrak A_{k,0}\|\theta-\beta\|_2^2/2.
\]
For the actual completion add
\[
 J(\theta)=(A_\beta-1)(\mathsf Q^{\mathrm{comp}}-\mathcal T_bf_\beta)\log r_\theta,
\]
whose Lipschitz constant is at most $\mathfrak A_{k,0}\epsilon$. The omitted
$\log A_\theta$ term has integral zero because the difference of
the two tail probabilities has zero mass; Lemma~\ref{supp:lem:mc-envelopes}
then bounds the derivative of $J$ along feasible segments.
Choose
\[
 s=\{2(1+\mathfrak A_{k,0}^2)\}^{-1}
 \le\min\{1/2,m_k/(m_k+\mathfrak A_{k,0})\},
\]
where the inequality uses $m_k\ge\mathfrak A_{k,0}^{-1}$, and compare the minimizer
of $\mathcal V_{\mathsf P}+D_{\mathrm{prox}}-J$ with
$z=(1-s)\beta+s\beta_*$. Strong convexity of $\mathcal V_{\mathsf P}$
cancels the quadratic upper bound on $D_{\mathrm{prox}}(\beta,z)$. Indeed, with $d=\|\beta-\beta_*\|_2$,
\[
 \begin{aligned}
 \mathcal V_{\mathsf P}(z)+D_{\mathrm{prox}}(\beta,z)
 &\le(1-s)\mathcal V_{\mathsf P}(\beta)
       +s\mathcal V_{\mathsf P}(\beta_*)\\
 &\quad+\tfrac12\{\mathfrak A_{k,0}s^2-m_ks(1-s)\}d^2\\
 &\le(1-s)\mathcal V_{\mathsf P}(\beta)
       +s\mathcal V_{\mathsf P}(\beta_*).
 \end{aligned}
\]
Optimality of $\theta$ for
$\mathcal V_{\mathsf P}+D_{\mathrm{prox}}-J$ and
$D_{\mathrm{prox}}(\beta,\theta)\ge0$ therefore yield
\[
 \mathcal V_{\mathsf P}(\theta)-\mathcal V_{\mathsf P}(\beta_*)
 \le(1-s)\{\mathcal V_{\mathsf P}(\beta)-\mathcal V_{\mathsf P}(\beta_*)\}
       +J(\theta)-J(z).
\]
Write $x^2=\mathcal V_{\mathsf P}(\theta)-\mathcal V_{\mathsf P}(\beta_*)$ and
$y^2=\mathcal V_{\mathsf P}(\beta)-\mathcal V_{\mathsf P}(\beta_*)$.
Constrained strong convexity implies
$\|\theta-\beta_*\|_2\le\sqrt{2/m_k}\,x$ and the corresponding
bound for $\beta$. Thus
\[
 x^2\le(1-s)y^2+\sqrt2\,\mathfrak A_{k,0}^{3/2}\epsilon(x+y).
\]
Set $\rho_{\mathrm{sqrt}}=\sqrt{1-s}\ge1/\sqrt2$. If
$x>\rho_{\mathrm{sqrt}}y$, factoring and
$x+\rho_{\mathrm{sqrt}}y\ge\rho_{\mathrm{sqrt}}(x+y)$ give
\[
 \begin{aligned}
 (x-\rho_{\mathrm{sqrt}}y)(x+\rho_{\mathrm{sqrt}}y)
 &\le\sqrt2\,\mathfrak A_{k,0}^{3/2}\epsilon(x+y),\\
 x-\rho_{\mathrm{sqrt}}y
 &\le\frac{\sqrt2\,\mathfrak A_{k,0}^{3/2}}
                {\rho_{\mathrm{sqrt}}}\epsilon.
 \end{aligned}
\]
If $x\le\rho_{\mathrm{sqrt}}y$ the same upper bound is immediate.
Since $\mathfrak A_{k,0}\ge2$ and the already fixed exponent satisfies
$p_{\mathrm{alg}}\ge5$,
\[
 \begin{aligned}
 \frac{s}{2}=\frac1{4(1+\mathfrak A_{k,0}^2)}
 &\ge\mathfrak A_{k,0}^{-5}\ge\mathfrak A_k^{-1},\\
 \frac{\sqrt2\,\mathfrak A_{k,0}^{3/2}}{\rho_{\mathrm{sqrt}}}
 &\le2\mathfrak A_{k,0}^{3/2}\le\mathfrak A_{k,0}^{3}
 \le\mathfrak A_k.
 \end{aligned}
\]
Thus $\rho_{\mathrm{sqrt}}\le1-s/2\le1-\mathfrak A_k^{-1}$,
which proves the stated square-root regret contraction with the
same fixed envelope. No differentiation of the center integral is needed.
\end{proof}

\subsection[Proof of Lemma S12]{Proof of Lemma~\ref{supp:lem:alg-local-reference}}
\label{supp:sec:proof-lemma-s12}

\begin{proof}
Restrict operators to the tangent space
$\mathcal E=\{h:\boldsymbol h_b^{\mathsf T}h=0\}$.
Define
\[
 K_a=s_{\beta_*}(a)\operatorname{Cov}_{\mathcal T_ap_*}\!\left(
             \frac{\Psi_k}{r_{\beta_*}}\right)
          +\frac{e_0e_0^{\mathsf T}}{A_{\beta_*}(A_{\beta_*}-1)}.
\]
The covariance decomposition over $(b,a]$ and $(a,\infty)$ gives
$0\preceq K_a\preceq K_b$. To check this order, put
$v(x)=\Psi_k(x)/r_{\beta_*}(x)$ and
$\lambda_a^*=s_{\beta_*}(a)/s_{\beta_*}(b)$. The law of total covariance,
applied to $\mathcal T_bp_*$ and its two restrictions, implies
\[
 s_{\beta_*}(b)\Cov_{\mathcal T_bp_*}(v)
 \succeq s_{\beta_*}(a)\Cov_{\mathcal T_ap_*}(v).
\]
The remaining rank-one term in $K_a$ is independent of $a$.
Also, with $\mathsf M_*=m_kI+K_b$,
\[
 \mathsf M_*^{-1/2}K_a\mathsf M_*^{-1/2}
 \preceq I-m_k\mathsf M_*^{-1}
 \preceq\left(1-\frac{m_k}{m_k+\|K_b\|}\right)I.
\]
Set
\begin{equation}
 \gamma_{0,k}=\frac{m_k}{m_k+\|K_b\|},\qquad
 \|\mathsf M_*^{-1/2}K_a\mathsf M_*^{-1/2}\|
                   \le1-\gamma_{0,k}.
 \label{supp:eq:local-metric-gap}
\end{equation}
The operator bounds, norm conversions, inverse metric gap, and twice
the perturbation-Lipschitz and localization constants below are bounded
by fixed powers of $\mathfrak A_{k,0}$. By the initial choice of
$p_{\mathrm{alg}}$, these bounds are covered by the same
$\mathfrak A_k$. The radii in \eqref{supp:eq:local-radii} therefore use
this fixed envelope; no later enlargement is made.

For a correct reference state the scaled refit criterion is
$L_{\mathsf P}(\theta)+M_a(\beta,\theta)$, where
\[
 M_a(\beta,\theta)
  =(A_\beta-1)\{\mathsf B_a^*+\lambda_a^*\mathcal T_af_\beta\}\log r_\theta
       -A_\beta\log A_\theta.
\]
Compute the derivatives in ambient coordinates and then restrict
the resulting operators to $\mathcal E$.
Temporarily put $A=A_{\beta_*}$,
$v_a=\int_a^\infty\phi\Psi_k$, and
$H_b=\int_b^\infty\phi\Psi_k\Psi_k^{\mathsf T}/r_{\beta_*}$.
Since $v_b=e_0-\boldsymbol h_b$ and $s_{\beta_*}(b)=A-1$,
\[
 \begin{aligned}
 \nabla_\theta M_a(\beta_*,\beta_*)&=v_b-e_0=-\boldsymbol h_b,\\
 D_\theta(-\nabla_\theta M_a)(\beta_*,\beta_*)&=H_b-e_0e_0^{\mathsf T}/A.
 \end{aligned}
\]
The first expression vanishes after restriction to $\mathcal E$.
On that space, $v_bv_b^{\mathsf T}$ and $e_0e_0^{\mathsf T}$ have
the same quadratic form, so the second expression equals
\[
 (A-1)\Cov_{\mathcal T_bp_*}(v)
       +\frac{e_0e_0^{\mathsf T}}{A(A-1)}=K_b.
\]
For the mixed derivative, differentiation of a normalized tail
expectation gives
\[
 D_\beta\{\mathcal T_af_\beta(v)\}\big|_{\beta_*}[h]
     =\Cov_{\mathcal T_ap_*}(v,v^{\mathsf T}h).
\]
Multiplication by $(A-1)\lambda_a^*=s_{\beta_*}(a)$ produces the
covariance term of $K_a$. Differentiating the factor $A_\beta-1$
and $-A_\beta\log A_\theta$ produces, on $\mathcal E$,
$e_0e_0^{\mathsf T}/\{A(A-1)\}$. Thus, with the minus sign,
the mixed derivative is $-K_a$. We have established
\[
 \begin{aligned}
 -\nabla_\theta M_a(\beta_*,\beta_*)&=0,\\
 D_\theta(-\nabla_\theta M_a)(\beta_*,\beta_*)&=K_b,\\
 D_\beta(-\nabla_\theta M_a)(\beta_*,\beta_*)&=-K_a.
 \end{aligned}
\]
The function $M_a(\beta,\cdot)$ need not be concave because
$-A_\beta\log A_\theta$ is convex in $A_\theta$.
Global localization follows from the quadratic growth of
$\mathcal V_{\mathsf P}+D_{\mathrm{prox}}(\beta_*,\cdot)$, using strong convexity of
$\mathcal V_{\mathsf P}$, $D_{\mathrm{prox}}\ge0$ and $D_{\mathrm{prox}}(\beta_*,\beta_*)=0$.

For the actual state,
$|\lambda-\lambda_a^*|=|F_D(a)|\le\mathfrak d$ and
\[
 \{\mathsf B+\lambda \mathcal T_af_\beta\}
       -\{\mathsf B_a^*+\lambda_a^*\mathcal T_af_\beta\}
 =D+(\lambda-\lambda_a^*)(\mathcal T_af_\beta-\mathcal T_ap_*).
\]
The weak-pairing bounds control the criterion-gradient perturbation
by $\mathfrak A_k(\mathfrak d+\nu)$. For example, the extra
term is the pairing of $D+E$ and
$(\lambda-\lambda_a^*)(\mathcal T_af_\beta-\mathcal T_ap_*)$
with a candidate tail score. The former is bounded by
$\mathfrak A_k(\mathfrak d+\nu)$; the latter is bounded by the same
quantity using $|\lambda-\lambda_a^*|\le\mathfrak d$ and the
uniform parameter diameter and tail-derivative bounds. The change from input $\beta_*$ to
$\beta$ has gradient at most $\mathfrak A_k\|\beta-\beta_*\|_2$.
Comparing any global maximizer with $\beta_*$ by quadratic growth
is made explicit as follows. If $J$ denotes the difference from
the reference maximization criterion,
optimality and $D_{\mathrm{prox}}(\beta_*,\theta)\ge0$ imply
\[
 \frac{m_k}{2}\|\theta-\beta_*\|_2^2
 \le J(\theta)-J(\beta_*)
 \le C_k\{\|\beta-\beta_*\|_2+\mathfrak d+\nu\}
                         \|\theta-\beta_*\|_2.
\]
Canceling the last norm when nonzero and converting metrics gives
\begin{equation}
 x\le \mathfrak A_k\{y+\mathfrak d+\nu\},\qquad
 x=\|\theta-\beta_*\|_{\mathsf M_*},\quad
 y=\|\beta-\beta_*\|_{\mathsf M_*}.
 \label{supp:eq:global-localization-radius}
\end{equation}
Conditions \eqref{supp:eq:local-input-domain} imply
$x\le R_k/2$ and $y\le R_k/(4\mathfrak A_k)$.

The center contribution can be treated without a center score.
Extend $\mathcal V_{\mathsf P}$ by $+\infty$ off $\mathcal D_{k,b}$ in
its affine hull, and denote the resulting strongly convex function
by $\mathcal J_{\mathsf P}$. For a global minimizer of
$\mathcal J_{\mathsf P}$ plus a differentiable tail term, its negative
tail gradient belongs to $\partial\mathcal J_{\mathsf P}$.
Indeed, the one-sided optimality inequality along each feasible
segment, followed by convexity of $\mathcal J_{\mathsf P}$, gives
the subgradient inequality. Also
$0\in\partial\mathcal J_{\mathsf P}(\beta_*)$.
Strong monotonicity of this subdifferential supplies the center
quadratic term $m_k\|\theta-\beta_*\|_2^2$. Specifically, if
$\Delta=\theta-\beta_*$ and $v_\theta\in\partial\mathcal J_{\mathsf P}(\theta)$
is the negative perturbed tail gradient, then
\[
 \langle v_\theta,\Delta\rangle
 =\langle v_\theta-0,\theta-\beta_*\rangle
 \ge m_k\|\Delta\|_2^2.
\]
This is the strong-monotonicity theorem for the subdifferential of
a strongly convex function, applied within the affine hull.
Taylor expansion is applied only to the tail term, whose derivatives
are bounded independently of upper center moments. Writing
$h=\beta-\beta_*$ and $F_a=-\nabla_\theta M_a$, the preceding
calculations give
\[
 F_a(\beta,\theta)=K_b\Delta-K_ah+\operatorname{Rem}_a(\beta,\theta),
\]
with remainder term bounded as $\|\operatorname{Rem}_a(\beta,\theta)\|_2 \le C_k \left(\|\Delta\|_2+\|h\|_2 \right)^2$.
The actual tail gradient adds a perturbation bounded by
$\mathfrak A_k(\mathfrak d+\nu)/2$, with the norm-conversion constants
already included in $\mathfrak A_k$. Strong monotonicity therefore gives
\[
 \Delta^{\mathsf T}(m_kI+K_b)\Delta
 \le\Delta^{\mathsf T}K_ah
       +|\langle\operatorname{Rem}_a(\beta,\theta),\Delta\rangle|
       +\tfrac12\mathfrak A_k(\mathfrak d+\nu)x.
\]
The metric gap bounds the first term by $(1-\gamma_{0,k})xy$;
the localized Taylor remainder gives
\[
 x^2\le(1-\gamma_{0,k})xy+\frac{\gamma_{0,k}}8x(x+y)
                         +\frac{\mathfrak A_k}{2}(\mathfrak d+\nu)x.
\]
Here the remainder coefficient is at most
$2\mathfrak A_kR_k\le\gamma_{0,k}/16$ with our radius choice, so the
weaker displayed bound applies. If $x=0$ the conclusion is immediate. Otherwise divide by $x$
and move its remainder term to the left to obtain
\[
 (1-\gamma_{0,k}/8)x
 \le(1-7\gamma_{0,k}/8)y
       +\tfrac12\mathfrak A_k(\mathfrak d+\nu).
\]
For $0<\gamma_{0,k}\le1$,
\[
 \frac{1-7\gamma_{0,k}/8}{1-\gamma_{0,k}/8}
       \le1-\gamma_{0,k}/2,
 \qquad \frac{1}{2(1-\gamma_{0,k}/8)}\le1.
\]
These two inequalities give the stated contraction coefficient and
perturbation constant.
Finally the explicit error threshold gives
\[
 (1-\gamma_k)r_k^{\mathrm{in}}
       +\mathfrak A_k\varepsilon_k^{\mathrm{loc}}
 \le(1-\gamma_k/2)r_k^{\mathrm{in}}<r_k^{\mathrm{in}}.
\]
This proves the contraction and its invariance assertion.
\end{proof}

\subsection[Proof of Lemma S13]{Proof of Lemma~\ref{supp:lem:mc-rank-freezing}}
\label{supp:sec:proof-lemma-s13}

\begin{proof}
Let $\widehat{\mathsf Q}_{n,m}=m^{-1}\sum_j\delta_{Y_{n,j}}$.
All expectations in the following batch calculation are conditional
on $\mathcal F_n^-$. Thus $\mathsf Q,m,\lambda$ and the reference
parameters are fixed throughout this calculation.
Symmetrization followed by the $L^2$ maximal inequality for
Rademacher partial sums gives
$\mathbb E\sup_x|F_{\widehat{\mathsf Q}_{n,m}}(x)-F_{\mathsf Q}(x)|\le4/\sqrt m$.
Indeed, after ordering the observations, the symmetrized supremum
is the maximum of the first $j$ Rademacher partial sums,
$0\le j\le m$. If $\varsigma_1,\ldots,\varsigma_m$ are
independent Rademacher variables, symmetrization and Doob's $L^2$
maximal inequality give explicitly
\[
 \E\|F_{\widehat{\mathsf Q}_{n,m}}-F_{\mathsf Q}\|_\infty
 \le\frac2m\E\max_{0\le j\le m}
                    \left|\sum_{i=1}^j\varsigma_i\right|
 \le\frac2m\,2\sqrt m=\frac4{\sqrt m}.
\]
Pointwise variance and Fubini give
\[
 \mathbb E\!\int(1+|x|)^{r_k}
       |F_{\widehat{\mathsf Q}_{n,m}}(x)-F_{\mathsf Q}(x)|\,dx
 \le\frac1{\sqrt m}
      \int(1+|x|)^{r_k}\sqrt{F_{\mathsf Q}(x)\{1-F_{\mathsf Q}(x)\}}\,dx.
 \]
Markov's inequality on the two tails bounds the last integral by
$2\{\int(1+|y|)^{2r_k+4}\mathsf Q(dy)\}^{1/2}$.
Indeed, write $M=\int(1+|y|)^{2r_k+4}\mathsf Q(dy)$.
For $x\ge0$ bound $1-F_{\mathsf Q}(x)$ by $M(1+x)^{-2r_k-4}$;
for $x<0$ use the analogous bound for $F_{\mathsf Q}(x)$.
The resulting integrand is at most $\sqrt M(1+|x|)^{-2}$, whose
integral is $2\sqrt M$.
Lemma~\ref{supp:lem:mc-envelopes} and
$\lambda/\sqrt m=\sqrt m/\tau_n\le1/\sqrt{\tau_n}$
prove \eqref{supp:eq:mc-noise-expectation}.

The batch has no ties almost surely.  The retained-block CDFs have
the exact representations
\[
 F_{\mathsf E_n}(x)=\min\{F_{\mathsf L_n}(x),\omega\},\qquad
 F_{\mathsf E_n^\circ}(x)=\min\{\lambda F_{\mathsf Q}(x),\omega\}.
 \]
Because the minimum with a fixed constant is $1$-Lipschitz,
their pointwise difference is bounded by
$|F_{\mathsf L_n}(x)-\lambda F_{\mathsf Q}(x)|$.
This proves \eqref{supp:eq:mc-block} in both components of
$\mathcal N_k$. In particular, the supremum part and the
weighted integral part are each bounded by the corresponding part
of $\mathcal N_k(\mathsf L_n-\lambda\mathsf Q)$; no separate empirical
process for the selected block is required.
Moreover,
$F_{\mathsf L_n}(\widehat a_n)=(h+1)/\tau_n=\omega+1/\tau_n$,
and therefore
\begin{equation}
 |\lambda F_{\mathsf Q}(\widehat a_n)-\omega|\le\nu_n.
 \label{supp:eq:mc-next-rank-mass}
\end{equation}
The lower sieve envelope gives
$(\mathcal T_a f_\beta)(x)\ge e^{-k^2}\phi(x)\ge\underline q_k$ on $I_k^+\cap(a,\infty)$.

If $\widehat a_n>C_{\mathrm{env}}(k+1)+1$,
while $a^\dagger\le C_{\mathrm{env}}(k+1)$,
the left side of \eqref{supp:eq:mc-next-rank-mass}
is at least $\lambda\underline q_k
\ge\eta\underline q_k$.
Thus the good event ensures
$\widehat a_n\in I_k^+$.
Integration of the density lower bound between $a^\dagger$ and
$\widehat a_n$ then proves \eqref{supp:eq:mc-quantile}. Specifically,
\[
 \eta\underline q_k|\widehat a_n-a^\dagger|
 \le\lambda|F_{\mathsf Q}(\widehat a_n)-F_{\mathsf Q}(a^\dagger)|
 \le\nu_n.
\]
This also covers $h=0$, where $a^\dagger=a$ and
$F_{\mathsf Q}(a)=0$.

We record two deterministic bounds used for the reference error.
For any mass-zero signed measure $D$ with the required moment,
\begin{equation}
 \mathcal N_k(D)
 \le 2\int(1+|x|)^{r_k+1}|D|(dx).
 \label{supp:eq:mc-tv-moment-upper}
\end{equation}
On $x>0$ write $F_D(x)=-D((x,\infty))$,
integrate this tail bound, and use the analogous argument on $x<0$.
For example, Tonelli's theorem gives
\[
 \int_0^\infty(1+x)^{r_k}|F_D(x)|\,dx
 \le\int_{y>0}\frac{(1+y)^{r_k+1}-1}{r_k+1}|D|(dy).
\]
The negative half-line has the same form, and the supremum term
is bounded by $|D|(\R)$; this proves
\eqref{supp:eq:mc-tv-moment-upper}.
Furthermore, uniformly over $a,a^{\prime}\in I_k^+$,
\begin{equation}
 \mathcal N_k(\mathcal T_ap^*-\mathcal T_{a^{\prime}}p^*)\le \mathfrak A_k|a-a^{\prime}|.
 \label{supp:eq:mc-tail-cutoff-lip}
\end{equation}
The normalization in this estimate is important: moving the cutoff
removes a small block and redistributes its mass over the surviving
tail, rather than simply deleting that block.
To see this when $a<a^{\prime}$, decompose the signed difference as
$(\mathcal T_ap^*)\mathbf1_{(a,a^{\prime}]}-
(\mathcal T_ap^*)((a,a^{\prime}])\mathcal T_{a^{\prime}}p^*$.
Use \eqref{supp:eq:mc-tv-moment-upper}, the compact density envelope,
and \eqref{supp:eq:mc-tail-envelopes}.
Consequently, replacing $\mathsf E_n^\circ,a^\dagger$ by $\mathsf E_n,\widehat a_n$
perturbs the reference error by at most $\mathfrak A_k\nu_n$.

For the population block put $\mathsf Q_*=\mathcal T_ap^*$.
The exact reference-error increment is
\begin{align*}
 I
 &:=\mathsf E_n^\circ+\lambda_+\mathcal T_{a^\dagger}p^*-\lambda \mathcal T_ap^*\\
 &=\lambda\left[
     (\mathsf Q-\mathsf Q_*)\mathbf1_{(a,a^\dagger]}
       -\{F_{\mathsf Q}(a^\dagger)-F_{\mathsf Q_*}(a^\dagger)\}\mathcal T_{a^\dagger}p^*
          \right].
\end{align*}
On the compact interval $(a,a^\dagger]$, the polynomial lower envelope,
the tail normalizers in \eqref{supp:eq:mc-tail-envelopes}, and
$\|f_\beta-f_{\beta^*}\|_1\le2\|\beta-\beta^*\|_2$
give
\begin{equation}
 |(\mathcal T_a f_\beta)(x)-(\mathcal T_ap^*)(x)|
 \le \mathfrak A_k\|\beta-\beta^*\|_2\,(\mathcal T_a f_\beta)(x).
 \label{supp:eq:mc-relative-local}
\end{equation}
The normalized ratios are
$(\mathcal T_a f_\beta)(x)=\phi(x)r_\beta(x)/s_\beta(a)$.
Here $|r_\beta-r_{\beta^*}|\le
\|\beta-\beta^*\|_2\|\Psi_k(x)\|_2$ and
$|s_\beta(a)-s_{\beta^*}(a)|\le\|\beta-\beta^*\|_2$.
The lower bounds for $r_\beta$ and $s_{\beta^*}(a)$
prove \eqref{supp:eq:mc-relative-local}. Using the displayed normalized-tail formulas, the calculation is
\[
 \begin{aligned}
 \frac{|(\mathcal T_af_\beta)(x)-(\mathcal T_ap^*)(x)|}
      {(\mathcal T_af_\beta)(x)}
 &\le\frac{s_\beta(a)}{s_{\beta^*}(a)}
             \frac{|r_\beta(x)-r_{\beta^*}(x)|}{r_\beta(x)}\\
 &\quad+\frac{|s_\beta(a)-s_{\beta^*}(a)|}{s_{\beta^*}(a)}\\
 &\le\mathfrak A_k\|\beta-\beta^*\|_2.
 \end{aligned}
\]
All bounds here are on the specified compact block interval.
Since $\mathsf Q((a,a^\dagger])=\omega/\lambda$,
\[
 \lambda\int_{(a,a^\dagger]}|\mathsf Q-\mathsf Q_*|
   \le \mathfrak A_k\omega\|\beta-\beta^*\|_2.
 \]
Apply \eqref{supp:eq:mc-tv-moment-upper} to $I$. Its compact block
and compensating tail have the moment bounds of
Lemma~\ref{supp:lem:mc-envelopes}.
Thus $\mathcal N_k(I)\le
\mathfrak A_k\omega\|\beta-\beta^*\|_2$.
The compensating tail has coefficient equal to the negative mass
of the block difference, whose absolute value is no larger than
that block's total variation. This is why the factor $\omega$
persists after adding the compensation and taking $\mathcal N_k$.
The triangle inequality proves \eqref{supp:eq:mc-debt}.

At the terminal block the exact identity is
$\widehat D_{n,+}=D+\lambda(\mathsf Q-\mathsf Q_*)+(\mathsf L_n-\lambda \mathsf Q)$.
A direct normalization bound gives
\[
 \int(1+|x|)^{r_k+1}|\mathsf Q-\mathsf Q_*|(dx)
 \le \mathfrak A_k\|\beta-\beta^*\|_2:
 \]
use Cauchy--Schwarz for
$\int(1+|x|)^{r_k+1}\phi|r_\beta-r_{\beta^*}|$
and the same tail-normalizer bounds.
Equation~\eqref{supp:eq:mc-tv-moment-upper} proves
\eqref{supp:eq:mc-terminal}.

Every selected-block and reference-error inequality above is pathwise
on the same incoming-history event. Hence their simultaneous use
with the refit does not assume independence between the refit and
the block retained from its batch.
\end{proof}

\subsection[Proof of Lemma S14]{Proof of Lemma~\ref{supp:lem:alg-quantitative-freezing}}
\label{supp:sec:proof-lemma-s14}

\begin{proof}
We first prove the empirical assertion. Use the main-text sampling boundaries
$a_{n,t}$ and let
$\xi_{n,t}=\tau_n^{-1}\sum_{s=1}^tM_{n,s}$ be the fraction
retained after iteration $t$ in the empirical algorithm. Put
$\lambda_{n,t}=1-\xi_{n,t}$ and $\omega_{n,t}=\xi_{n,t}-\xi_{n,t-1}$.
For a nonterminal state define
\[
 D_{n,t}=\mathsf B_{n,t}+\lambda_{n,t}\mathcal T_{a_{n,t}}p_*-\mathcal T_bp_*,\qquad
 \mathfrak d_{n,t}=\mathcal N_k(D_{n,t}),\qquad
 e_{n,t}=\|\widehat\beta_n^{(t)}-\beta_*\|_{\mathsf M_*}.
\]
The signed measure $D_{n,t}$ has mass zero, and
\begin{equation}
 F_{D_{n,t}}(a_{n,t})
 =\frac{1-F_*(a_{n,t})}{1-F_*(b)}-\lambda_{n,t}.
 \label{supp:eq:alg-cutoff-debt}
\end{equation}
Indeed, $\mathsf B_{n,t}(( -\infty,a_{n,t}])=1-\lambda_{n,t}$,
$\mathcal T_{a_{n,t}}p_*$ has no mass there, and
\[
 (\mathcal T_bp_*)(( -\infty,a_{n,t}])
 =1-\frac{1-F_*(a_{n,t})}{1-F_*(b)}.
\]
Subtracting these masses gives the identity. It converts the weak
reference error into direct control of the sampling boundary.
For a population block with
$\mathsf Q=\mathcal T_af_\beta$, $\mathsf Q_*=\mathcal T_ap_*$ and
$\lambda\mathsf Q((a,a^\dagger])=\omega$, direct subtraction gives
\begin{equation}
 D_{\mathrm{new}}-D_{\mathrm{old}}
 =\lambda(\mathsf Q-\mathsf Q_*)\mathbf1_{(a,a^\dagger]}
       +\{\lambda\mathsf Q_*((a,a^\dagger])-\omega\}\mathcal T_{a^\dagger}p_*.
 \label{supp:eq:alg-debt-identity}
\end{equation}
The relative-density and moment bounds in
Lemma~\ref{supp:lem:mc-rank-freezing} give, also for an empirical block,
\begin{equation}
 \mathfrak d_{n,t}\le\mathfrak d_{n,t-1}
      +\mathfrak A_k\omega_{n,t}e_{n,t-1}+\mathfrak A_k\nu_{n,t},
 \qquad t<T.
 \label{supp:eq:alg-debt-noisy}
\end{equation}
The same $\nu_{n,t}$ controls the fresh-batch error in the refit.
The coefficient in the block-error term is $e_{n,t-1}$ because both
the block and the refit batch are generated from the incoming fit.
The refit $\widehat\beta_n^{(t)}$ is not treated as independent of
that same batch.

Use the initial interval-envelope choice, so that $\mathfrak a_k$
belongs to $I_k$. Keep the fixed $\mathfrak A_k$ and the metric, gap,
and radii in Lemma~\ref{supp:lem:alg-local-reference} unchanged.
Define explicit margins
\begin{equation}
 \begin{aligned}
 d_k^{\mathrm{stop}}&=\min\{\eta/4,
                          \varepsilon_k^{\mathrm{loc}}/2\},\\
 \nu_k^{\mathrm{stop}}&=\min\{\varepsilon_k^{\mathrm{loc}}/2,
                  \eta\underline q_k/2,\mathfrak A_k^{-3}/4\}.
 \end{aligned}
 \label{supp:eq:bootstrap-margins}
\end{equation}
After a deterministic burn-in of $J_k$ stages chosen below, the
localization region for an incoming state is
\begin{equation}
 \begin{split}
 &b\le a_{n,t-1}\le\mathfrak a_k,\qquad
   \lambda_{n,t-1}\ge\eta,\qquad
   \mathfrak d_{n,t-1}\le d_k^{\mathrm{stop}},\\
 &e_{n,t-1}\le r_k^{\mathrm{in}}
                       \quad\text{when }t-1\ge J_k.
 \end{split}
 \label{supp:eq:bootstrap-region}
\end{equation}
Before $J_k$, only the cutoff and reference-error restrictions are
imposed. At a reached stage require
$\nu_{n,t}\le\nu_k^{\mathrm{stop}}$, and for $t<T$ require
$\omega_{n,t}\le \mathfrak A_k^{-3}/4$. In the local phase these conditions give
$\mathfrak d_{n,t-1}+\nu_{n,t}\le\varepsilon_k^{\mathrm{loc}}$, which
verifies the complete input domain \eqref{supp:eq:local-input-domain}.

The cutoff is checked before applying the next-stage bounds.
If $\mathfrak d_{n,t-1}\le\eta/4$ and $\lambda_{n,t-1}\ge\eta$,
\eqref{supp:eq:alg-cutoff-debt} makes the reference tail ratio at
$a_{n,t-1}$ at least $3\eta/4$. Its value at $\mathfrak a_k$ is at
most $\eta/4$. The reference conditional density has upper bound
$\mathfrak A_k$, and the sampling density has lower bound $\mathfrak A_k^{-1}$ on
this interval. The reference block between $a_{n,t-1}$ and
$\mathfrak a_k$ has mass at least $\eta/2$. Its density upper bound
therefore implies
\[
 \mathfrak a_k-a_{n,t-1}\ge\frac{\eta}{2\mathfrak A_k}.
\]
The sampling density lower bound and $\lambda_{n,t-1}\ge\eta$ then
give sampling mass at least $\eta^2/(2\mathfrak A_k^2)$ on that
interval. Since the fixed envelope satisfies $\mathfrak A_k\ge2/\eta^2$,
\[
 \lambda_{n,t-1}(\mathcal T_{a_{n,t-1}}f_{\widehat\beta_n^{(t-1)}})
                   ((a_{n,t-1},\mathfrak a_k])\ge \mathfrak A_k^{-3}.
\]
The inequalities for $\omega_{n,t}$ and $\nu_{n,t}$ imply that the
population block quantile and the next empirical live order statistic
are below $\mathfrak a_k$. The quantile requirement and the
small-noise event of Lemma~\ref{supp:lem:mc-rank-freezing} therefore
hold at every reached nonterminal stage.
For the empirical boundary, set
$\mathsf Q=\mathcal T_{a_{n,t-1}}f_{\widehat\beta_n^{(t-1)}}$,
the incoming sampling law. The mass identity
\eqref{supp:eq:mc-next-rank-mass} gives
\[
 \lambda_{n,t-1}F_{\mathsf Q}(a_{n,t})
 \le\omega_{n,t}+\nu_{n,t}\le\mathfrak A_k^{-3}/2,
\]
strictly below the available mass at $\mathfrak a_k$. This checks
the outgoing cutoff without assuming in advance that it lies in
the localization interval.

\textbf{Burn-in.}
As long as the cutoff restrictions hold, density lower bounds and
the order-statistic mass identity give, for a fixed integer $p$,
\[
 a_{n,t}-b\le \mathfrak A_k^p\left\{t/T+\sum_{s\le t}\nu_{n,s}\right\}.
\]
To obtain this bound, apply the lower sampling-density envelope at
each reached stage:
\[
 a_{n,s}-a_{n,s-1}
 \le\mathfrak A_k^p(\omega_{n,s}+\nu_{n,s}).
\]
Summing, and using $\sum_{s\le t}\omega_{n,s}\le Ct/T$, proves the
claim, including the initial identity $a_{n,0}=b$.
In this proof fixed powers $\mathfrak A_k^p$ are kept explicit.
A common exponent $p$ may be increased finitely many times while
collecting these bounds, but is fixed before $\Gamma_k$ is chosen;
$\mathfrak A_k$ itself remains unchanged.
The incoming completed tail
$\mathsf Q_{n,t}^{\mathrm{comp}}=\mathsf B_{n,t-1}+\lambda_{n,t-1}
 \mathcal T_{a_{n,t-1}}f_{\widehat\beta_n^{(t-1)}}$ consequently obeys
\[
 \mathcal N_k(\mathsf Q_{n,t}^{\mathrm{comp}}-\mathcal T_bf_{\widehat\beta_n^{(t-1)}})
 \le \mathfrak A_k^p\left\{t/T+\sum_{s<t}\nu_{n,s}\right\}.
\]
On $(b,a_{n,t-1}]$ use its bounded, retained mass and model mass.
Above $a_{n,t-1}$ the remaining difference is a scalar multiple of
a truncated model law. The tail moments convert these bounds into
$\mathcal N_k$ bounds.
For clarity, with $f=f_{\widehat\beta_n^{(t-1)}}$, $a=a_{n,t-1}$,
and $\lambda=\lambda_{n,t-1}$, the exact decomposition is
\[
 \begin{aligned}
 \mathsf Q_{n,t}^{\mathrm{comp}}-\mathcal T_bf
 &=\mathsf B_{n,t-1}-(\mathcal T_bf)\ind_{(b,a]}\\
 &\quad+\{\lambda-(\mathcal T_bf)((a,\infty))\}\mathcal T_af.
 \end{aligned}
\]
The compact part has total variation at most
$\xi_{n,t-1}+C_k(a-b)$; the last scalar is its negative total mass.
Equation~\eqref{supp:eq:mc-tv-moment-upper} and the tail moments
then give the claimed weak-norm bound.
Lemma~\ref{supp:lem:alg-global-regret} gives geometric decrease of
square-root regret plus this perturbation and the current batch
error. To spell out the summation, let
\[
 y_t=\{\mathcal V_{\mathsf P}(\widehat\beta_n^{(t)})
                   -\mathcal V_{\mathsf P}(\beta_*)\}^{1/2},
 \qquad \rho_{\mathrm{burn},k}=1-\mathfrak A_k^{-1}.
\]
At every reached burn-in stage,
\[
 y_t\le\rho_{\mathrm{burn},k} y_{t-1}
    +\mathfrak A_k^p\left\{t/T+\sum_{s\le t}\nu_{n,s}\right\}.
\]
Iteration and $\sum_{j\ge0}\rho_{\mathrm{burn},k}^j=\mathfrak A_k$ give, after increasing
$p$,
\[
 y_t\le\rho_{\mathrm{burn},k}^ty_0+
       \mathfrak A_k^p\left\{t/T+\sum_{s\le t}\nu_{n,s}\right\},
 \qquad e_{n,t}\le\mathfrak A_k^p y_t.
\]
The last inequality is the constrained quadratic-growth bound
followed by conversion to $\mathsf M_*$. Choose $1\le J_k\le \mathfrak A_k^p$ so the unperturbed regret puts the
iterate within $r_k^{\mathrm{in}}/8$ of $\beta_*$ in the metric.
Such a choice exists because the initial regret, inverse global gap,
metric conversions and inverse input radius are bounded by fixed
powers of $\mathfrak A_k$. Put
$\varepsilon_n=\sum_{t=1}^T\nu_{n,t}$. The preceding recursion and
metric conversion give, with a common sufficiently large fixed $p$,
\[
 e_{n,J_k}\le r_k^{\mathrm{in}}/8
       +\mathfrak A_k^p(J_k/T+\varepsilon_n),\qquad
 \max_{t\le J_k}\mathfrak d_{n,t}
             \le \mathfrak A_k^p(T^{-1}+\varepsilon_n).
\]
The second inequality follows from \eqref{supp:eq:alg-debt-noisy}
and the global diameter bound, absorbing $J_k$ into the fixed power.
Consequently, the sufficient burn-in size condition is
\[
 \mathfrak A_k^p(J_k/T+\varepsilon_n)
       <\min\{r_k^{\mathrm{in}}/8,d_k^{\mathrm{stop}}/2\}.
\]
It implies $e_{n,J_k}\le r_k^{\mathrm{in}}/4$ and keeps the
reference error strictly below $d_k^{\mathrm{stop}}$ throughout
burn-in. If a first exit occurred there, the cutoff calculation
would apply to its incoming state and these bounds would also
keep its outgoing state within the required margins. The additional
inequality in \eqref{supp:eq:bootstrap-size-check} below enforces
this burn-in condition for every allowed $T$ and total noise.

\textbf{Local recursion.}
At every reached local stage,
\[
 \begin{aligned}
 e_{n,t}&\le(1-\gamma_k)e_{n,t-1}
                         +\mathfrak A_k\mathfrak d_{n,t-1}+\mathfrak A_k\nu_{n,t},\qquad t\le T,\\
 \mathfrak d_{n,t}&\le\mathfrak d_{n,t-1}
                         +\mathfrak A_k\omega_{n,t}e_{n,t-1}+\mathfrak A_k\nu_{n,t},\qquad t<T.
 \end{aligned}
\]
Let $\overline\omega_n=\max_{t<T}\omega_{n,t}\le C/T$.
For $J=J_k$ and $\rho_{\mathrm{loc},k}=1-\gamma_k$, the first recursion gives
\[
 e_{n,s}\le\rho_{\mathrm{loc},k}^{s-J}e_{n,J}
       +\mathfrak A_k\sum_{j=J+1}^s
               \rho_{\mathrm{loc},k}^{s-j}(\mathfrak d_{n,j-1}+\nu_{n,j}).
\]
Summing over $J\le s<t$ and exchanging the finite sums yields
\[
 \sum_{s=J}^{t-1}e_{n,s}
 \le\frac{e_{n,J}}{\gamma_k}
    +\frac{\mathfrak A_k}{\gamma_k}
           \sum_{j=J}^{t-2}\mathfrak d_{n,j}
    +\frac{\mathfrak A_k}{\gamma_k}\varepsilon_n.
\]
Meanwhile the second recursion, summed over stages, is
\[
 \mathfrak d_{n,t}\le\mathfrak d_{n,J}
       +\mathfrak A_k\overline\omega_n
                    \sum_{s=J}^{t-1}e_{n,s}
       +\mathfrak A_k\varepsilon_n.
\]
Summing the geometric convolution in the first inequality and
inserting it into the second gives
\[
 \mathfrak d_{n,t}\le\mathfrak d_{n,J_k}
       +\frac{\mathfrak A_k\overline\omega_n}{\gamma_k}e_{n,J_k}
       +\mathfrak A_k^p\varepsilon_n
       +\frac{\mathfrak A_k^2\overline\omega_n}{\gamma_k}
                       \sum_{j=J_k}^{t-1}\mathfrak d_{n,j}.
\]
By the discrete Gronwall inequality, a recurrence
$d_t\le a+c\sum_{j=J}^{t-1}d_j$ satisfies
$d_t\le a(1+c)^{t-J}\le ae^{c(t-J)}$. Here
$\gamma_k^{-1}\le\mathfrak A_k^p$ and
\[
 cT=\frac{\mathfrak A_k^2\overline\omega_nT}{\gamma_k}
        \le C\mathfrak A_k^2/\gamma_k\le\mathfrak A_k^p.
\]
Therefore, with $H_k=\mathfrak A_k^p\exp(\mathfrak A_k^p)$ for a
sufficiently large fixed $p$, we obtain
\[
 \max_{J_k\le t<T}\mathfrak d_{n,t}\le H_k(T^{-1}+\varepsilon_n),\qquad
 e_{n,t}\le(1-\gamma_k)^{t-J_k}e_{n,J_k}
                              +H_k(T^{-1}+\varepsilon_n).
\]
All preceding estimates are applied only to the stopped path.
Choose $C_3$ in $\Gamma_k=\exp\{\exp(C_3(k+1)^3)\}$ so that
\begin{equation}
 \begin{gathered}
 \Gamma_k^2\ge2J_k,\qquad
 C\Gamma_k^{-2}\le \mathfrak A_k^{-3}/4,\qquad
 \Gamma_k^{-2}\le\nu_k^{\mathrm{stop}},\\
 \mathfrak A_k^p(J_k+1)\Gamma_k^{-2}
       <\min\{r_k^{\mathrm{in}}/8,d_k^{\mathrm{stop}}/2\},\\
 2H_k\Gamma_k^{-2}
 <\min\{d_k^{\mathrm{stop}}/2,r_k^{\mathrm{in}}/4\}.
 \end{gathered}
 \label{supp:eq:bootstrap-size-check}
\end{equation}
This is possible because all inverse radii and margins are bounded
by fixed powers of $\mathfrak A_k$, and $H_k\le\exp(\mathfrak A_k^{p+1})$ after
fixing a sufficiently large $p$ before this choice. For $T\ge\Gamma_k^2$
and $\varepsilon_n\le\Gamma_k^{-2}$, the added burn-in inequality gives
\[
 \mathfrak A_k^p(J_k/T+\varepsilon_n)
 \le\mathfrak A_k^p(J_k+1)\Gamma_k^{-2}
 <\min\{r_k^{\mathrm{in}}/8,d_k^{\mathrm{stop}}/2\}.
\]
Thus the burn-in reaches the local input region without an exit.
The local bounds then keep
$\mathfrak d_{n,t}<d_k^{\mathrm{stop}}$ and $e_{n,t}<r_k^{\mathrm{in}}$.
The cutoff calculation keeps $a_{n,t}<\mathfrak a_k$.
Each reached stage therefore remains in
\eqref{supp:eq:bootstrap-region}, closing the first-exit induction.

The final refit uses live mass $\lambda_{n,T-1}\ge\eta$ and satisfies
the same local inequality. There is no nonterminal block of this
size in $\overline\omega_n$. For $T\ge\Gamma_k^2$, the decaying
term $(1-\gamma_k)^{T-J_k}e_{n,J_k}$ is bounded by $\Gamma_k/T$,
after increasing $C_3$ once in its initial choice.
For example, $T\ge2J_k$ gives
\[
 T(1-\gamma_k)^{T-J_k}e_{n,J_k}
 \le T e^{-\gamma_kT/2}e_{n,J_k}
 \le\frac{2e_{n,J_k}}{e\gamma_k},
\]
and the final constant is bounded by a fixed power of
$\mathfrak A_k$, hence absorbed into $\Gamma_k$. Metric conversion
and the bound on $H_k$ give \eqref{supp:eq:alg-terminal-bound}.
Finally, for feasible $\theta,\beta$,
\[
 \|f_\theta-f_\beta\|_1
 \le A_\theta^{-1}\int\phi(x)|r_\theta(x) - r_\beta(x)| \: dx
       +A_\theta^{-1}|A_\theta-A_\beta|
 \le2\|\theta-\beta\|_2.
\]
The first inequality follows by writing
$r_\theta/A_\theta-r_\beta/A_\beta
=(r_\theta-r_\beta)/A_\theta
+r_\beta(A_\theta^{-1}-A_\beta^{-1})$ and using
$\int\phi r_\beta=A_\beta$. The second uses
$A_\theta\ge1$, $|A_\theta-A_\beta|\le\|\theta-\beta\|_2$, and
Hermite orthonormality with Cauchy--Schwarz:
\[
 \int\phi(x)|r_\theta(x) - r_\beta(x)| \: dx
 \le\|r_\theta-r_\beta\|_{L^2(\phi)}=\|\theta-\beta\|_2.
\]
The terminal retention occurs after the last refit, so it leaves the
proved density bound unchanged.

For the population assertion, apply the same deterministic estimates to
$\beta_{\mathrm{pop}}^{(t)}$ and the population states, using exact retained
block measures, cumulative fractions $\xi_t=(1-\eta)t/(T-1)$ for $t<T$
and $\xi_T=1$, and zero simulation discrepancies. The same bounds give
$\|\beta_{\mathrm{pop}}^{(T)}-\beta_*\|_2\le\Gamma_k/T$ and hence
$\|f_{k,\mathrm{pop}}^{(T)}-p_*\|_1\le2\Gamma_k/T$.
\end{proof}

\section{Technical lemmas of Theorem 3}
\label{supp:sec:lemmas-theorem-3}

For $j=0,1$, write
$\mathbb G^o_{j,n}(t)=n_{j,n}^{-1}\sum_{i\in\mathcal H_{j,n}}
\mathbf1\{p^o_{n,i}\le t\}$ and define
$\widehat{\mathbb G}_{j,n}$ using the fitted values. Let
$\mathbb F_{1,n}$ be the signal empirical CDF and put
$z_{n,U}(t)=\overline F_{0,n,U}^{-1}(1-t)$ for $0<t<1$.

\begin{lemma}[Compatibility with a moving retained center]
\label{supp:lem:compatibility}
Under Conditions~\ref{supp:ass:factor}--\ref{supp:ass:center}, for
almost every $U=u$ and every fixed $x<-b_{\max}(u)$,
\begin{equation}
 \mathbb F_{1,n}(x)\xrightarrow{\Pp (\cdot \mid U=u)}0,\qquad
 \frac1{n_{1,n}}\sum_{i\in\mathcal H_{1,n}}\Pp (Z_{n,i}\le x \mid U=u)\to0.
 \label{supp:eq:compatibility-left}
\end{equation}
\end{lemma}

\begin{lemma}[Signal fluctuations, equicontinuity]
\label{supp:lem:signal-law}
Under Condition~\ref{supp:ass:signal},
\begin{equation}
 \overline G_{1,n,U}(t)=\frac1{n_{1,n}}\sum_{i\in\mathcal H_{1,n}}
 \left[1-K\!\left(
 \frac{z_{n,U}(t)-\mu_{n,i}-\ell_{n,i}^{\mathsf T}U}{\sigma_{n,i}}
 \right)\right].
 \label{supp:eq:signal-row-p}
\end{equation}
Also, the average absolute signal copula correlation satisfies
\begin{equation}
 D_{1,n}:=\frac1{n_{1,n}^2}\sum_{i,j\in\mathcal H_{1,n}}|(R_{1,n}^{V})_{ij}|
 \le\frac{\|R_{1,n}^{V}\|_F}{n_{1,n}}
 \le\sqrt{\lambda_{\max}(R_{1,n}^{V})/n_{1,n}}\longrightarrow0.
 \label{supp:eq:signal-correlation}
\end{equation}
Under Condition~\ref{supp:ass:signal} and the nondegenerate null model,
\begin{equation}
 \E\big[\|\mathbb G_{1,n}^o-\overline G_{1,n,U}\|_\infty \mid U \big]
       \le C D_{1,n}^{1/3}\longrightarrow0.
 \label{supp:eq:signal-fluctuation}
\end{equation}
For each fixed $u$ and compact $J\subset(0,1)$, the functions
$\overline G_{1,n,u}$ are equicontinuous on $J$: a common modulus
$\omega_{u,J}(h)\to0$ as $h\downarrow0$ bounds all their increments.
\end{lemma}

\begin{lemma}[Pooled calibration]
\label{supp:lem:null-calibration}
Under the null model,
\begin{equation}
 \frac1{n_{0,n}}\sum_{i\in\mathcal H_{0,n}}
                \Pp (p_{n,i}^o\le t \mid U)=t,\qquad 0\le t\le1,
 \label{supp:eq:finite-average-calibration}
\end{equation}
and pathwise
\begin{equation}
 \|\mathbb G_{0,n}^o-\operatorname{id}\|_\infty
 \le\|\mathbb F_{0,n}-\overline F_{0,n,U}\|_\infty=\epsilon_{0,n}.
 \label{supp:eq:null-p-calibration}
\end{equation}
Under Conditions~\ref{supp:ass:factor}--\ref{supp:ass:residual},
$\epsilon_{0,n}\to0$ in conditional probability by
Lemma~\ref{supp:lem:cdf-stat}.
\end{lemma}

\begin{lemma}[Pathwise CDF transfer]
\label{supp:lem:testing-transfer}
Let $\epsilon_{\mathrm{fit},n}=\|\widehat F_n-\overline F_{0,n,U}\|_\infty$.
For every realization,
\begin{equation}
 \max_i|\widehat p_{n,i}-p_{n,i}^o|\le\epsilon_{\mathrm{fit},n},\qquad
 \mathbb G_{j,n}^o(t-\epsilon_{\mathrm{fit},n})
 \le\widehat{\mathbb G}_{j,n}(t)
 \le\mathbb G_{j,n}^o(t+\epsilon_{\mathrm{fit},n}),\quad j=0,1,
 \label{supp:eq:testing-sandwich}
\end{equation}
with the usual empirical-CDF extensions outside $[0,1]$.  In particular,
\begin{equation}
 \|\widehat{\mathbb G}_{0,n}-\operatorname{id}\|_\infty
                  \le\epsilon_{0,n}+\epsilon_{\mathrm{fit},n}.
 \label{supp:eq:fitted-null-calibration}
\end{equation}
If $\epsilon_{\mathrm{fit},n}\to0$ in $\Pp (\cdot \mid U)$-probability and
Lemma~\ref{supp:lem:signal-law} applies, then for every compact
$J\subset(0,1)$,
\[
 \sup_{t\in J}|\widehat{\mathbb G}_{1,n}(t)
                         -\overline G_{1,n,U}(t)|\xrightarrow{\Pp (\cdot \mid U)}0.
\]
\end{lemma}

\begin{lemma}[Separated-crossing BH stability and rank preservation]
\label{supp:lem:local-bh}
Fix $u$ and continuous population CDFs $\mathsf H_{n,u}$ whose last
crossings are bounded away from zero and $\alpha$. Suppose
Condition~\ref{supp:ass:crossing} holds at this factor realization.
If empirical CDFs $\mathbb G_n$ of vectors in $[0,1]^n$ obey
\[
 \sup_{t\in J}|\mathbb G_n(t)-\mathsf H_{n,u}(t)|\to0
 \quad\text{for every compact }J\subset(0,1),
\]
then their BH thresholds differ from $t_{n,u}^\star$ by $o(1)$.
The implication also holds in probability.

For the fitted and oracle vectors in the main text, with
$\widehat R_n=|\widehat{\mathcal R}_n|$ and $R_n^o=|\mathcal R_n^o|$,
\begin{equation}
 \frac{|\widehat{\mathcal R}_n\triangle\mathcal R_n^o|}{n}
 =\frac{|\widehat R_n-R_n^o|}{n}
 =\frac{|\widehat t_n-t_n^o|}{\alpha}.
 \label{supp:eq:rank-set-identity}
\end{equation}
\end{lemma}

\section{Proof of Lemmas S15--S19}
\label{supp:sec:proofs-lemmas-theorem-3}

\subsection[Proof of Lemma S15]{Proof of Lemma~\ref{supp:lem:compatibility}}
\label{supp:sec:proof-lemma-s15}

\begin{proof}
Fix a factor realization $u$ where the conditions hold.
For $x<-b_{\max}(u)$, every signal at most $x$ is retained when
$b_n>x$. Lemma~\ref{supp:lem:cdf-stat} makes this event have
probability tending to one. Therefore
\[
 \mathbb F_{1,n}(x)
 \le(n/n_{1,n})\chi_n+\ind\{b_n\le x\}\xrightarrow{\Pp (\cdot \mid U=u)}0.
\]
For every $\epsilon>0$, boundedness gives
\[
 0\le\E[\mathbb F_{1,n}(x) \mid U=u ]
 \le\epsilon+\Pp \{\mathbb F_{1,n}(x)>\epsilon \mid U=u\}\longrightarrow\epsilon.
\]
Letting $\epsilon\downarrow0$ shows that this expectation tends to
zero. Linearity of expectation identifies it as $n_{1,n}^{-1}\sum_i\Pp (Z_{n,i}\le x \mid U=u)$; no independence is needed.
\end{proof}

\subsection[Proof of Lemma S16]{Proof of Lemma~\ref{supp:lem:signal-law}}
\label{supp:sec:proof-lemma-s16}

\begin{proof}
The signal margins are continuous and the moving null CDF is strictly
increasing, so \eqref{supp:eq:signal-row-p} is exactly
$\overline G_{1,n,u}(t)=\E [\mathbb G_{1,n}^o(t) \mid U=u ]$.
To verify the spectral bound, put $m=n_{1,n}$ and
$R=R_{1,n}^{V}$. Cauchy--Schwarz and positive semidefiniteness give
\[
 \begin{aligned}
 D_{1,n}&\le m^{-1}\|R\|_F,\\
 \|R\|_F^2
 &=\sum_{j=1}^m\lambda_j(R)^2
 \le\lambda_{\max}(R)\sum_{j=1}^m\lambda_j(R)
 =m\lambda_{\max}(R),
 \end{aligned}
\]
since the correlation matrix has trace $m$. This proves
\eqref{supp:eq:signal-correlation} from
\eqref{supp:eq:signal-spectrum}.

Each indicator is a measurable function of one standard Gaussian
coordinate $V_{n,i}$.  The Hermite covariance inequality from the
proof of Lemma~\ref{supp:lem:cdf-stat} gives
\[
 \operatorname{Var}_u\{\mathbb G_{1,n}^o(t)\}\le D_{1,n}/4.
\]
The average sample-size-specific CDF is continuous, with values zero and one at
the endpoints.  A probability quantile grid, monotonicity, Chebyshev,
and a union bound give
\[
 \Pp \{\|\mathbb G_{1,n}^o-\overline G_{1,n,u}\|_\infty
                    >2\epsilon \mid U=u\}\le C D_{1,n}\epsilon^{-3}.
\]
Integrating the bound truncated at one proves
\eqref{supp:eq:signal-fluctuation}.  Equation~\eqref{supp:eq:signal-correlation} supplies
its vanishing right side.

A continuous proper CDF $K$ is uniformly continuous on $\R$. Let
$\omega_K$ be its modulus.  On a compact interval $J\subset(0,1)$,
Lemma~\ref{supp:lem:target-stability} gives
$|z_{n,u}(s)-z_{n,u}(t)|\le C_{u,J}|s-t|$ uniformly in $n$.
Every summand of \eqref{supp:eq:signal-row-p} therefore has modulus at most
$\omega_K(C_{u,J}|s-t|/\sigma_*)$, independently of $\mu_{n,i}$.
This proves equicontinuity. Namely, with
$\omega_K(h)=\sup_{|x-y|\le h}|K(x)-K(y)|$,
\[
 |\overline G_{1,n,u}(s)-\overline G_{1,n,u}(t)|
 \le\omega_K(C_{u,J}|s-t|/\sigma_*)=:\omega_{u,J}(|s-t|).
\]
The bound is independent of the signal locations, even when some
of them diverge with $n$.
\end{proof}

\subsection[Proof of Lemma S17]{Proof of Lemma~\ref{supp:lem:null-calibration}}
\label{supp:sec:proof-lemma-s17}

\begin{proof}
For $0<t<1$, let $z=z_{n,U}(t)$. Strict monotonicity of the
pooled null CDF gives the exact event identity
\[
 \{p_{n,i}^o\le t\}
 =\{1-\overline F_{0,n,U}(Z_{n,i})\le t\}
 =\{Z_{n,i}\ge z\}.
\]
Every conditional null margin is continuous, so averaging yields
\[
 \frac1{n_{0,n}}\sum_i\Pp (Z_{n,i}\ge z \mid U)
       =1-\overline F_{0,n,U}(z)=t.
\]
This is an average over the heterogeneous conditional null margins;
it does not require each individual oracle $p$-value to be uniform.

Pathwise,
$\mathbb G_{0,n}^o(t)=1-\mathbb F_{0,n}(z-)$.
Since $\overline F_{0,n,U}$ is continuous, taking a left limit in the
supremum definition gives
$|\mathbb F_{0,n}(z-)-\overline F_{0,n,U}(z)|\le\epsilon_{0,n}$.
Consequently
\[
 |\mathbb G_{0,n}^o(t)-t|
 =|1-\mathbb F_{0,n}(z-)-\{1-\overline F_{0,n,U}(z)\}|
 \le\epsilon_{0,n}.
\]
The left limit correctly includes any tied observations at $z$ in
the right-tail rejection event.
This proves \eqref{supp:eq:null-p-calibration}.
Finite observations and a positive pooled density give the endpoint
identities.  Lemma~\ref{supp:lem:cdf-stat} proves convergence.
\end{proof}

\subsection[Proof of Lemma S18]{Proof of Lemma~\ref{supp:lem:testing-transfer}}
\label{supp:sec:proof-lemma-s18}

\begin{proof}
Subtract the fitted and oracle $p$-values coordinatewise.
For each $i$, on the same realization of both the data and the fit,
\[
 |\widehat p_{n,i}-p_{n,i}^o|
 =|\widehat F_n(Z_{n,i})-\overline F_{0,n,U}(Z_{n,i})|
 \le\epsilon_{\mathrm{fit},n}.
\]
The supremum definition gives the maximum error bound and the
inclusions
\[
 \{p_{n,i}^o\le t-\epsilon_{\mathrm{fit},n}\}\subseteq
 \{\widehat p_{n,i}\le t\}\subseteq
 \{p_{n,i}^o\le t+\epsilon_{\mathrm{fit},n}\}.
\]
These prove the sandwich.  The null calibration lemma gives
\eqref{supp:eq:fitted-null-calibration}, using the clipped identity CDF at
probability endpoints. For instance, writing
$\epsilon=\epsilon_{\mathrm{fit},n}$, the upper side is
\[
 \widehat{\mathbb G}_{0,n}(t)
 \le\mathbb G_{0,n}^o(t+\epsilon)
 \le\min(t+\epsilon,1)+\epsilon_{0,n}
 \le t+\epsilon+\epsilon_{0,n}.
\]
The lower side follows from
$\max(t-\epsilon,0)-\epsilon_{0,n}\ge t-\epsilon-\epsilon_{0,n}$.
The argument is pathwise, so the fit need not be independent of the
observations to which it is applied.
For compact $J\subset(0,1)$, choose a larger compact $J'$ containing
a neighborhood of $J$.  When $\epsilon_{\mathrm{fit},n}$ is smaller than that width,
the fitted signal error on $J$ is at most
\[
 \|\mathbb G_{1,n}^o-\overline G_{1,n,u}\|_{\infty,J'}
                 +\omega_{u,J'}(\epsilon_{\mathrm{fit},n}).
\]
The first term vanishes by the signal fluctuation bound, and the
second by its uniform modulus. The event that
$\epsilon_{\mathrm{fit},n}$ exceeds the chosen neighborhood width
has probability tending to zero. Thus the local bound applies with
conditional probability tending to one; composition of the
modulus with an $o_{\Pp (\cdot \mid U=u)}(1)$ argument is also $o_{\Pp (\cdot \mid U=u)}(1)$.
\end{proof}

\subsection[Proof of Lemma S19]{Proof of Lemma~\ref{supp:lem:local-bh}}
\label{supp:sec:proof-lemma-s19}

\begin{proof}
Fix a sufficiently small $\epsilon>0$. The crossing bounds put all
required intervals in a common compact subset of $(0,1)$.
If the local empirical error is less than half the negative margin,
no BH-feasible grid point at or above $t_{n,u}^\star+\epsilon$
exists. Indeed, a positive BH threshold belongs to the grid
$\{\alpha r/n:1\le r\le n\}$ and is feasible precisely when the
empirical CDF there is at least $r/n$. On the right-margin interval,
\[
 \mathbb G_n(t)-t/\alpha
 \le\{\mathsf H_{n,u}(t)-t/\alpha\}
       +|\mathbb G_n(t)-\mathsf H_{n,u}(t)|
 \le-m_{u,\epsilon}/2<0.
\] Choose $s_n\in[t_{n,u}^\star-\epsilon,t_{n,u}^\star]$
with population margin at least half the positive margin. For
sufficiently small empirical error, $\mathbb G_n(s_n)>s_n/\alpha$.
Let $r_n=\#\{i:p_{n,i}\le s_n\}$. Then
$r_n/n=\mathbb G_n(s_n)>s_n/\alpha$, so
$p_{n,(r_n)}\le s_n<\alpha r_n/n$. The BH threshold is at least
$\alpha r_n/n>s_n$. These two bounds prove convergence to the
moving crossing and also its probability version.
In particular, on the event where the empirical error is smaller
than $m_{u,\epsilon}/4$, the selected $s_n$ has a strictly positive
empirical margin and
\[
 t_{n,u}^\star-\epsilon\le s_n<t_n<t_{n,u}^\star+\epsilon.
\]
The deterministic implication gives the probability version by
bounding the failure probability by that of the empirical-error
event. No interpolation of the empirical CDF or assumption about
absence of ties is used.

Finally the fitted and pooled null CDFs are strictly increasing.
Both vectors order the observations in decreasing order of $Z_{n,i}$.
BH rejection sets are prefixes of this common ordering, and tied
observations are rejected together. The latter follows from
maximality of the BH rank: if a tied coordinate after rank $R$ also
satisfied $p_i\le\alpha R/n$, its later rank would satisfy the BH
inequality as well, contradicting the definition of $R$. The sets are nested, so their
symmetric difference has size $|\widehat R_n-R_n^o|$.
The identity $t=\alpha R/n$ proves
\eqref{supp:eq:rank-set-identity}.
\end{proof}